\documentclass[twocolumn, astrosymb]{aastex631}
\usepackage{longtable}
\newcommand{\eg}{$e.g.,$  }
\newcommand{\ie}{$i.e.,$  }
\newcommand{\kms}{km\,s$^{-1}$ }
\newcommand{\kmsno}{km\,s$^{-1}$}

\newcommand{\teff}{T$_{\rm eff}$ }
\newcommand{\logg}{log\,$g$ }

\accepted{ApJ}
\shorttitle{C-19 GHOST Story}
\shortauthors{Venn et al.}
\begin{document}

\title{GHOST Commissioning Science Results. V: GHOST spectral analysis of the outermost stars in the extremely metal-poor stream C-19}
%% you may use \\ to force a line break 
%% you can include a footnote in the title.

\correspondingauthor{Kim Venn}
\email{kvenn@uvic.ca}

\author[0000-0003-4134-2042]{Kim A. Venn}
\affiliation{Department of Physics and Astronomy, University of Victoria, 3800 Finnerty Road, Victoria, BC, V8P 5C2, Canada}

\author[0000-0002-8129-5415]{Zhen Yuan}
\affiliation{School of Astronomy and Space Science, Nanjing University, Nanjing, Jiangsu, 210093, China}
\affiliation{Key Laboratory of Modern Astronomy and Astrophysics,Nanjing University, Ministry of Education, Nanjing, Jiangsu, 210093, China}
\affiliation{Universit\'e de Strasbourg, CNRS, Observatoire astronomique de Strasbourg, UMR 7550, F-67000 Strasbourg, France}

\author[0000-0002-1349-202X]{Nicolas F. Martin}
\affiliation{Universit\'e de Strasbourg, CNRS, Observatoire astronomique de Strasbourg, UMR 7550, F-67000 Strasbourg, France}
\affiliation{Max-Planck-Institut f\"{u}r Astronomie, K\"{o}nigstuhl 17, D-69117 Heidelberg, Germany}

\author[0009-0001-1099-1384]{Daria Zaremba}
\affiliation{Department of Physics and Astronomy, University of Victoria, 3800 Finnerty Road, Victoria BC V8P 5C2, Canada}
\affiliation{Department of Engineering, Kyiv School of Economics, 3 Mykoly Shpaka St., Kyiv, 03113, Ukraine}
%\affiliation{Department of Physics and Engineering, Kyiv School of Economics, Kyiv, Ukraine}

\author[0009-0001-1871-2001]{Anya Dovgal}
\affiliation{Department of Physics and Astronomy, University of Victoria, 3800 Finnerty Road, Victoria BC V8P 5C2, Canada}

\author[0000-0002-9252-6899]{Felipe Gran}
\affiliation{Observatoire de la Côte d'Azur, CNRS, Université Côte d'Azur, Laboratoire Lagrange, Nice, France}
\affiliation{Gemini Observatory/NSF's NOIRLab, Casilla 603, La Serena, Chile}
\affiliation{Instituto de Astrof\'isica, Pontificia Universidad Cat\'olica de Chile, Av.\ Vicu\~na Mackenna 4860, 782-0436 Macul, Santiago, Chile}

\author[0000-0003-2969-2445]{Christian R. Hayes}
\affiliation{Space Telescope Science Institute, 3700 San Martin Drive, Baltimore, MD, 21218, USA}
\affiliation{NRC Herzberg Astronomy \& Astrophysics, 5071 West Saanich Road, Victoria, BC V9E 2E7, Canada}

\author[0000-0002-3795-0005]{Vanessa Hill}
\affiliation{Observatoire de la Côte d'Azur, CNRS, Université Côte d'Azur, Laboratoire Lagrange, Nice, France}

\author[0000-0002-8077-4617]{Tadafumi Matsuno}
\affiliation{Astronomisches Rechen-Institut, Zentrum f{\"u}r Astronomie der Univ.\
%Universit{\"a}t 
Heidelberg, M{\"o}nchhofstrasse 12-14, 69120, Heidelberg, Germany}

\author[0000-0003-4666-6564]{Alan W. McConnachie}
\affiliation{NRC Herzberg Astronomy \& Astrophysics, 5071 West Saanich Road, Victoria, BC V9E 2E7, Canada}
\affiliation{Department of Physics and Astronomy, University of Victoria, 3800 Finnerty Road, Victoria BC V8P 5C2, Canada}

\author[0000-0002-1715-1257]{Madeleine McKenzie}
\affiliation{Observatories of the Carnegie Institution for Science, 813 Santa Barbara St., Pasadena, CA 91101}

\author[0000-0002-5781-9645]{Martin Montelius}
\affiliation{Kapteyn Astronomical Institute, University of Groningen, Landleven 12, NL-9747, AD Groningen, The Netherlands}

\author[0000-0003-3862-5076]{Julio F. Navarro}
\affiliation{Department of Physics and Astronomy, University of Victoria, 3800 Finnerty Road, Victoria BC V8P 5C2, Canada}
\affiliation{KU-KIST Green School, Korea University}

\author[0000-0002-3182-3574]{Federico Sestito}
\affiliation{Department of Physics and Astronomy, University of Victoria, 3800 Finnerty Road, Victoria BC V8P 5C2, Canada}
\affiliation{Centre for Astrophysics Research, University of Hertfordshire, Hatfield, AL10 9AB, UK}

\author[0000-0003-4501-103X]{Else Starkenburg}
\affiliation{Kapteyn Astronomical Institute, University of Groningen, Landleven 12, NL-9747, AD Groningen, The Netherlands}

\author[0000-0002-0544-2217]{Anke Ardern-Arentsen}
\affiliation{Institute of Astronomy, University of Cambridge, Madingley Road, Cambridge, CB3 0HA, UK}

\author[0000-0002-6551-4294]{Giuseppina Battaglia}
\affiliation{Instituto de Astrofísica de Canarias, Calle Vía Láctea s/n, 38206 La Laguna, Santa Cruz de Tenerife, Spain }
\affiliation{Universidad de La Laguna, Avda. Astrofísico Francisco Sánchez 38205 La Laguna, Santa Cruz de Tenerife, Spain }

\author[0000-0002-7667-0081]{Raymond G. Carlberg}
\affiliation{Department of Astronomy \& Astrophysics, University of Toronto, Toronto ON, M5S 3H4 Canda}

\author[0000-0003-2239-7988]{Sebastien Fabbro}
\affiliation{NRC Herzberg Astronomy \& Astrophysics, 5071 West Saanich Road, Victoria, BC V9E 2E7, Canada}

\author[0000-0001-9256-5516]{Morgan Fouesneau}
\affiliation{Max-Planck-Institut f\"{u}r Astronomie, K\"{o}nigstuhl 17, D-69117 Heidelberg, Germany}

\author[0000-0002-3292-9709]{Rodrigo A. Ibata}
\affiliation{Universit\'e de Strasbourg, CNRS, Observatoire astronomique de Strasbourg, UMR 7550, F-67000 Strasbourg, France}

\author[0000-0002-6194-043X]{Michael J. Ireland}
\affiliation{Research School of Astronomy and Astrophysics, Australian National University, Canberra ACT 2601}

\author[0000-0002-9655-1063]{Pascale Jablonka}
\affiliation{Laboratory of Astrophysics, École Polytechnique Fédérale de Lausanne (EPFL), Observatoire de Sauverny, 1290 Versoix, Switzerland}
\affiliation{LIRA, Observatoire de Paris, Université PSL, CNRS, France}

\author[0000-0002-4350-7632]{Jaclyn Jensen}
\affiliation{Department of Physics and Astronomy, University of Victoria, 3800 Finnerty Road, Victoria BC V8P 5C2, Canada}
\affiliation{Department of Physics and Astronomy, Dartmouth College, 6127 Wilder Laboratory, Hanover, NH 03755, USA}

\author[0000-0002-4641-2532]{Venu Kalari}
\affiliation{Gemini Observatory/NSF's NOIRLab, Casilla 603, La Serena, Chile}

\author[0000-0002-4343-0487]{Chiaki Kobayashi}
\affiliation{Centre for Astrophysics Research, University of Hertfordshire, Hatfield, AL10 9AB, UK}

\author[0000-0002-1014-0635]{Georges Kordopatis}
\affiliation{Observatoire de la Côte d'Azur, CNRS, Université Côte d'Azur, Laboratoire Lagrange, Nice, France}

\author[0000-0002-4295-8773]{Carmela Lardo}
\affiliation{Dipartimento di Fisica e Astronomia, Università degli Studi di Bologna, Via Gobetti 93/2, 40129, Bologna, Italy}

\author[0000-0002-9225-5822]{Stephanie Monty}
\affiliation{New Mexico State University, Department of Astronomy, 1320 Frenger Mall, Las Cruces, NM 88003-8001, USA}

\author{John S. Pazder}
\affiliation{NRC Herzberg Astronomy \& Astrophysics, 5071 West Saanich Road, Victoria, BC V9E 2E7, Canada}
\affiliation{Department of Physics and Astronomy, University of Victoria, 3800 Finnerty Road, Victoria BC V8P 5C2, Canada}

\author[0000-0003-4479-1265]{Vinicius M. Placco}
\affiliation{NSF NOIRLab, Tucson, AZ 85719, USA}

\author[0000-0003-4945-0056]{Rubén Sánchez-Janssen}
\affiliation{Isaac Newton Group of Telescopes, 38700 Santa Cruz de La Palma, Canary Islands, Spain}

\author[0000-0001-8589-4055]{Chris Simpson}
\affiliation{Gemini Observatory/NSF NOIRLab, 670 N A'ohoku Pl  Hilo, HI 96720, USA}

\author[0000-0002-6946-8280]{Simon E. T. Smith}
\affiliation{Department of Physics and Astronomy, University of Victoria, 3800 Finnerty Road, Victoria BC V8P 5C2, Canada}

\author[0000-0002-7507-5985]{Akshara Viswanathan}
\affiliation{Department of Physics and Astronomy, University of Victoria, 3800 Finnerty Road, Victoria BC V8P 5C2, Canada}

\author[0000-0001-5412-869X]{Sara Vitali}
\affiliation{INAF - Osservatorio Astrofisico di Torino, Strada Osservatorio 20, I-10025 Pino Torinese (TO), Italy}
\affiliation{Instituto de Astrofísica de Canarias, Calle Vía Láctea s/n, 38206 La Laguna, Santa Cruz de Tenerife, Spain }
\affiliation{Universidad de La Laguna, Avda. Astrofísico Francisco Sánchez 38205 La Laguna, Santa Cruz de Tenerife, Spain }

\author{Fletcher Waller}
\affiliation{Department of Physics and Astronomy, University of Victoria, 3800 Finnerty Road, Victoria BC V8P 5C2, Canada} 

\author[0000-0001-8713-0366]{Long Wang}
\affiliation{School of Physics and Astronomy, Sun Yat-sen University, Daxue Road, Zhuhai, 519082, China}
\affiliation{CSST Science Center for the Guangdong-Hong Kong-Macau Greater Bay Area, Zhuhai, 519082, China}

\author{Zhen Wang}
\affiliation{School of Physics and Astronomy, Sun Yat-sen University, Daxue Road, Zhuhai, 519082, China}

\author[0000-0002-8808-4282]{Siyi Xu}
\affiliation{Gemini Observatory/NSF NOIRLab, 670 N A'ohoku Pl  Hilo, HI 96720, USA}

\author{Greg Burley}
\affiliation{NRC Herzberg Astronomy \& Astrophysics, 5071 West Saanich Road, Victoria, BC V9E 2E7, Canada}

\author[0000-0002-9020-5004]{Kristin Chiboucas}
\affiliation{Gemini Observatory/NSF NOIRLab, 670 N A'ohoku Pl  Hilo, HI 96720, USA}

\author[0000-0001-9796-2158]{Emily Deibert}
\affiliation{Department of Physics \& Astronomy, University of Waterloo, 200 University Avenue West, Waterloo, Ontario N2L 3G1, Canada}
\affiliation{Waterloo Centre for Astrophysics, University of Waterloo, Waterloo, Ontario N2L 3G1, Canada}

\author[0000-0001-9716-5335]{Ruben Diaz}
\affiliation{Gemini Observatory/NSF's NOIRLab, Casilla 603, La Serena, Chile}

\author[0009-0006-5048-0972]{Manuel Gomez-Jimenez}
\affiliation{Gemini Observatory/NSF's NOIRLab, Casilla 603, La Serena, Chile}

\author[0000-0002-6633-7891]{Kathleen Labrie}
\affiliation{Gemini Observatory/NSF NOIRLab, 670 N A'ohoku Pl  Hilo, HI 96720, USA}

\author[0000-0001-8205-9441]{Steven Margheim}
\affiliation{NSF–DOE Vera C. Rubin Observatory/NSF NOIRLab, Casilla 603, La Serena, Chile}

\author[0000-0001-5528-7801]{J. Gordon Robertson}
\affiliation{Australian Astronomical Optics, Macquarie University, 9 Wally's Walk, NSW 2109, Australia}
\affiliation{Sydney Institute for Astronomy, School of Physics, University of Sydney, NSW 2006, Australia}

\author[0000-0001-7518-1393]{Roque Ruiz-Carmona}
\affiliation{Gemini Observatory/NSF's NOIRLab, Casilla 603, La Serena, Chile}

\author[0000-0003-1033-4402]{Joanna Thomas-Osip}
\affiliation{Gemini Observatory/NSF's NOIRLab, Casilla 603, La Serena, Chile}

% \author[]{}
% \affiliation{}

\begin{abstract}

The C-19 stellar stream is the most metal-poor stellar structure yet identified.  
We present a detailed analysis of high-resolution spectroscopic observations for 14 members of the C-19 stream, spanning more than 100$^\circ$ on the sky. 
This includes observations for 7 new members observed with the Gemini/GHOST high-resolution spectrograph, some during its commissioning.  
Precision radial velocity measurements were used to ascertain their membership and to search for binary stars. 
Chemical abundances or upper limits for 16 elements, ranging from carbon to europium, are derived from model atmospheres.
The mean iron abundance from this dataset of 14 members of the C-19 stream is [Fe/H] = $-3.25 \pm0.05$ (1DNLTE; or $-3.38 \pm0.06$ 1DLTE).
The C-19 members have very similar chemical compositions (1DNLTE) to extremely metal-poor field stars in the Milky Way (MW) halo. 
The potassium (K) and 
neutron-capture elements (Sr, Ba) in the C-19 stars compare better to MW halo stars of similar metallicities than to stars in the nearby ultra faint dwarf galaxies.
Although earlier data suggested variations in some elemental abundances (C, Na, Mg), 
our analysis indicates that the observed scatter is consistent with measurement errors. Only aluminum shows a statistically significant dispersion. 
An examination of one- and two-component models of the 14 elements in the 14 members of C-19 in this analysis favours  a single stellar population model.
%
%Several of the spectra analysed in this study were taken as part of the GHOST commissioning observations, testing faint observation limits 
The spectra taken as part of GHOST commissioning were also used to test the faint observation limits 
(G $<17$) and binning effects with the single IFU observing mode.

\end{abstract}


\section{Introduction} 
\label{sec:intro}

From the measured positions and motions of nearly two billion stars in the Milky Way measured by the Gaia mission \citep{Gaia16, Gaia18, Gaia21, GaiaDR3}, the number of known stellar streams has increased by an order of magnitude from $\sim$10 to $\sim$100 \citep{Malhan2018, Ibata14, Ibata21, TLi2022, Bonaca2025NewA}. The precision of the Gaia data has further enabled detailed stream kinematics and orbits to be calculated, constraining the shape and mass of the Milky Way potential \citep{Errani2015, Bovy2016, Garavito-C2021}.
These stellar streams are the tidal remnants of very low-mass dwarf galaxies, globular clusters, or a combination of the two, many of them brought into the Milky Way (MW) halo by more massive progenitor galaxies \citep{Forbes2010, Leaman2013, Deason2015, Kruijssen19, Ibata2020, Vasiliev2020}. Thus, studies of the stellar populations in streams can also be used to identify their origins and the characteristics of their progenitor hosts, which contributed to building the Milky Way halo.
%, and which we can only otherwise glimpse as faint high redshift smudges with \textit{JWST/NIRSpec} \citep{MLi2023SMACS, Curti2024JWST}.

Amongst the many newly identified stellar streams, one stands out as quite unique: C-19.  This stellar stream was first identified by \cite{Martin2022} from a comparison of Gaia EDR3 data with the Pristine survey's 
\citep{Starkenburg2017b}
photometric metallicities.  C-19 appeared as a stream of faint (V$\sim$16.5) extremely metal-poor stars, stretching across $\sim10$ degrees on the sky, well below the Galactic plane (l = 100 to 106 deg, b = $-$40 to $-$27 deg).  This discovery was immediately followed up with high- and medium-resolution spectra taken with the Gemini GRACES and GTC OSIRIS spectrographs \citep{MartinVenn2022}.  
From 9 stars, the mean metallicity and dispersion were both found to be extremely low, [Fe/H]$=-3.34 \pm$0.06 with negligible dispersion.  Such a low metallicity and dispersion is consistent with a recently disrupted globular cluster; however, the mean metallicity is much lower than any globular cluster identified in the MW (157 GCs in the MW; \citealt{Harris10}) or in any other galaxy (1928 GCs in 28 galaxies; \citealt{Beasley19}).
Detailed chemical abundances for three C-19 member stars observed with Gemini/GRACES spectrograph further showed clear variations in [Na/Mg], which was interpreted as perhaps signaling the presence of multiple populations \cite[c.f.,][]{Bastian2018}.
Also, \citet{Yuan2022} found one new member of C-19 located $\sim30$ degrees away from the main body, 
with 
[Al/Mg] variations and [Ba/Fe] abundances consistent with stars in globular clusters.

\begin{table*}
\caption{C-19 stellar stream targets and two standard stars analysed in this paper and six C-19 members from the literature. Data from the Gaia DR3 database includes sourceID, RA, DEC, G, and BP-RP. Extinction values A$_G$(SFD) are from \citet{Schlafly11}.
%, which are $\leq$A$_G$(Gaia~DR3).  
Three heliocentric distances are calculated (see text, and table comments below).
}
\label{tab:targets}
\hspace{-2.3 cm}
\begin{tabular}{lccccccccc}
\toprule
Name      & Gaia DR3            &  RA    &   DEC   & Gaia     & Gaia & A$_{G}$ & DIST & DIST & DIST  \\
 &  sourceID & (J2016) & (J2016) & G & BP$-$RP & (SFD) & (isoch) & (astrom) & (orbit)  \\
  & & & & & &  & (kpc) & (kpc) & (kpc)  \\
\hline
\textbf{This Paper:} \\
C19A        & 2605574384366803968 &  346.2032242 & $-$10.6965325 & 15.88 & 1.05 & 0.10  & 
$23.6^{+1.2}_{-1.3}$ & $16.4^{+46.5}_{-0.7}$ &  16.8  \\
C19B        & 6559328209695612544 &  327.6864168 & $-$49.3800051 & 13.74 & 1.34 & 0.07 &
$19.9^{+0.5}_{-0.6}$ & $11.1^{+2.8}_{-1.1}$  & 17.0  \\
C19C        & 6600784780223506944 &  339.9024531 & $-$32.7936701 & 13.47 & 1.38 & 0.04 &
$18.6^{+0.1}_{-0.2}$ & $16.6^{+14.1}_{-1.8}$  & 17.0  \\
C19D        & 6594796290142997376 &  335.4728668 & $-$38.3530749 & 14.51 & 1.18 & 0.04 &
$20.9^{+0.8}_{-0.7}$ & $16.6^{+28.4}_{-1.4}$ & 17.0  \\
C19E        & 2641204161744171392 &  353.4703215 & $-$00.9700050 & 14.85 & 1.27 & 0.14 &
$25.4^{+0.9}_{-0.9}$ & $7.7^{+3.5}_{-0.8}$ & 18.0  \\
C19F   & 2658115921889849472 & 350.1429640 & +02.5838145 & 17.09 & 1.01 & 0.15  & $29.3^{+2.1}_{-2.0}$ & $8.9^{+35.9}_{-0.2}$ & 18.2  \\
C19G   & 2760807387346283648 & 351.3102187 & +07.9633947 & 16.39 & 1.09 & 0.25  & $24.3^{+1.6}_{-1.4}$ & $12.0^{+43.2}_{-0.3}$ & 18.0   \\
C19N & 2288313499629002624 & 298.7454500 & +74.6564189 & 14.08 & 1.43 & 0.39 & $17.8^{+0.6}_{-0.6}$ &  $16.7^{+9.4}_{-1.8}$ & 14.0   \\
HD122563  & 3723554268436602240 & 210.6318350 & +09.6857827 & 5.87 & 1.22 & 0.07  &
$0.32$ & $0.32$ & ...  \\
HD222925  & 6487799171512458624 &  356.3248257 & $-$61.9123391 & 8.85 & 0.82 & 0.06  & 
$0.45$ & $0.45$ & ...  \\
%HE2340-6036 & 6488378098744387968 & 355.9213860 & $-$60.3229441 & 12.54 & 1.19 & ... & $8.1^{+0.3}_{-0.3}$ & $5.6^{+0.5}_{-0.3}$ & ...     \\
\\
\textbf{Literature:} \\
%P3553224 
C19J & 2865251577418971392 & 355.3224059	& +27.5993569 & 14.19 & 1.37 & 0.24  & $20.7^{+0.6}_{-0.6}$ & $15.3^{+14.6}_{-1.6}$ &  18.0 \\
%P3547701 
C19K & 2868052548930201984 &  354.7701575 & +30.2509843 & 15.69 & 1.18 & 0.25  & $24.5^{+1.2}_{-1.2}$ & $16.1^{+45.8}_{-0.6}$  &  18.0  \\
%P3551327
C19L  & 2865368434887899008 &  355.1326831 & +27.9819596 & 15.39 & 1.25 & 0.39  & $20.7^{+0.9}_{-1.0}$ & $10.7^{+14.9}_{-1.0}$ &   18.0  \\
%P3552755 
C19M & 2865256628300500352 &  355.2755506 & +27.7483341 & 15.81 & 1.16 & 0.25  & $24.3^{+1.5}_{-1.1}$ & $14.2^{+39.4}_{-0.7}$ &  18.0  \\
%J2338+0902 
C19W & 2758373652717936640 & 354.5731253 & +09.0353924 & 14.65 & 1.38 & 0.36  & $21.9^{+0.7}_{-0.7}$ & $17.0^{+33.1}_{-1.3}$ & 18.0  \\
%P3549615   
C19Y & 2866151046649496832 &  354.9615142 & +28.4659616 & 14.28 & 1.58 & 0.56 & $21.3^{+0.1}_{-0.6}$ & $19.4^{+37.4}_{-1.5}$   & 18.0  \\
\\
\hline
\vspace{-1ex}
\end{tabular}
%\begin{tablenotes}
Distance calculations are described in Section~\ref{sec:dists}.  D(astrom) and D(isoch) astrometric distances are from the Bayesian analysis method developed by \citet{Sestito19}, with and without including metal-poor stellar isochrones from MIST/MESA models.  D(orbit) are distances from an orbit calculation per target (see text) assuming the core of C-19 is anchored at D= 18 kpc; uncertainties are $\Delta$D(orbit)$\approx^{+1.0}_{-0.5}$ kpc. 

%\end{tablenotes}
\end{table*}

On the other hand, the kinematic properties of C-19 
are similar to those expected from disrupting dwarf galaxies.
The C-19 stream width is $\sim180$ pc (at a heliocentric distance of 20 kpc) and has a velocity dispersion $\sigma_v \sim 6-8$ km\,s$^{-1}$ \citep{MartinVenn2022, Yuan2022, Yuan2025}, 
both substantially larger than the tidal debris expected for a disrupted globular cluster on C-19's orbit. 
N-body simulations of the tidal disruption of a 100 pc King-model stellar component embedded in a 20 \kms cuspy cold dark matter halo by
\citet{Errani2022} suggest that the C-19 progenitor is more consistent with a dark matter-dominated stellar system, i.e., a dwarf galaxy (total mass $\sim$ a few x $10^8$ M$_\odot$). 
\cite{Errani2022} proposed an unconventional model where the C-19 progenitor is a dark matter-dominated stellar system with an embedded GC, or a GC-like abundance pattern.

\begin{figure}
\begin{center}
    \setlength{\unitlength}{1cm}
    \includegraphics[width=\linewidth]{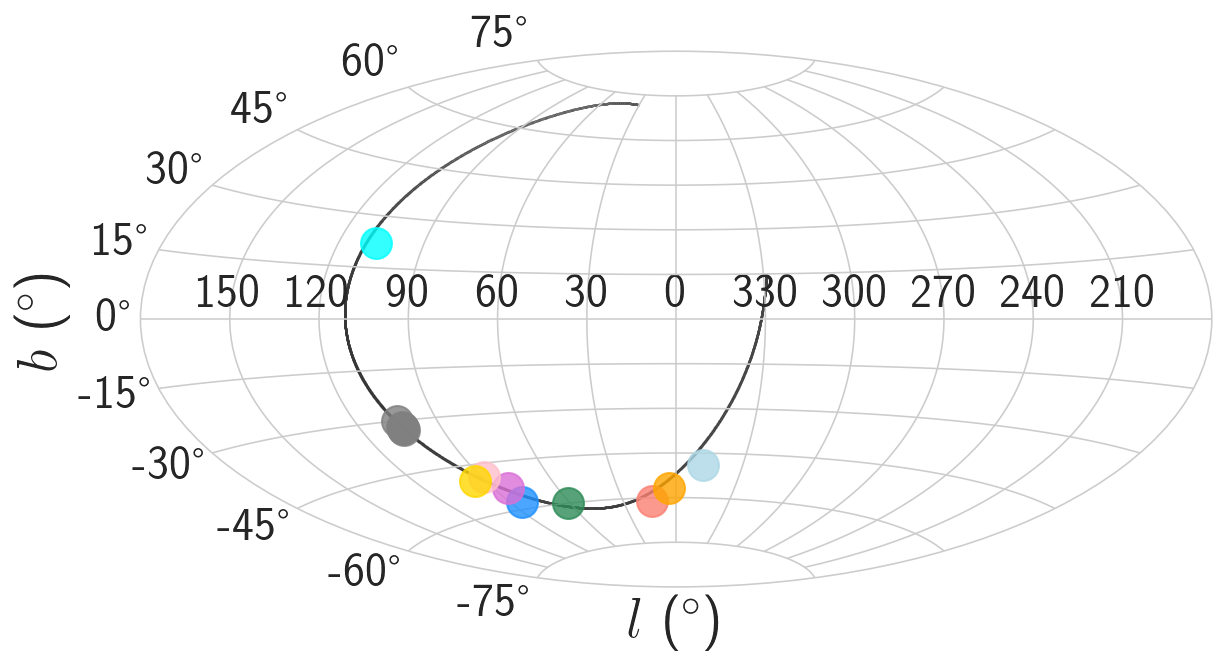}
    \includegraphics[width=\linewidth]{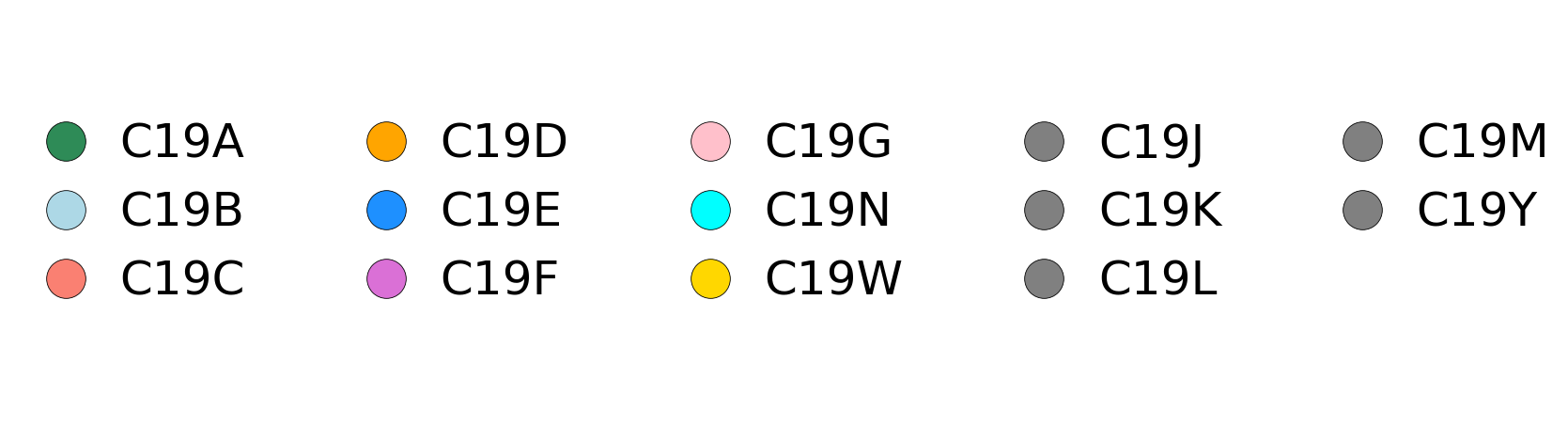}
    \vskip -3mm
    \caption{Galactic (l, b) coordinates of the 9 new C-19 stream members with detailed chemical abundances in this paper; see Table~\ref{tab:targets} for identification details.  
    %Stars from this paper are coloured points: C19A=seagreen, C19B=light blue, C19C=salmon, C19D=orange, C19E=dodgerblue, C19F=orchid, C19G=pink, C19J=grey, C19N=cyan, C19W=gold.
    An additional 5 stars with chemical analyses in the core of C-19 are shown in grey.  The black line is an orbit for C-19 as described in \cite{Yuan2025}.
}
    \label{fig:lb}
\end{center}
\end{figure}

\begin{figure}
\begin{center}
    \setlength{\unitlength}{1cm}
    \includegraphics[width=8cm]{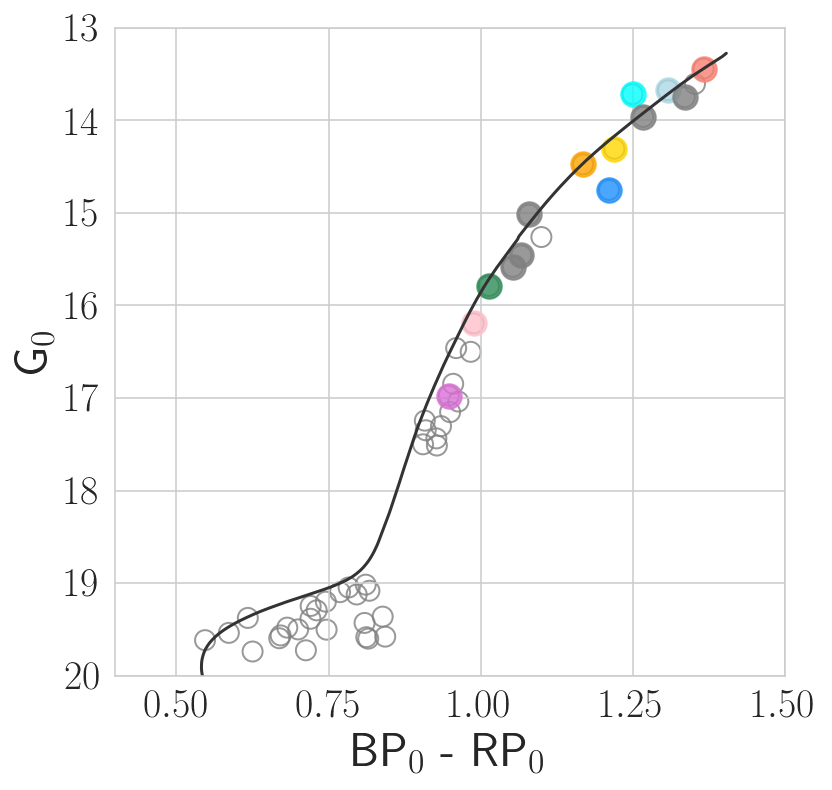}
    \caption{The Gaia DR3 photometry colour-magnitude diagram \citep{GaiaDR3} for members of the C-19 stellar stream. The filled circles are the same as those identified in Fig.~\ref{fig:lb}. Open circles are other likely members of C-19 from \citet{Yuan2025}. The black line is a BASTI-IAC isochrone \citep{Hidalgo2018, Pietrinferni2021} for an old (14 Gyr), extremely metal-poor ([Fe/H]$=-3.2$), 
    $\alpha$-enhanced stellar population, anchored at d=18 kpc.
}
    \label{fig:cmd}
\end{center}
\end{figure}

In this paper, we analyse the high-resolution spectra of 14 members (8 newly identified) in the outskirts of the C-19 stellar stream, shown on-sky as an Aitoff projection in Fig.~\ref{fig:lb} and on the Gaia colour-magnitude diagram (CMD) in Fig.~\ref{fig:cmd}.
Details of the observations, target selections, and literature values are in Section~\ref{sec:obs}.
The stellar parameters and model atmospheres analyses in this paper are described in Section~\ref{sec:atmos}.
The spectral lines analysis in 1DLTE and 1DNLTE are described in Section~\ref{sec:lines}, 
and the chemical abundances presented in Section~\ref{sec:chem}.
The chemical abundances for the members of C-19 are compared to metal-poor stars in the Milky Way halo and nearby ultra-faint dwarf galaxies in Section~\ref{sec:disc}.

\section{GHOST Observations} 
\label{sec:obs}

The Gemini High Resolution Optical Spectrograph (GHOST) is a fiber-fed spectrograph on the Gemini South telescope that provides simultaneous wavelength coverage from 348 to 1061 nm, with optimal performance between 363 to 950 nm.  It can observe up to two objects simultaneously using IFUs, at high spectral resolution (R $>50,000$); however, we only use single object mode in this project as our C-19 targets are widely spread across the sky.
Science commissioning of GHOST was carried out over several nights between June 2022 and February 2023
(see details in \citealt{McConnachie2024} and \citealt{Kalari2024}).
The GHOST spectra for four of our C-19 targets were taken during science commissioning, and three more during two Gemini Fast Turnaround programs (Dec 2023 and Sept 2024); Table~\ref{tab:targets}.  
Sample spectra are shown in the Appendix, see Fig.~\ref{fig:C19A} to Fig.~\ref{fig:C19G}.

\subsection {Target Selection}
\label{sec:targets}

Two complementary stream-searching algorithms, 
\texttt{STARGO} \citep{Yuan18} and \texttt{STREAMFINDER}
\citep{Ibata21}, were used with the low-metallicity samples from the 
\textit{Pristine Survey} and the
Gaia XP spectro-photometric catalogues \citep{Martin2024} to search for new C-19 members over the whole sky.
New members of the C-19 stellar stream have been identified in its extended tail by \citet{Yuan2025}. 
Twelve new members (G$<17$), spread over more than
100 degrees on sky, were confirmed from velocity and metallicity information from
VLT/UVES and
Magellan/MIKE
spectroscopy \citep{Yuan2025}.

In this paper, we analyse new Gemini/GHOST spectra for 8 members in the outskirts of the C-19 stream, in addition to 2 standard stars observed
with GHOST.
Using the same analysis techniques (described below), we also reanalyse a Subaru/HDS spectrum for one target (P2987454, observed by \citealt{Yuan2025}). 
These stars with high-resolution spectral analyses are listed in Table~\ref{tab:targets}, 
where we also include five stars in the core of C-19 analysed previously  \citep{MartinVenn2022, Yuan2022, Jeong23} and one star found serendipitously in the literature from metal-poor surveys of the MW halo taken with Subaru/HDS \citep[J2338+0902 = C19W;][]{Aoki13}.  

All other spectral analyses of C-19 members have been carried out with medium resolution spectroscopy, \ie 
GTC/OSIRIS spectra for five stars on the lower RGB \citep{MartinVenn2022} and VLT/Xshooter spectra of 12 faint subgiants \citep{Bonifacio2025}, or with such low SNR high resolution spectra that chemical abundances could not be determined \citep{Yuan2025}.
Together, those analyses focused on radial velocity and metallicity studies, with one study including [Mg/Fe] estimates; \cite{Bonifacio2025} found 
hints of a potentially large [Mg/Fe] dispersion (typical of globular clusters) which contrasts with the relatively high velocity dispersion of the stream.

\subsection{GHOST Commissioning Data } 

\begin{table*}
\caption{GHOST and literature observations summary (GHOST commissioning data file details in the Appendix, \textbf{including exposure times}), including observing date, spectral resolution (R), radial velocities, and SNR per pixel at two wavelengths ($\approx$450 and 650 nm).  
%Information for other C-19 stars and standard stars is from the literature.
}
\label{tab:obs}
\hspace{-2.4 cm}
\begin{tabular}{llrcrrcccr}
\hline
Target &  INST & Obs.Date &  R & RV$_{HRS}$ & RV$_{GaiaDR3}$ & RV$_{Yuan}$     &   SNR & SNR   & REF \\
& & &  & (\kms) & (\kms) & (\kms) & 450\,nm & 650\,nm  \\
\hline

\textbf{This Paper:} \\
C19A & GHOST & June 2022 & 56000 & $-132.8\pm0.2$  & ... & $-133.5\pm0.4$ & 32 & 40  & (1,2) \\ 
C19B & GHOST & Sept 2022 &  56000 & $+35.4\pm0.1$  & $36.1\pm2.6$  & \, $+32.4\pm0.2$ & 27 & 40 & (1,3) \\
C19C  & GHOST & Sept 2022 & 56000 & $-41.3\pm0.1$  & $-42.1\pm2.6$ & \, $-41.2\pm2.0$ & 30 & 55 & (1,3) \\
C19D  & GHOST & Sept 2022 & 56000 & $-19.6\pm0.2$  & $-14.0\pm5.8$  & \, $-22.1\pm2.0$ & 27 & 40  & (1,3) \\
C19E  & GHOST & Dec 2023 & 56000 & $-134.2\pm0.2$  & $-127.6\pm8.3$  & $-134.5\pm0.7$ & 54 & 60  & (1,3) \\
C19F  & GHOST & Sept 2024 & 56000 & $-166.4\pm0.3$ & ...  & $-172.5\pm0.8$ & 16 & 32 & (1,3) \\
C19G  & GHOST & Sept 2024 & 56000 & $-178.4\pm0.1$ & ... & $-178.6\pm0.6$ & 24 & 51  & (1,3) \\
C19N & Subaru/HDS & Sept 2022 & 90000 & $-134.3\pm0.7$ & $-135.0\pm3.9$ & $-134.3\pm0.7$ & 22 & 60 & (1,3) \\
HD122563 & GHOST & June 2022 & 56000 & $-26.3\pm0.1$   & $-26.1\pm0.1$ & ... & 300 & 370  & (1) \\
HD222925 & GHOST & June 2022 & 56000 & $-38.5\pm0.1$   & $-38.2\pm0.2$ & ... & 300 & 370 & (1) \\
%HE2340-6036 & VLT/UVES & 2024 & 40 K & 212.4 \kv{$\pm??$} & $212.1\pm1.8$ & 70 & 175 & (1) \\
%{\it HE2340-6036*} & {\it Mag/MIKE} & {\it Sept 2014} & 0.7"x5" & 212.4 & 212.1 & 74 & ... \\
\\
\textbf{Literature:} \\
%P3553224 
C19J & GRACES & Sept 2018 & 68000 & $-182.3$ & $-180.4\pm3.7$  & ... &  ... & 77 & (4)\\
--- & VLT/UVES &  Oct 2021 & 40000 & ... & $-180.4\pm3.7$ & $-183.6\pm0.3$  & ... & 32 & (2) \\
%J2338+0902 
C19K & GRACES & Jan 2021 & 6000K & $-186.7\pm2.2$ & ... & ... &   ... & 50  & (5) \\
%P3547701  
C19L & GRACES & Dec 2020 & 68000 & $-194.4\pm2.0$ & ... & ... &  ... & 100  & (5) \\
%P3551327 
C19M & GRACES & Jan 2021 & 68000 & $-197.3\pm2.1$ & ... & ... & ... & 55   & (5) \\
%P3552755 
C19W & Subaru/HDS & July 2008 & 90000  & $-156.8\pm1.7$ & $-155.8\pm5.1$ &  
 ... & 30 & ... & (6) \\
%P3549615 
C19Y & VLT/UVES &  Oct 2021 & 40000 & ... & $-190.4\pm5.3$ & $-190.6\pm0.6$ &   ... & 9  & (2) \\
%HD122563    & CFHT/ESP  &  archive  &  42 K  &   ...    & $-26.1\pm0.1$ & ... &  ...  &  608   & (10) \\
HD122563  & Mag/MIKE & Feb 2009 & 70000 &  $-25.7$ & $-26.1\pm0.1$ & ...  & 165 & 540 & (7) \\
---  & McD/HET & May 2008 & 70000 & $-26.1$  & --- & ...  & 165 & 540 & (7) \\
---  & McD/Tull & Feb 2009 & 70000 &  $-26.5$  & --- & ...  & 165 & 540 & (7) \\
--- & McD/HRS  &  1997      &  60000  & ...   & --- & ... &  125  &  250   & (8) \\
HD222925    & Mag/MIKE & Sept 2017 & 70000 & $-38.9\pm0.6$ & $-38.2\pm0.2$ & ... & 500 & 700 & (9) \\
%HE2340-6036 & Mag/MIKE & Sept 2014 & $\sim$50 K & $212.4\pm??$ & $212.1\pm1.8$ & 74 & ... & (9) \\
\\
\hline
\vspace{-2ex}
\end{tabular}

References:
(1) This paper;
(2) \citet{Yuan2022};
(3) \citet{Yuan2025};
(4) \citet[][= P3553224]{Jeong23};
(5) \citet{Martin2022};
(6) \citet[][= J2338+0902]{Aoki13};
(7) \citet{Roederer2014};
(8) \citet{Afsar2016}; 
(9) \citet{Roederer22}.
%(10) \citet{Casamiquela2025};
%*RV$_{Yuan}$ for C-19 stars are from \citet{Yuan2025}, with the exception of C19A from \citet{Yuan2022} 

\end{table*}

GHOST spectra for four targets were taken during commissioning (C19A-D; June/Sept 2022). These were used to help test and develop the now excellent GHOST data reduction pipeline, \texttt{GHOSTDR} \citep{Ireland2018, Hayes2022, Simpson2024}.
As such, some steps were a reduction challenge, e.g., the early decisions and importance in the selection of the slit viewing image, as our objects were often too faint to be used directly. 
Instead, the slit viewer image of a brighter standard star was used, taken with the same observing mode set up during the commissioning run.  
The saturation effects on the red chip were also under investigation, and forced us to exclude some images, even when the blue chip exposures were ideal.
These observations were also used to test guiding on faint objects, though at ${\rm G}<17$ we had a few problems in guiding on the C19 targets.
These spectra also contributed to the development of the wavelength solution (see Section~\ref{sec:rvs}).

The GHOST data files (science, flat, arc, biases, and slitviewer) are listed in the Appendix.
A final minor issue can be seen in the early naming of these data files, and also minor clocking issues were also corrected during the post-processing stages (for June 2022 data only).
After several iterations, the 1D extracted spectra from GHOST commissioning are used in this analysis for four stars, C19A-D.

\subsection{GHOST Fast Turnaround Data}

Data from two Gemini Fast Turnaround programs\footnote{Gemini Fast Turnaround programs; GS-2023B-FT-206, GS-2024B-FT-202.} are included in this analysis, for three targets, C19E-G.
Fast Turnaround was requested as these newly discovered members along the C-19 stream were confirmed, and/or required higher SNR spectra (C19G; see Appendix).

Two of the greatest advantages of the new Gemini/GHOST spectrograph are (1) the GHOST data reduction pipeline package (\citealt{Ireland2018, Labrie19, Labrie2023, Simpson2024}), and (2) that the US National Gemini Office \citep{Placco2023} provides reduced data products for all GHOST programs. The spectra are reduced and extracted via the Gemini/DRAGONS pipeline 
\textit{automatically} every day \citep{Placco2024}. Thus, GHOST spectra are available in the Gemini Observatory Archive as reduced, extracted, wavelength-corrected, barycentric-corrected, and order-combined spectra. These 1D spectra require only a few final steps, e.g., continuum normalization and radial velocity corrections, for chemical abundance analyses as in this paper.  We include the key calibration files used for our stars in the Appendix,
and note that this information is available through the Gemini Observatory Archive (as Provenance and History under the [H] prompt in the Archive).

\subsection{Radial Velocities (RVs)}
\label{sec:rvs} 

Radial velocities (RVs) for all C-19 members were determined from cross-correlation of their GHOST spectra with a synthetic metal-poor template of similar stellar parameters and broadened to the GHOST resolution. These were also checked by cross-correlation with the GHOST spectrum for the standard RGB star, HD~122563.
Three wavelength regions were examined independently (400-440, 440-530, 530-700 nm).
The final RV is the average of these three sections, and the RV uncertainty $\sigma$(RV) is taken as their standard deviation. 
RVs per target are provided in Table~\ref{tab:obs}. 

The CaII triplet 860 nm lines were also examined, but often offset from the otherwise excellent and consistent solution of all other lines with $\lambda < 700$ nm.  This is likely due to difficulties in the data reduction steps, where there are significant telluric lines and fewer calibration lines in our wavelength arcs.  The offset was typically $<1$ km\,s$^{-1}$. We discard our Gemini/GHOST CaT radial velocity results in this analysis. 
\citealt{Yuan2025} determined RVs only from the CaT lines in similar VLT/UVES and Subaru/HDS spectra.   

\begin{figure}
\begin{center}
    \setlength{\unitlength}{1cm}
    \includegraphics[width=0.45\textwidth]{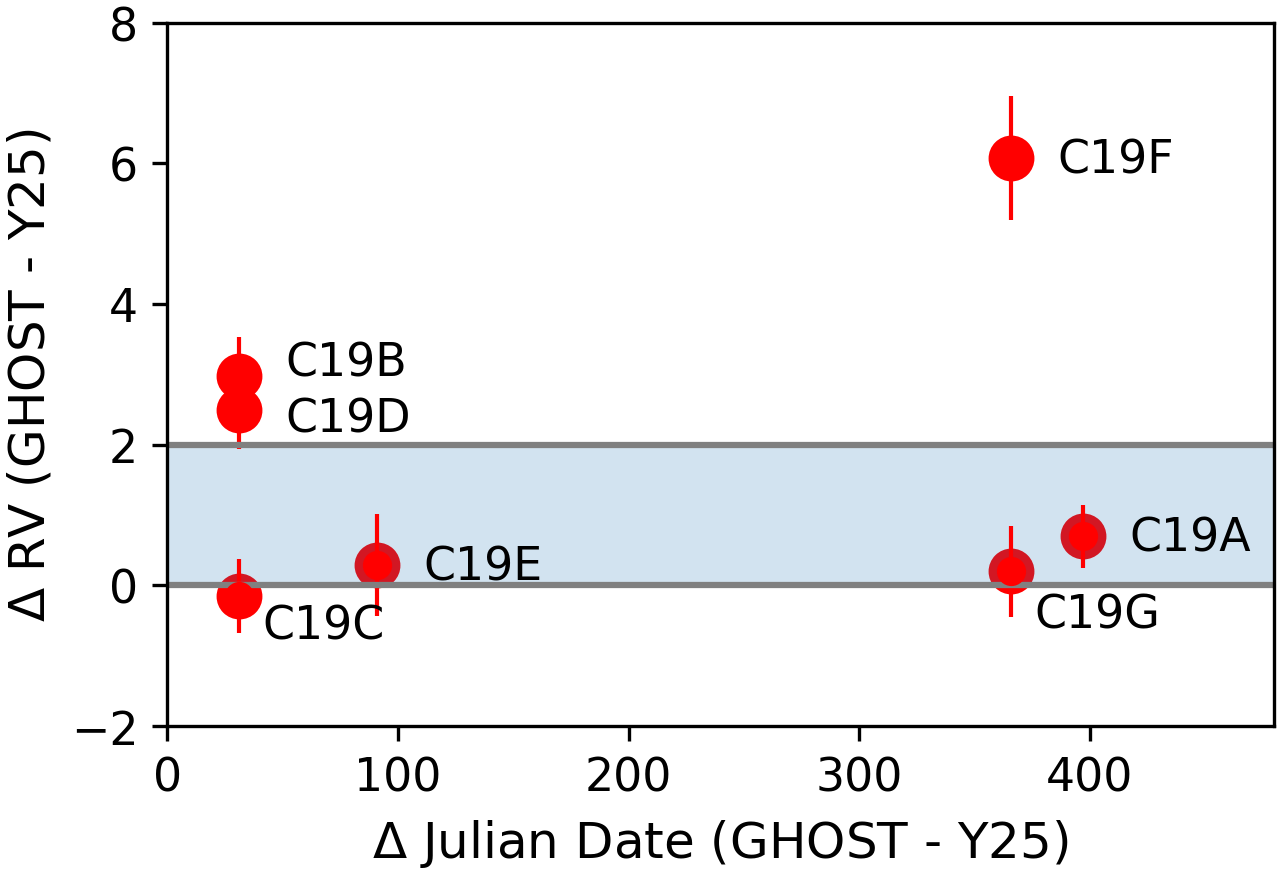}
    \caption{GHOST precision RV measurements vs those from CaT lines by \citet[][=Y25]{Yuan2022, Yuan2025}.
    Three stars show RV offsets larger than the estimated systemic errors in Y25 ($\sim$ 2~km\,s$^{-1}$, blue bar).  As C19B and C19D were observed only 1 month apart, they are unlikely to have had large RV changes, implying the CaT measurement errors may be slightly larger. 
    The large RV variation of C19F may be an indication of binarity. 
    }
    \label{fig:rvcomps}
\end{center}
\end{figure}

A comparison of our RV measurements with those in the Gaia DR3 catalogue \citep{GaiaDR3} are provided in Table~\ref{tab:obs}. In Fig.~\ref{fig:rvcomps}, we also compared our high precision RV measurements to the lower precision results based on the CaT lines \textit{only} from VLT/UVES and Subaru/HDS spectra in \citet{Yuan2025}.  For one star (C19F), the result is in stark disagreement with ours. We note the spectrum used by \citet{Yuan2025} had very low SNR ratio, and suggest that both their RV and metallicity values for C19F are poorly constrained.  Nevertheless, the offset in RV is so large that it suggests C19F might be a candidate binary system.

\subsection{C19N Subaru/HDS}

One star in this analysis was observed at the Subaru telescope using the High Dispersion Spectrograph (HDS).  Details of the exposure are described in \cite{Yuan2025} and summarized in Table~\ref{tab:obs}.  Its spectral quality (SNR and resolution) are sufficient for a detailed chemical analysis, carried out for the first time in this paper and using the same methods as for the stars observed with GHOST.

\subsection{Distances}
\label{sec:dists}

Heliocentric distances to our targets have been calculated from three analyses: astrometric, isochrone, and orbit.   

Astrometric distances are from the Bayesian analysis method developed by \citet{Sestito19}, without the use of metal-poor stellar isochrones. 
Astrometric distances were initially estimated from the Gaia DR3 data \citep{GaiaDR3}. A posterior probability distribution function was obtained by multiplying a Gaussian likelihood on the parallax, shifted by the zero-point offset \cite{Lindegren21}.
A second solution that we call the isochrone distance follows a similar method as used for our astrometric distances, but also incorporates a set of extremely metal-poor MESA/MIST isochrones \citep{choi2016mesa, Dotter2016}, a prior on the Galactic stellar density distribution, and a prior on the age of the metal-poor stars. The results were not very satisfactory, as some distances were unrealistically large (given the magnitude of our stars and large uncertainties on their parallax measurements), which we attribute to the known systematics and poor constraints of the most metal-poor isochrones \cite[e.g.,][]{Heiter2015, Karovicova20}.  Both the astrometric and isochrone distances are summarized per target in Table~\ref{tab:targets}. 

Lastly, we calculated distances assuming our targets are on the same orbit as the core of C-19 when anchored at 18 kpc \citep{MartinVenn2022, Yuan2025}.
Orbital parameters were determined using \texttt{galpy} \citep{Bovy15} and a Galactic halo stellar density distribution prior, i.e., \texttt{MWPotential2014} with an increased dark matter halo mass of $1.2 \times 10^{12}\rm M_{\odot}$  \citep{BlandHawthorn16}. 
Uncertainties on the orbital parameters were derived from a Monte Carlo simulation which draws 1000 times from a Gaussian distribution on the uncertainties in distance, proper motions, and our radial velocities. 
Orbits are listed in Table~\ref{tab:targets} and described further in the Appendix (i.e., Fig.~\ref{fig:orbits}). The estimated uncertainties in these distances are $\sim^{+1.0}_{-0.5}$ kpc.

For the (nearby) standard stars, distances in Table~\ref{tab:targets} are calculated from the zero-point corrected parallax measurements from Gaia DR3, with and without isochrones.

\section{Model Atmospheres Analysis}
\label{sec:atmos}

\begin{figure}
\begin{center}
    \setlength{\unitlength}{1cm}
    \includegraphics[width=0.45\textwidth]{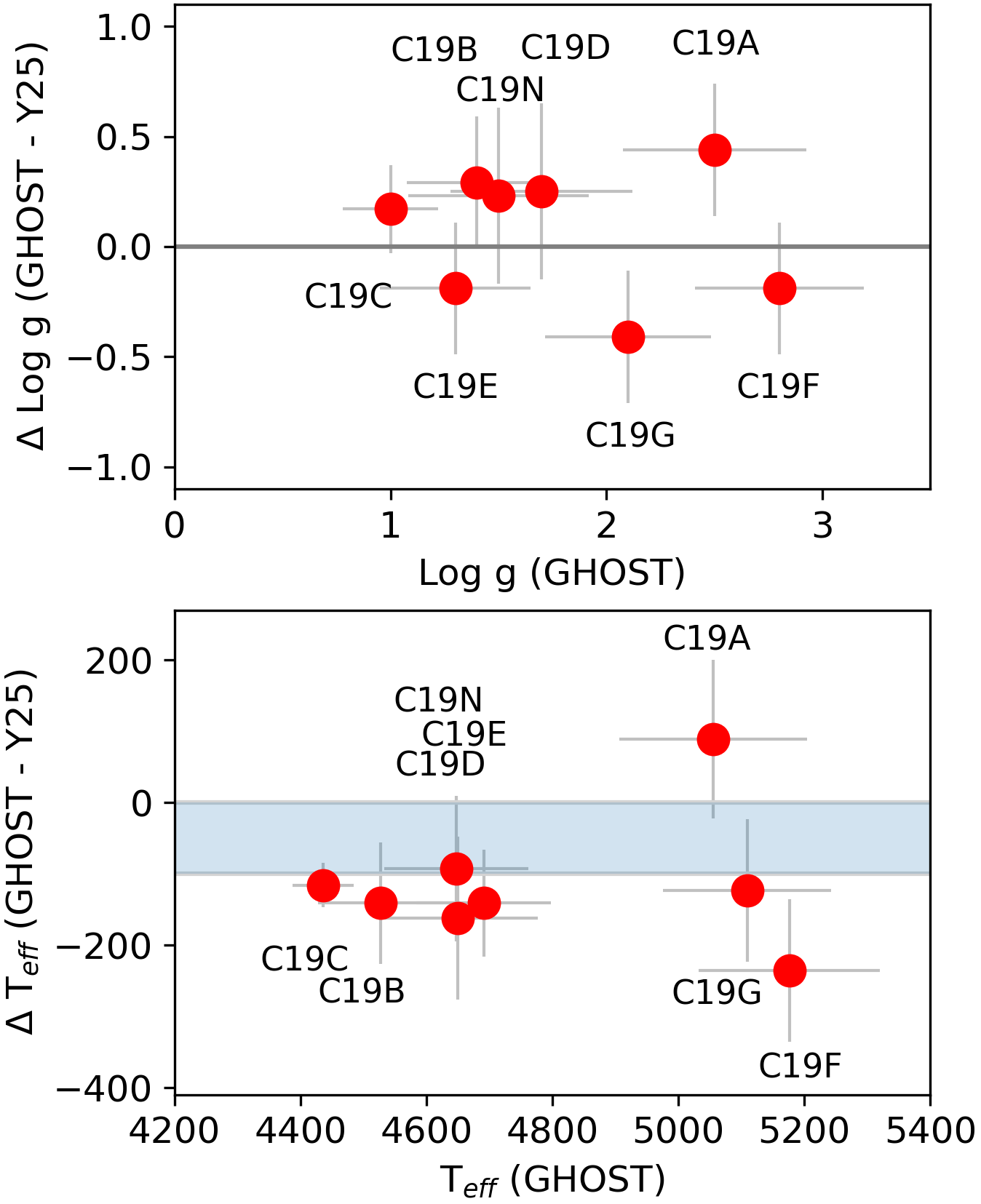}
    \caption{Comparison of stellar parameters between our GHOST results and those from \citet[][=Y25]{Yuan2025}.  We notice their temperatures are hotter (about +100~K shown as the blue bar, and up to 250~K), which is typical when comparing to methods that use metal-poor isochrones. Only C19A differs as it was analysed by \citet{Yuan2022} using iron lines. Surface gravities are generally in good agreement.
    }
    \label{fig:tgcomps}
\end{center}
\end{figure}

Chemical abundances are determined in this paper from a classical model atmosphere analysis of the spectral features in each star.
Model atmospheres from MARCS \citep{Gustafsson08} were adopted, particularly the OSMARCS spherical models, given that all the targets are giants, with \logg$<3.5$.
The 1D LTE radiative transfer code 
MOOG\footnote{MOOG (2019) at \url{http://www.as.utexas.edu/~chris/moog.html}} 
\citep[][]{Sneden73, Sobeck11} 
was used to analyse the stellar spectra.
Corrections due to non-LTE effects were applied from the literature (see Section \ref{sec:NLTE}).

\begin{table*}
\caption{Stellar parameters for C-19 members and standard stars observed with GHOST, and those from the literature.  
}
\label{tab:params}
\hspace{-1.8 cm}
\begin{tabular}{llclccccr}
\hline
Target & \,\,\,\,\, T$_{\rm eff}$* & log$g$* & \,\,\,\, \teff & \logg  & $\xi$ & [Fe/H] & [Fe/H] & REFS* \\
 & \, (phot) & (phot) & \,\, (spec) & (spec)      &       & LTE    & NLTE  \\
 & \,\,\,\, (K) & (cgs) & \,\,\,\,\, (K) & (cgs)        & \, (\kmsno) & (dex)  & (dex) \\
\hline

\textbf{This Paper:} \\

C19A &  $5055\pm105$ & $2.0\pm0.2$ & 
    $5055\pm111$ & $2.5\pm0.3$ & $1.5\pm0.1$ & $-3.40\pm0.16$ &  $-3.28$ & (1) \\ 
C19B & $4527\pm76$ \, & $0.8\pm0.2$ &   
    $4527\pm85$ \, &  $1.4\pm0.3$ & $2.0\pm0.1$ & $-3.46\pm0.19$ & $-3.35$  & (1)  \\
C19C  & $4436\pm72$ \, & $0.7\pm0.2$ &
    $4436\pm31$ \, &  $1.0\pm0.2$ & $2.5\pm0.2$ & $-3.57\pm0.15$ & $-3.48$ & (1)  \\
C19D  &  $4754\pm95$ \, &  $1.2\pm0.2$ &  
    $4754\pm114$ \, & $1.7\pm0.4$ & $2.3\pm0.1$ & $-3.26\pm0.23$  & $-3.16$  & (1) \\
C19E  & $4691\pm94$ \, &  $1.3\pm0.2$ & 
    $4691\pm75$ \, & $1.3\pm0.3$ & $2.2\pm0.1$ & $-3.55\pm0.20$ & $-3.44$ & (1) \\
C19F  & $5176\pm106$  &  $2.4\pm0.2$ & 
    $5176 \pm100$ & $2.8\pm0.3$ & $2.8\pm0.2$ & $-3.36\pm0.19$  & $-3.23$  & (1) \\
C19G  & $5109\pm111$ &  $2.1\pm0.2$ & 
     $5109 \pm100$ & $2.1\pm0.3$ & $2.0\pm0.2$ &  $-3.38\pm0.22$ & $-3.18$ & (1) \\
C19N & $4647\pm77$ & $0.9\pm0.2$ & $4647\pm102$ & $1.5\pm0.4$ & $2.9\pm0.2$ & $-3.50\pm0.21$ &  $-3.39$ & (1)  \\
%C19N* & $4740\pm52$ & $1.3\pm0.1$ & ... & ... & ... & ... & (8) \\
\textit{HD122563} & $4615\pm69$ &  $1.3\pm0.2$ &  
    $4615\pm32$ \, &  $1.4\pm0.1$ & $2.3\pm0.1$ & $-2.99\pm0.09$ &  $-2.88$  & (1,9) \\ 
\textit{HD222925} & $5582\pm142$ & $2.5\pm0.1$ &
     $5582\pm49$ \, & $2.5\pm0.1$ & $2.1\pm0.1$ & $-1.58\pm0.08$ &  $-1.43$ & (1) \\
%\textit{HE2340-6036} & $4756\pm91$ & $1.4\pm0.6$ & $4756\pm84$ & $2.0\pm0.3$ & $2.0\pm0.1$ & $-3.5\pm0.2$  & $-3.4\pm0.2$  & (1)  \\

\\
\textbf{Literature:} \\
%P3553224 
C19J & $4603\pm86$ & $1.0\pm0.2$ & $4569\pm100$ & 0.95  & 2.21 &  $-3.5\pm0.2$  & $-3.3$ & (2)    \\
--- & $4603\pm86$ & $1.0\pm0.2$ & $4628\pm46$ & $1.0\pm0.2$  & 2.3 &  $-3.2\pm0.1$  & ... & (3)    \\
%P3547701 
C19K &  $4974\pm103$ & $1.7\pm0.2$ & $4928\pm100$ & $1.8\pm0.1$ & $2.1\pm0.1$ & $-3.2\pm0.2$ & $-3.1$ &  (4)   \\
%P3551327 
C19L &  $4955\pm84$ \, & $1.6\pm0.1$ & $4881\pm100$ & $1.6\pm0.1$ & $2.2\pm0.1$ & $-3.3\pm0.2$ & $-3.1$ &  (4)  \\
%P3552755 
C19M &  $4982\pm105$ & $1.8\pm0.2$ & $4958\pm100$ & $1.9\pm0.1$ & $2.1\pm0.1$ &  $-3.2\pm0.2$ &  $-3.0$ &  (4)   \\
%J2338+0902 
C19W & $4698\pm81$ \, & $1.2\pm0.2$ & $4900\pm150$ & $1.9\pm0.5$ & 1.5 & $-3.1\pm0.4$ &  ... &  (5) \\
%P3549615 
C19Y & $4527\pm77$ & $0.8\pm0.2$ & $4446\pm100$ & $0.87$  & 2.19 &  $-3.5\pm0.2$  & $-3.3$ & (2) \\

%\textit{HD122563} & $4615\pm69$* &  $1.3\pm0.2$ & $4642\pm35$  & $1.32\pm0.03$  &  $1.85\pm0.06$  & $-2.68\pm0.05$ & ...  & (3) \\ 
\textit{HD122563} & $4615\pm69$ & $1.3\pm0.2$ &
$4500\pm34$   & $0.55\pm0.15$   & $1.95\pm0.06$  & ... & $-2.9$ & (6,9) \\
%---                &    ...          & ...     & $4754\pm29$  & $1.36\pm0.15$  &  $2.2\pm0.6$  & $-2.7\pm0.3$ & $-2.6\pm0.1$ & (8) \\ 
---                &    ...          & ...     & $4500$  & $0.8$  &  2.2  & $-2.9\pm0.2$ & ...  & (7) \\ 
\textit{HD222925} & $5582\pm142$ & $2.5\pm0.1$  & $5636\pm103$ & $2.5\pm0.2$ & $2.2\pm0.2$  & $-1.7\pm0.1$ & $-1.6$  & (8) \\
%\textit{HE2340-6036} & $4756\pm91$ & $1.4\pm0.6$ & $4689\pm100$ & $1.9\pm0.3$ & $1.9\pm0.1$ &   $-3.6\pm0.2$ & ... & (9)  \\
\\
   
\hline
\vspace{-2ex}
\end{tabular}

\textbf{
*Stellar parameters from photometric analyses are calculated for all stars. Spectroscopic parameters are adopted for our analysis of the GHOST spectra.}
C-19 targets are assumed to be located at 18$\pm2$ kpc \citet{MartinVenn2022,Yuan2025}. Distances for the standard stars HD122563 and HD222925 are from their astrometric solutions ($0.32\pm0.01$ and $0.45\pm0.01$, respectively). Spectroscopic logg values were adjusted from the photometric values by analysing FeI and FeII (in both LTE and NLTE).  [Fe/H] are the unweighted mean of the FeI and FeII line abundances. 

*REFS: 
%$^*$References for recent chemical abundance analyses of other potential members and standard stars:
(1) Gemini/GHOST from this paper; 
(2) VLT/UVES spectra by \citet{Yuan2022} 
%including C19A = C-19\_346.20-10.70, 
%P3549615, C19Y = Pristine\_354.96+28.47, and
%P3553224, C19J = Pristine\_355.32+27.59;
%(3) CFHT ESPaDOnS archival spectra by \citet{Casamiquela2025};
(3) C19J (P3553224, J2341=LAMOST-J234117.38+273557.7)  also observed with Gemini/GRACES by \citet{Jeong23};
(4) Gemini/GRACES by \citet{Martin2022};
(5) C19W (SDSS J233817.55+090207.5) observed with Subaru/HDS by \citet{Aoki13};
(6) Mag/MIKE + McD/HET + McD/Tull for HD122563 by \citet{Roederer2014};
(7) HD122563 observed at McDonald Observatory by \citet{Afsar2016};
(8) Magellan/MIKE for HD222925 by \cite{Roederer2018};
%(8) Recent updates by \citet{Mittal2025};
% (9) VLT/UVES spectrum analysed here for comparison with similar from Magellan/MIKE by \citet{Andales2024}.
(9) For HD122563, \teff\  adopted from  \citealt{Giribaldi2023} (see text).

\end{table*}

\subsection{Stellar Parameters \& Standard Stars}

Stellar parameters for extremely metal-poor stars (EMP; [Fe/H] $<-3$) are notoriously challenging.  A variety of methods can produce different results, which are critical to the precision and reliability of a chemical abundance analysis \cite[\eg see][]{Karovicova20, Giribaldi2023}.  
For this reason, we have included two standard stars, HD122563 and HD222925.  These standard stars were analyzed using the same methods as the C-19 stars, to test our calibrations, and provide the basis for a line-by-line differential analysis (below).  

All stellar parameters in this paper are initially determined using Gaia DR3 data, following the procedure developed by \cite{Sestito2023PIGSV}.
Effective temperatures (T$_{\rm eff}$) are found using the colour-temperature calibrations from Gaia photometry from \citet[][hereafter MB2020]{Mucciarelli2020}. This calibration was selected based on the inclusion of very metal-poor stars (from \citealt{GonzalezHernandez2009}). When calculating these photometric temperatures (T$_{\rm phot}$), we adopted their calibration for giant stars, the Gaia DR3 colour BP-RP \citep{GaiaDR3}, extinction values from \cite{Schlafly11, Schlafly18} with reddening conversions to Gaia colours from \cite{Evans18}, and an initial metallicity; [Fe/H]=$-3.4$ for all C-19 stars \citep{MartinVenn2022} or [Fe/H]=$-2.8$ and $-1.5$ for HD122563 and HD222925, respectively \citep{Giribaldi2023, Roederer2018}.
For all C-19 targets and standard stars, these initial T$_{\rm phot}$ values were further scrutinized using the iron spectral lines, \ie minimizing the slope in the \ion{Fe}{1} line abundances vs excitation potentials ($\chi$ in eV).  However, no significant changes were necessary.
NLTE corrections (discussed below) to the \ion{Fe}{1} line abundances did not significantly affect these results when tested in HD122563.
Resonance and low excitation ($\chi<1.2$ eV) \ion{Fe}{1} lines with good measurements were kept in the analysis. Keeping those lines improves our sampling of the line-forming stellar atmosphere layers, which can be important for the differential analyses, and when other chemical species are sampled only from resonance or low excitation lines. 

When using this method, the colour dominates the uncertainty in T$_{\rm phot}$.  While the colours 
are generally quite precise in the Gaia DR3 database, the reddening values can be more uncertain,  
e.g., the reddening value for the bright standard star HD122563 is unrealistic and results in a T$_{\rm phot}$  about $\sim$150~K larger\footnote{We note that a recent re-analysis of metal-poor stars by \citet{Mittal2025} has this hotter temperature for HD122563, thus we do not adopt those results for reference in this paper.}
than found from interferometric measurements, Hydrogen line profile fitting, and 3D atmosphere analyses \citep[\eg see][]{Karovicova20, Giribaldi2023}.  
Although HD122563 is bright (G=5.9), it is well below the Gaia bright star limit in G (G$\sim$3); but it is unclear what the limits are for G$_{BP}$ and/or G$_{RP}$.
Thus, we adopt \teff=4615~K for HD122563 from \citet{Giribaldi2023} throughout this analysis.

Surface gravities (log~$g$) were determined using the Stefan-Boltzmann equation \cite[\eg see][]{Venn2017, Kraft2003,Bonifacio2025}. This method requires G and A$_G$, the solar bolometric magnitude (M$_{\rm bol}$ = 4.74) and bolometric corrections for Gaia DR3 photometry \cite[EDR3;][]{Lindegren21}, the colour (or \teff, from above), and a distance. 
We adopt the C-19 anchor distance of 18 kpc for all stars in this analysis, though we note that the orbit does imply small differences (see Section~\ref{sec:dists}).  These small differences mainly affect the \logg values and are taken into account in the uncertainties in gravity (see Table~\ref{tab:params}). 

For each step, we perform a Monte Carlo on all the input parameters to estimate the uncertainties on the effective temperature and surface gravity. The input parameters are randomized within 1$\sigma$ using a Gaussian distribution, except for the stellar mass. The stellar mass is treated with a flat prior from 0.5 to 0.8 M$_\odot$, which is consistent with the mass of old RGB stars.
Our final temperature and surface gravity values are listed in Table~\ref{tab:params}.  
A comparison of the parameters that we determine to those published previously for C-19 member stars found in the literature shows good agreement, $\lesssim 1\sigma$.
We also compare our stellar parameter results to those from an isochrone fitting method used by \cite{Yuan2025} in Fig.~\ref{fig:tgcomps}, 
where we notice their temperatures are offset by $\sim+100$~K. Stellar parameter offsets (in \teff and/or log$g$) are typical when spectroscopic parameters are compared to those from metal-poor isochrones \cite[e.g.,][]{Vandenberg2008, Tayar2017, Venn2020, Giribaldi2023}.

\begin{figure}
\begin{center}
    \setlength{\unitlength}{1cm}
    \includegraphics[width=8.5 cm,height=5.5 cm]{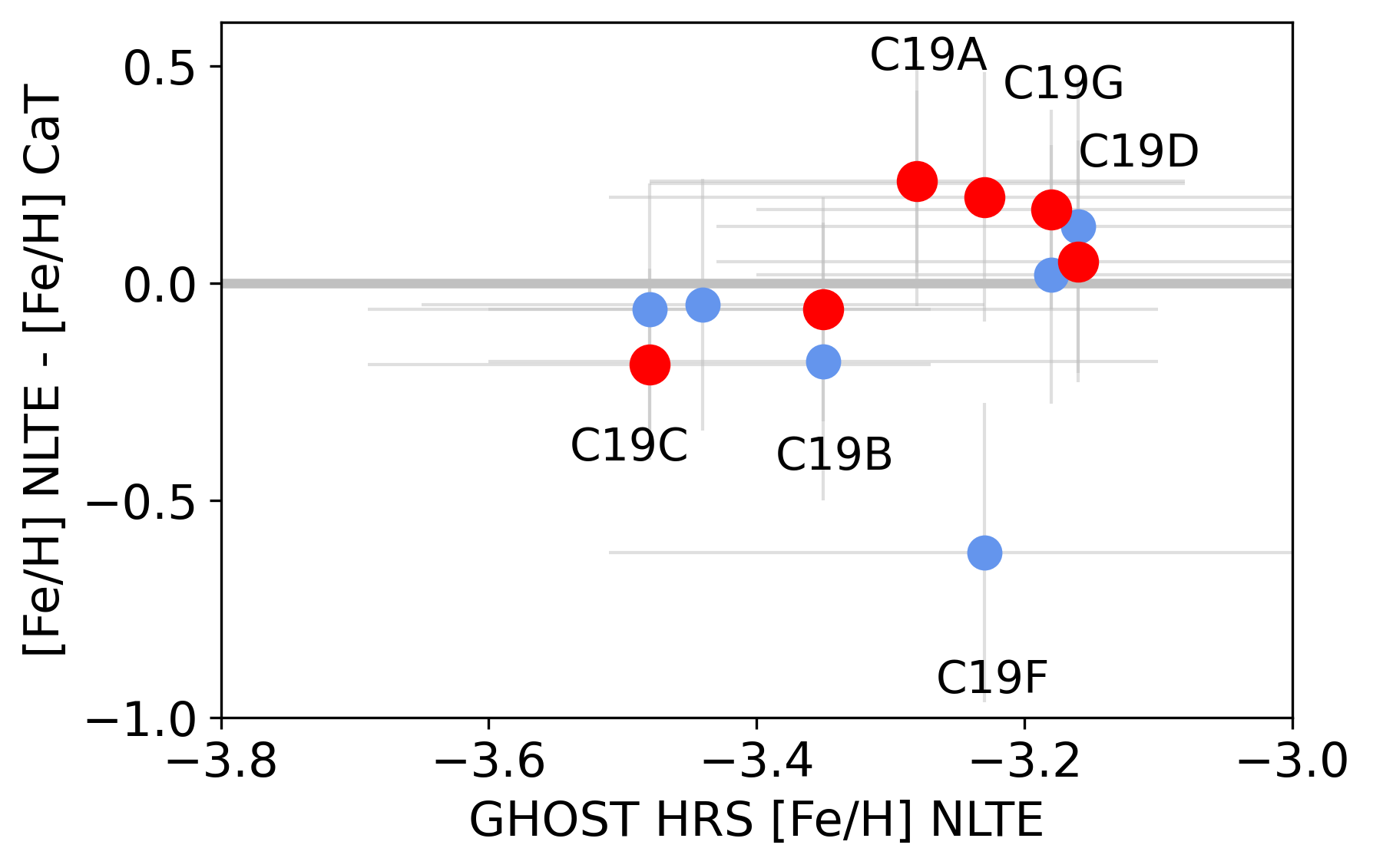}
    \caption{Comparison of the metallicity measurements for the C19 members.  
    The GHOST HRS [Fe/H] (1DNLTE) values are compared to metallicities inferred from the CaT lines per star (using the calibration by \citealt{Carrera13}). 
    Metallicities entirely from GHOST data (red points) are compared to the CaT metallicities presented in \cite{Yuan2025} (\citealt{Yuan2022}: C19A).  Only C19F shows a significant offset, also seen in its radial velocity measurement in Fig.~\ref{fig:rvcomps}.
    }
    \label{fig:ironcomps}
\end{center}
\end{figure}

Initial metallicities for the stellar parameter determinations were taken as [Fe/H]=$-3.4$ for the C-19 stream targets (and from the literature for the standard stars). 
This value compares well to the metallicity estimated from the
Ca II triplet (CaT) lines for
these targets from \cite{Yuan2025}.
We also calculate CaT metallicities from our GHOST spectra.  The CaT spectral lines are fitted with a Voigt profile to each component in each star (reported in the Appendix, see Table~\ref{tab:otherlines}).
We then derived the metallicity using the relation between metallicity, total equivalent width, and absolute I band magnitude, as calibrated by \cite{Carrera13}.  
The metallicity uncertainties are dominated by the uncertainty in distance. We propagated the distance uncertainty using 1000 Monte Carlo realizations, resulting in a 1$\sigma$ metallicity uncertainty of $<$0.1 dex.
Our CaT metallicities are compared to those from \cite{Yuan2025} and also our final NLTE iron abundances in Fig.~\ref{fig:ironcomps}.
Our final NLTE corrected iron metallicities (described below) are
provided in Table~\ref{tab:params}
and used to generate the final
model atmospheres per star.

Microturbulence ($\xi$) was initially determined using the formula for red giants from \citet{Mashonkina2017}, and adjusted if required to remove any slope in the \ion{Fe}{1} line abundances (see below) with equivalent widths. 
These adjustments were small (typically $0.1-0.2$ km\,s$^{-1}$).
As the slope can be sensitive to NLTE corrections, 
we examined both LTE and NLTE \ion{Fe}{1} abundances to assess the impact of NLTE corrections, but found negligible differences.

\subsection{Spectral Lines Analysis}
\label{sec:lines}

As the goals of this paper are a detailed comparison of the spectra of highly probable members of the C-19 stellar stream, we have been extremely careful in the selection and use of the line lists. 
Spectral lines were initially selected from a collection of metal-poor halo star studies, such as \cite{Kielty2021, Yong2021, Lucchesi22, Roederer22}.  
Atomic data was updated with the \texttt{linemake}\footnote{The \texttt{linemake} catalogue includes laboratory atomic data (transition probabilities, hyperfine and isotopic substructures) published by the Wisconsin Atomic Physics and the Old Dominion Molecular Physics groups. These lists and accompanying line list assembly software have been developed by C. Sneden and V. Placco, available at https://github.com/vmplacco/linemake. }
atomic and molecular line database \citep{Placco21}.

For iron, nearly 100 spectral lines were selected (82 \ion{Fe}{1}, 10 \ion{Fe}{2}; see Table~\ref{tab:felines}), i.e., lines which are fairly isolated, in spectral regions with sufficient SNR, and weak equivalent widths (EW$<150$ m\AA) to reduce sensitivities to stellar parameters and non-Gaussian profiles.   
Measurements were initially made by fitting Gaussian profiles, followed by measurements of the area under the continuum to account for Lorentzian line wings for lines stronger than $\sim$100 m\AA.
All of the iron lines are available in our standard star HD122563, most in our second standard star HD222925, and most are also in the C-19 targets; the only exceptions are 3 \ion{Fe}{1} lines that are ideal in most of the C-19 targets but too strong (EW $\sim200$ m\AA) in the standard stars.

For other elements, EWs were measured as above for iron (see Table~\ref{tab:otherlines}), and spectral lines with depths below $40\%$ continuum or EW$>150$ m\AA\ were discarded (i.e., strong lines that may form well away from the continuum region in the model atmosphere, and can be extremely sensitive to stellar parameter uncertainties). Exceptions were made if they were the only lines available of the element and well behaved, i.e., abundances in good agreement with other weak lines of the same element (e.g., NaD, \ion{Mg}{1}), especially in our standard stars. 
When a slightly stronger line has been kept, the EW has been noted in italics in our line list (Tables~\ref{tab:felines} and \ref{tab:otherlines}), and we use those lines with caution. 

Isotopic and hyperfine structure were taken into account for the odd-Z elements \ion{Sc}{2}, \ion{Mn}{1}, 
and \ion{Co}{1}, as well as the heavy elements \ion{Ba}{2} and \ion{Eu}{2}.
The corrections for barium and europium were calculated with the r-process isotopic ratios from \cite{Sneden2008}.
These effects were not taken into account for \ion{Sr}{2}, as the line formation is dominated by the even-Z isotope $^{88}$Sr  \citep[e.g.,][]{Mishenina2019, Prantzos2020}.

Spectrum syntheses were necessary for carbon from the G-band near 4300 \AA. Each synthetic spectrum was broadened to match the observed spectra using a Gaussian smoothing kernel with FWHM=0.14 to 0.17 \AA. 
This was found to be a good match to the thermal broadening of the spectral lines in these RGB stars, whether 1x4, 2x4, or 2x8 pixel binning was chosen during the GHOST observations. For C specifically, we adopt the $^{12}$C/$^{13}$C = 5, typical for an evolved star \citep{Gratton00}.

\subsection{NLTE corrections}
\label{sec:NLTE}

Departures from Local Thermodynamic Equilibrium (LTE) due to the radiation field in metal-poor red giants are known to impact the statistical equilibrium solution for some elements and isotopes.  These non-LTE (NLTE) effects can be large,
depending on the elements, stellar parameters, and spectral line strength,
especially for resonance lines
(e.g., $\Delta$log(X/H)$>0.2$), including lines of \ion{Fe}{1}.  

To investigate the impact of NLTE corrections on our \ion{Fe}{1} and other element abundances,  we examined three databases for their recommended corrections.  These databases include: 
(1) the MPIA database\footnote{MPIA NLTE corrections: \url{http://nlte.mpia.de.}  for \ion{Fe}{1} and \ion{Fe}{2} from \citet{Bergemann2012}, \ion{Mg}{1} from \citet{Bergemann2017}, \ion{Ca}{1} from \citet{Mashonkina2017}, \ion{Ti}{1} and \ion{Ti}{2} from \citet{Bergemann2011}, and \ion{Cr}{1} from \citet{Bergemann2010b}.};
(2) the INSPECT database\footnote{INSPECT NLTE corrections: \url{http://inspect-stars.com} for \ion{Na}{1} from \citep{Lind2012}.}; and
(3)  a database provided by L.~Mashonkina\footnote{Mashonkina NLTE corrections: \url{https://spectrum.inasan.ru/nLTE2/} for Sr and Ba from \cite{Mashonkina2019}}.
NLTE corrections for \ion{K}{1} and \ion{Al}{1} are taken from the literature, i.e., \citet{Reggiani2017} and \citet{Lind2022}, respectively.
We also adopt a mean +0.5 correction for A(\ion{Ni}{1})  abundances based on the recent and first NLTE analysis for \ion{Ni}{1} in EMP stars by \citet{Storm2025}.

%%%% STUB TABLE OF ABUNDANCES -- Full Table as Machine Readable Table (Abund.Master.All)

\begin{table}
\caption{Chemical abundances for C-19 members from this study and the literature, including measurement errors ($\sigma$) and stellar parameters uncertainties (``epar"). Carbon abundances (from CH) are 1DLTE only, reported with and without corrections for stellar evolution effects. Upper limits are noted with UL*.  HD12* is a label for the shorter line list analysis of HD122563. Full table available online.}
\label{tab:abunds_stub}
\hspace{-1.5 cm}
\begin{tabular}{lrrrrrr}
\hline
Elem & A(X) & $\sigma$ & epar & N & [X/Fe] & [X/Fe] \\
     &      &     &      &   & (LTE)  & (NLTE) \\
\hline
\textbf{HD12*} \\
CH  & 5.23 & 0.1 & 0.2 & 1 & $-0.21$ & $-0.32$ \\
CH$_{\rm corr}$  & 5.23 & 0.1 & 0.2 & 1 & $0.31$ & $0.20$ \\
\ion{Na}{1} & 3.37 & 0.01 & 0.07 & 1 & 0.11 & $-0.27$ \\
\ion{Mg}{1} & 5.31 & 0.1 & 0.04 & 4 & 0.69 & 0.73 \\
\ion{Al}{1} & 3.20 & 0.35 & 0.09 & 2 & $-0.27$ & $-0.43$ \\
\ion{Si}{1} & 5.09 & 0.1 & 0.05 & 1 & 0.56 & 0.49 \\
\ion{K}{1} & 2.72 & 0.1 & 0.04 & 1 & 0.67 & 0.23 \\
\ion{Ca}{1} & 3.70 & 0.11 & 0.04 & 11 & 0.34 & 0.43 \\
\ion{Sc}{2} & 0.10 & 0.12 & 0.05 & 8 & $-0.07$ & ... \\
\ion{Ti}{1} & 2.17 & 0.14 & 0.05 & 10 & 0.20 & 0.67 \\
\ion{Ti}{2} & 2.43 & 0.18 & 0.05 & 25 & 0.46 & 0.41 \\
\ion{Cr}{1} & 2.17 & 0.17 & 0.07 & 6 & $-0.49$ & $-0.14$ \\
\ion{Mn}{1} & 1.82 & 0.1 & 0.11 & 3 & $-0.63$ & $-0.63$ \\
\ion{Fe}{1} & 4.49 & 0.14 & 0.08 & 76 & $-3.01$ & $-2.88$ \\
\ion{Fe}{2} & 4.66 & 0.10 & 0.04 & 10 & $-2.84$ & $-2.84$ \\
\ion{Co}{1} & 1.82 & 0.10 & 0.08 & 3 & $-0.19$ & 0.31 \\
\ion{Ni}{1} & 3.10 & 0.1 & 0.09 & 3 & $-0.14$ & ... \\
\ion{Sr}{2} & -0.27 & 0.1 & 0.08 & 2 & $-0.16$ & $-0.25$ \\
\ion{Ba}{2} & -1.51 & 0.1 & 0.05 & 4 & $-0.71$ & 
$-0.91$ \\
Eu\_UL* & $-2.34$ & 0.1 & 0.05 & 2 & 0.12 & ... \\ [2ex]
\textbf{C19A} \\
CH  & 5.43 & 0.15 & 0.22 & 1 & 0.40 & 0.28 \\
CH$_{\rm corr}$  & 5.43 & 0.15 & 0.22 & 1 & 0.41 & 0.29 \\
\ion{Na}{1} & 3.01 & 0.15 & 0.13 & 2 & 0.17 & $-0.16$ \\
\ion{Mg}{1} & 4.42 & 0.38 & 0.11 & 5 & 0.22 & 0.26 \\
\ion{Al}{1} & 2.59 & 0.36 & 0.14 & 2 & $-0.46$ & $-0.19$ \\
\ion{Si}{1} & 4.60 & 0.15 & 0.13 & 2 & 0.49 & 0.36 \\
\ion{Ca}{1} & 3.35 & 0.25 & 0.12 & 8 & 0.41 & 0.48 \\
\ion{Sc}{2} & $-0.18$ & 0.15 & 0.12 & 6 & 0.07 & ... \\
\ion{Ti}{1} & 2.02 & 0.19 & 0.13 & 6 & 0.47 & 0.99 \\
\ion{Ti}{2} & 2.22 & 0.32 & 0.12 & 21 & 0.67 & 0.58 \\
\ion{Cr}{1} & 1.78 & 0.24 & 0.18 & 5 & $-0.46$ & 0.05 \\
\ion{Mn}{1}* & 1.11 & 0.15 & 0.15 & 3 & $-0.92$ & $-0.81$ \\
\ion{Fe}{1} & 4.09 & 0.20 & 0.12 & 66 & $-3.41$ & $-3.28$ \\
\ion{Fe}{2} & 4.17 & 0.16 & 0.09 & 5 & $-3.33$ & $-3.33$ \\
\ion{Co}{1} & 1.36 & 0.15 & 0.14 & 4 & $-0.23$ & 0.42 \\
\ion{Ni}{1} & 2.26 & 0.38 & 0.14 & 3 & $-0.56$ & ... \\
\ion{Sr}{2} & $-0.84$ & 0.18 & 0.15 & 2 & $-0.31$ & $-0.47$ \\
\ion{Ba}{2} & $-1.94$ & 0.15 & 0.14 & 2 & $-0.72$ & $-0.63$ \\
Eu\_UL* & $-1.30$ & 0.15 & 0.0 & 1 & 1.58 & ... \\ [2ex]
\hline
\vspace{-2ex}
\end{tabular}
*Eu\_UL = Upper limits for \ion{Eu}{2} calculated from $\lambda$4129 \AA. 
*\ion{Mn}{1} include correction described in Appendix~\ref{app:mn}.
\end{table}

\section{Chemical Abundances}
\label{sec:chem}

The average LTE and NLTE chemical abundances for all C-19 targets and standard stars are provided in Table~\ref{tab:abunds_stub}.
This includes the abundances determine here from our GHOST spectra for C19A-C19G, and the Subaru/HDS spectrum for C19N, using our stellar parameters in Table~\ref{tab:params}, and the abundances collected from the literature for C19J-Y using the published stellar parameters.  We do not recalculate the stellar abundances for the literature stars from our stellar parameters for four reasons: (1) the spectral wavelength regions for the literature stars can be quite different (e.g., GRACES spectra for C19J-M are only useful above 480 nm), (2) the SNR of some data sets are far too low (e.g., the VLT UVES data for C19Y), (3) we have no access to the spectra (i.e., the Subaru spectrum of C19W), and (4) the published stellar parameters for the literature stars are within 1$\sigma$ of our calculations.
Chemical abundances are compared to the Sun using the solar abundances from \citet{Asplund09} and standard notation\footnote{Standard notation: [X/Y] = log n(X)/n(Y)$_*$ $-$ log n(X)/n(Y)$_\odot$, where n(X) and n(Y) are column densities (in cm$^{-2}$).}.

The chemical abundance uncertainties are determined from the measurement errors and the impact of stellar parameter uncertainties, added in quadrature. 
The measurement errors ($\sigma$ in Table~\ref{tab:abunds_stub}) include uncertainties in the continuum placement, the local SNR, the quality of the atomic data, and the line-to-line scatter when the number of lines $N >1$.  
We adopt a minimum $\sigma = 0.1$ for standard stars (due to their higher SNR spectra) and $\sigma = 0.15$ for C19 targets, as representative of the data quality.
Abundance errors induced by the stellar parameter uncertainties are calculated by jostling the \teff, \logg, and $\xi$ within their uncertainties, as listed in Table~\ref{tab:params}.
These three are added in quadrature per species and reported as ``epar" in Table~\ref{tab:abunds_stub}.
An example of the abundance errors due to the individual stellar parameter uncertainties per species is provided in Appendix~\ref{app:errors} for two sample stars (HD122563 and C19A).
The total abundance error per species is determined from the error in the mean abundance (= $\sigma/\sqrt{N}$) and the stellar parameters errors (epar) added in quadrature.

\subsection{Iron (Metallicity)}

The iron abundances are used as a proxy for the metallicity in the model atmospheres analysis.  
Iron is determined for all C-19 targets from $23-80$ lines of \ion{Fe}{1} and $2-7$ \ion{Fe}{2} lines in each star. Both LTE and NLTE average abundances are reported here, where [Fe/H] $\pm\sigma({\rm Fe})$ is the average and standard deviation of the individual lines.  The precision in the mean of [Fe/H] is excellent ($\le0.03$, determined as $\sigma$(Fe)/$\sqrt{\rm N}$).

To test our analysis methods, the iron results for the standard stars are compared to both abundances in the literature, as well as iron from a larger line list, i.e., lines in the standard stars that are not seen or used in the C-19 targets due to the higher SNR available for the spectra of these bright stars. 
Both standard stars were compared to results from \cite{Roederer2014, Roederer2018}, and any differences can be directly attributed to small differences in the adopted atmospheric parameters (see Section~\ref{sect:diffabu}). 

In general,
the agreement in the standard star LTE abundances is excellent from line to line and when compared to previous analyses in the literature.

\subsection{Carbon}
\label{sec:carb}

Carbon has been calculated by fitting portions of the CH $G$-band (4290--4315 \AA) in our two standard stars.  Our results are in excellent agreement with those from \citet{Roederer2014, Roederer2018}, adopting $^{\rm 12}$C/$^{\rm 13}$C = 5 for stars on the upper RGB.  Carbon is depleted during the normal course of stellar evolution, suggesting the natal abundances may have been higher.  Our measured carbon abundances and those corrected for stellar evolution effects are reported in Table~\ref{tab:abunds_stub}. The corrections were calculated using the carbon correction calculator\footnote{A calculator for carbon corrections due to stellar evolutionary stage are available at https://vmplacco.github.io.} from \citet{Placco14}.

\begin{figure}
\begin{center}
    \setlength{\unitlength}{1cm}
    \includegraphics[width=8.5cm]{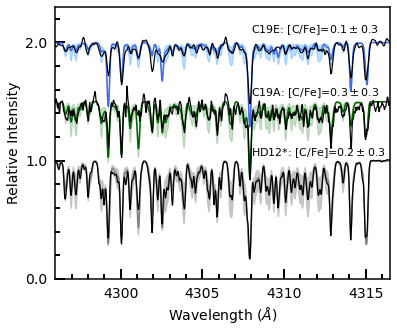}

    \caption{Syntheses of the CH G-band observed in HD122563 (HD12* label) and two representative C-19 targets, C19A/hotter and C19E/cooler (black lines). The best fit [C/Fe] synthesis are shown for each star (coloured or grey lines), including filled areas between sample syntheses offset by the given uncertainties. The abundances are corrected for stellar evolution effects using the carbon correction calculator by \citet{Placco14}
    }
    \label{fig:csynth}
\end{center}
\end{figure}

\begin{figure}
\begin{center}
    \setlength{\unitlength}{1cm}
    \includegraphics[width=8.5cm]{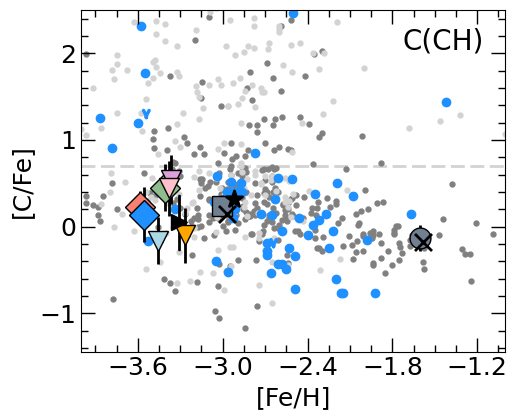}
    \caption{[C/Fe] values (diamonds) or upper limits (inverted triangles) for the individual C-19 stars, shown as large coloured symbols, as in Fig.~\ref{fig:lb}. The black right-facing triangle represents the literature result for C19J from \citet{Yuan2022}.
    GHOST spectral analysis of the standard stars (HD122563 and HD222925) are the large grey symbols, and in excellent agreement with literature (black X's and star symbols; see labels in Fig.~\ref{fig:abus_LTE}). Comparison stars in the MW halo (small grey dots) and UFD galaxies (small blue dots) are shown for reference (see text for references).
    Evolutionary corrections have been included from \citet{Placco14}. None of our selected targets or stars from the literature associated with C-19 are C-rich. 
    }
    \label{fig:cfe}
\end{center}
\end{figure}

Three C-19 stars have sufficient SNR in the CH band and carbon abundances for direct measurements (see syntheses for C19A and C19E in Fig.~\ref{fig:csynth}; the third is C19C).  Upper limits are determined for the remainder of the C-19 stream stars. 
The carbon measurements are typically sub-solar, however the stellar evolution corrections raise them to slightly above solar (shown in Fig.~\ref{fig:cfe}). 
No members of C-19 are found to be C-enhanced, i.e., no C-19 star has [C/Fe]$>0.7$ (the C-rich definition according to \citealt{Aoki07}).  
While many EMP stars in the MW are found to be C-rich at low metallicity
\citep{yong2013CEMP, Placco14}, many are also C-normal, as shown in Fig.~\ref{fig:cfe}.
The lack of CEMP stars may be a selection effect, as we used the narrow-band Pristine survey photometry for the original target selections. Such filters are focused on the Ca H\&K region, which are known to be affected by pollution from the neighbouring CH molecular absorption features, effectively biasing a sample against C-enhanced metal-poor stars \citep{Arentsen2022, Arentsen2025, Martin2024, Sestito2024, Placco2025}.
We discuss the lack of CEMP stars in our sample further in Section~\ref{sect:ufdcomp}.

\subsection{Alpha Elements (Mg, Si, Ca) and Ti}

Alpha elements are even-Z nuclei produced primarily through successive helium-capture reactions during the carbon-, neon- and silicon-burning stages of massive star evolution, and through the $\alpha$-rich freeze-out during core collapse supernovae.  

Magnesium abundances are determined from 2-7 lines of \ion{Mg}{1} from across the spectrum.  The mean NLTE corrections for Mg are similar to those for \ion{Fe}{1} such that [Mg/Fe] are not significantly affected.
Silicon is from the analyses of only 2 \ion{Si}{1} blue lines (at 390.5, 410.2 nm).
Calcium is from up to 11 weak lines of \ion{Ca}{1} from across the spectrum, which yield very homogeneous results, both line to line in a single star, and also from star to star in the C-19 stream.
Titanium is determined from up to 10 lines  \ion{Ti}{1} and 25 lines of \ion{Ti}{2}.  
As above for Mg, the NLTE corrections for \ion{Si}{1}, \ion{Ca}{2}, and \ion{Ti}{2} lines are similar to those for \ion{Fe}{1}, such that the [Si/Fe], [Ca/Fe], and [\ion{Ti}{2}/Fe] ratios show a very small star-to-star scatter.
The NLTE corrections for \ion{Ti}{1} appear to be too large, and result in poor agreement with metal-poor stars in the Galaxy, as well as poor Ti ionization balance in the C-19 targets and standard stars.  
Recently calculated Ti NLTE corrections across a grid of stellar parameters, including extremely metal-poor stars, yield slightly smaller corrections for \ion{Ti}{1} that would improve our Ti ionization balance \citep{Mallinson2024}; however, these corrections were not yet available for our individual stellar analyses.

\subsection{Odd-Z Elements (Na, Al, K, Sc)}

Odd-Z elements are important indicators of core collapse supernova yields, as the difference in the energetic requirements for $\alpha$ particle capture versus neutron capture produces a noticeable odd-even effect in the predicted yields \citep{HegerWoosley2010, Takahashi18}. 
Odd-(Z) elements are also important tracers, as their yields depend on the initial neutron excess, hence the initial metallicity and rotation
\citep{Limongi2018}.
In metal-poor MW stars, an odd-even effect is both seen and predicted, where odd-Z elements have noticeably lower abundances than even-Z elements (e.g., \citealt{Kobayashi2006}).

Sodium abundances in our C-19 targets are from the two strong \ion{Na}{1}~D resonance lines near 590~nm.  There is a large dispersion in the C-19 star-to-star [\ion{Na}{1}/Fe] abundances.  This appears to be real, and reinforced by the C-19 [Na/Fe] values from stars in the literature.  NLTE corrections were adopted from INSPECT \citep{Lind2012}. 
We also examined the more recent \ion{Na}{1}~D corrections presented by \citet{Lind2022}, which are approximately twice as large.  Interestingly, \citet{Lind2022} also analyse HD~122563, allowing a direct comparison of our results. We find excellent agreement in the measured line equivalent widths, stellar parameters, and LTE abundances. 
However, applying their larger NLTE corrections results in a Na abundance for HD~122563 that is poorly matched to those of the other Milky Way standard stars. We therefore retain the INSPECT corrections for consistency across our analysis.

Aluminum is from 2 resonance lines of \ion{Al}{1} in the blue ($<$400 nm).  
NLTE corrections from the grid provided by \citet{Lind2022} range from 0 to +0.4.
The dispersion in the [Al/Fe] abundances in the C-19 stars is larger than seen in the $\alpha$-elements, for either the LTE or NLTE abundance ratios (discussed below).

We also determine abundances for the odd-Z elements potassium and scandium. Potassium is measured from 1-2 resonance lines of \ion{K}{1} near 7700~\AA\ whenever possible as this region is less affected by telluric features.
NLTE corrections are fairly consistent for these stars on the RGB ($\sim -0.3$) when interpolated from the grid provided by \citet{Reggiani2017}.
Scandium is measured from up to 8 \ion{Sc}{2} subordinate lines, analysed with hyperfine structure and isotopic corrections using the atomic data provided by \texttt{linemake} (see above).  
Only 1DLTE Sc abundances are reported here as we did not find NLTE corrections for EMP RGB stars that are publicly available in the literature.

\subsection{Iron-Peak Elements (Cr, Mn, Co, Ni)}

The iron-peak elements in extremely metal-poor stars are generally expected to have formed in core collapse supernova \citep{HegerWoosley2010, Kobayashi20}.  
The specific yields of the iron-group elements depend sensitively on the mass, energy, and mixing that occurs during the supernova explosion.
However, the metallicities of the
C-19 targets are low enough to also warrant consideration of theoretical yields from Population III events (see Section~\ref{sec:disc}).

In addition to iron itself, we determine the chemical abundances for 4 other iron-group elements: chromium, manganese, cobalt, and nickel. 
Chromium abundances  are from 6 lines of \ion{Cr}{1}, including 3 resonance lines and 3 subordinate lines.  Manganese is primarily from the 3 resonance lines of \ion{Mn}{1} near 4030~\AA, which yield consistent line abundance results. 
Cobalt is from up to 4 blue subordinate lines of \ion{Co}{1}, including hyperfine structure and isotopic corrections, and provide line abundances that are in good agreement with one another.
Nickel is from 3 lines of \ion{Ni}{1}, two in the far blue (3807 and 3858 \AA) and one at 5476~\AA; we do find lower abundances from the far blue lines in C-19 targets, which may be due to their lower SNR as the Ni line abundances in our standard stars are in excellent agreement. 

For Cr and Co, the NLTE corrections to the C-19 targets and HD122563 abundances are quite large (+0.5 to +0.8; see Table~\ref{tab:abunds_stub}), and consistent across the available spectral lines. The resulting NLTE [Cr/Fe] and [Co/Fe] abundances are larger than similar stars in the MW. 
Unfortunately, we do not have specific NLTE line corrections for Ni; however, the first NLTE analysis of Ni was recently published by \citet{Storm2025}, who found large NLTE corrections in [Ni/Fe] up to +0.5 dex in EMP stars. 
We cautiously adopt this value, though their stellar parameters were hotter and less luminous, i.e., for stars closer to the main sequence turn off than the red giant branch.

Only for Mn do our abundance ratios for the C-19 stars not resemble those from HD122563. 
Abundances for \ion{Mn}{1} include hyperfine structure, isotopic corrections, and NLTE corrections.
A similar large scatter in \ion{Mn}{1} abundances from these same resonance lines was seen for EMP stars in Sculptor by \citet{Skuladottir2024}. 
%They find 3 Scl stars where [Mn/Fe] could be anti-correlated with temperature; however, we do not find a relation with temperature here (i.e., our hotter stars have higher Mn, and our cooler stars have both high and low Mn).
%
%Unfortunately, we are unable to observe any other (weak) Mn lines in our spectra to test the validity of our analysis of these resonance lines, including in the standard stars.
%
A deeper examination of Mn I is presented in Appendix~\ref{app:mn}, where we examine an extremely high SNR ($>500$) spectrum of the bright EMP standard star BPS~CS22952-0015 (Gaia~DR3 2440816964991039232).
This star has similar atmospheric parameters to the C-19 targets, and the high SNR spectrum means that it has measurements of both the strong Mn I resonance lines and weak Mn I subordinate lines. We show that the line formation of the strong Mn I resonance lines breaks down in this standard star; thus, the C-19 stars will also be affected, but our (slightly higher metallicity) standard stars are not. 
We do not include our [Mn/Fe] measurements for the C-19 targets in the rest of this analysis.

\subsection{Neutron-capture Elements (Sr, Ba, Eu)}

Neutron-capture elements in metal-poor stars are primarily formed through rapid-neutron capture in a variety of core collapse supernovae and neutron-star mergers.  Later contributions from slow-neutron capture may also occur during thermal-pulsing in AGB stars.  The ratio of [Ba/Fe] produced in core-collapse supernovae (CCSNe) is expected to provide an r-process Ba floor, with contributions from the other processes building above that.  To ascertain these different nucleosynthetic sources, often it is necessary to observe a pure r-process element (usually Eu) and examine the ratios (e.g., [Ba/Eu] and [Sr/Ba]).

Strontium is determined from the strong \ion{Sr}{2} resonance lines at 4077 and 4215 \AA.  Their abundances are in good agreement with one another throughout our analysis. 
Barium is determined from up to 4 \ion{Ba}{2} lines, including 2 resonance lines and two subordinate lines of \ion{Ba}{2} across the spectrum.
Hyperfine structure and isotopic corrections are calculated for barium using the r-process isotopic mixture \citep{Sneden2008}.
NLTE corrections are small for both elements ($<0.2$), as also seen in  calculations by \citet{Mashonkina2019}.

We target the strong \ion{Eu}{2} line near 4129~\AA, but were only able to measure upper limits in the C-19 targets and HD122563.  We note that our (LTE) calculation for [Eu/Fe] in the r-process rich standard star HD222925 is in excellent agreement with \citet{Roederer2018}.

\subsection{Differential Abundances (wrt HD122563)} 
\label{sect:diffabu}

\begin{figure*}
\begin{center}
    \setlength{\unitlength}{1cm}
    \includegraphics[width=16cm]{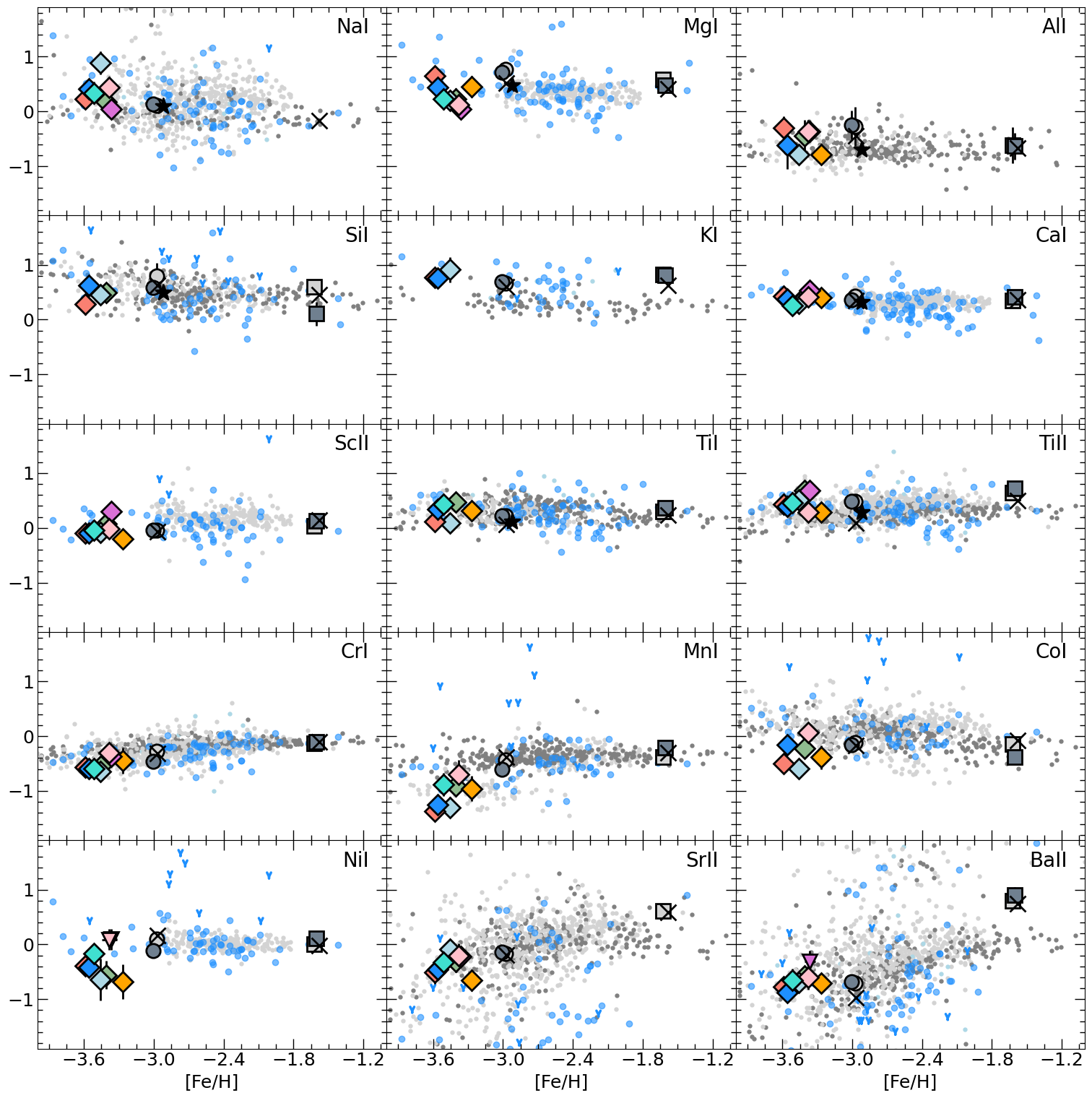}
    \\[0.15cm]
    \makebox[0pt][l]{\hspace{-7cm}\includegraphics[width=17cm]{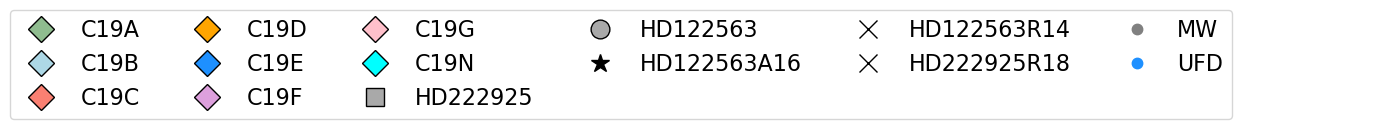}}
    \caption{1DLTE [X/Fe] chemical abundances for 14 members of C-19 (diamond shapes), including some upper limits (downward pointing triangles).  Our abundances for two standard stars are also shown, including HD1222563 near [Fe/H]=$-3$ (large grey circles) and HD222925 near [Fe/H]=$-1.7$ (large grey squares). Abundances
    from the literature are also shown for the standard stars, including HD122563 by \citet{Afsar2016} and \citet{Roederer14}, and HD222925 by \citet{Roederer2014, Roederer2018}; see Legend.  
    These are compared to Galactic standards (from the SAGA database as light grey dots, and \citealt{Li2018} as dark grey dots), and ultra faint dwarf galaxies (blue dots; see text for references).  Corrections for 1DNLTE effects are shown in Fig.~\ref{fig:abus_NLTE}, and we note that the \ion{Mn}{1} analysis is unreliable in both LTE and NLTE (see discussion in Appendix~\ref{app:mn}.)
    }
    \label{fig:abus_LTE}
\end{center}
\end{figure*}

Differential abundances were calculated with respect to two standard stars from our curated line list. 
This is a key component of our analysis as we have selected spectral lines that are available in the fainter C-19 stars and compared those mean abundances to the abundances that could be determined from a broader selection of spectral lines in the brighter standard stars.
Thus, the abundances for HD122563 were calculated in two ways: (1) using the C-19 line list (dark grey markers in Figs.~\ref{fig:abus_LTE} and \ref{fig:abus_NLTE} and labeled in Table~\ref{tab:abunds_stub} as ``HD12*"), and (2) using as many lines as possible from our HD122563 GHOST spectrum (light grey markers in Figs.~\ref{fig:abus_LTE} and \ref{fig:abus_NLTE} and labeled in Table~\ref{tab:abunds_stub} as ``HD122563").  The latter is a similar line list to \citet{Roederer14} for lines above 4000~\AA.  

Examination of the LTE and NLTE [X/Fe] abundances for HD122563 are shown in Figs.~\ref{fig:abus_LTE} and \ref{fig:abus_NLTE}. Some of our abundances are slightly offset from those in \citet{Roederer2014} (=R14, listed in Table~\ref{tab:params_sums}), primarily due to differences in the stellar parameters and updates in the available NLTE corrections.   
This is particularly clear when comparing our 1DLTE abundances in Fig. \ref{fig:abus_LTE}, where all of our abundances are in near perfect agreement with \citet{Roederer14}.

As a sanity check, we also examine the 1DLTE abundances for HD122563 from an analysis of both optical and near-IR spectra taken at the McDonald Observatory by \citet{Afsar2016}.
Generally there is excellent agreement with R14 and our results.
Their 1DLTE sodium abundance is [Na/Fe]=+0.1, which is in excellent agreement with our results; this element was not examined by R14. 
We notice a small difference in [Al/Fe] between the 1DLTE abundances for HD122563 between R14 and \citet{Afsar2016}.  Our results are in excellent agreement with the former, while the latter is only $\sim$0.2 dex lower. 
The only line in common between ours and both of their analyses is AlI 3961, for which we all have nearly identical EWs.  Thus, we attribute the Al abundance differences to the inclusion of \ion{Al}{1} lines in the near-IR by \citet{Afsar2016}.

The offsets in our HD122563 abundances compared to R14 are noted in Table~\ref{tab:params_sums}.
In general, inspection of Fig.~\ref{fig:abus_NLTE} shows that our differential abundances (C-19$-$Offset) place the C-19 chemical abundances in excellent agreement with other MW halo stars; including for \ion{Ti}{1}, \ion{Co}{1}, and \ion{Ba}{2}.  

Finally, our \ion{Mn}{1} analyses are neglected due to challenges in the line formation of the resonance lines, discussed further in Appendix~\ref{app:mn}.

\subsection{C-19 stars in the literature}
\label{sect:lit}

\begin{figure*}
\begin{center}
    \setlength{\unitlength}{1cm}
    \includegraphics[width=16cm]{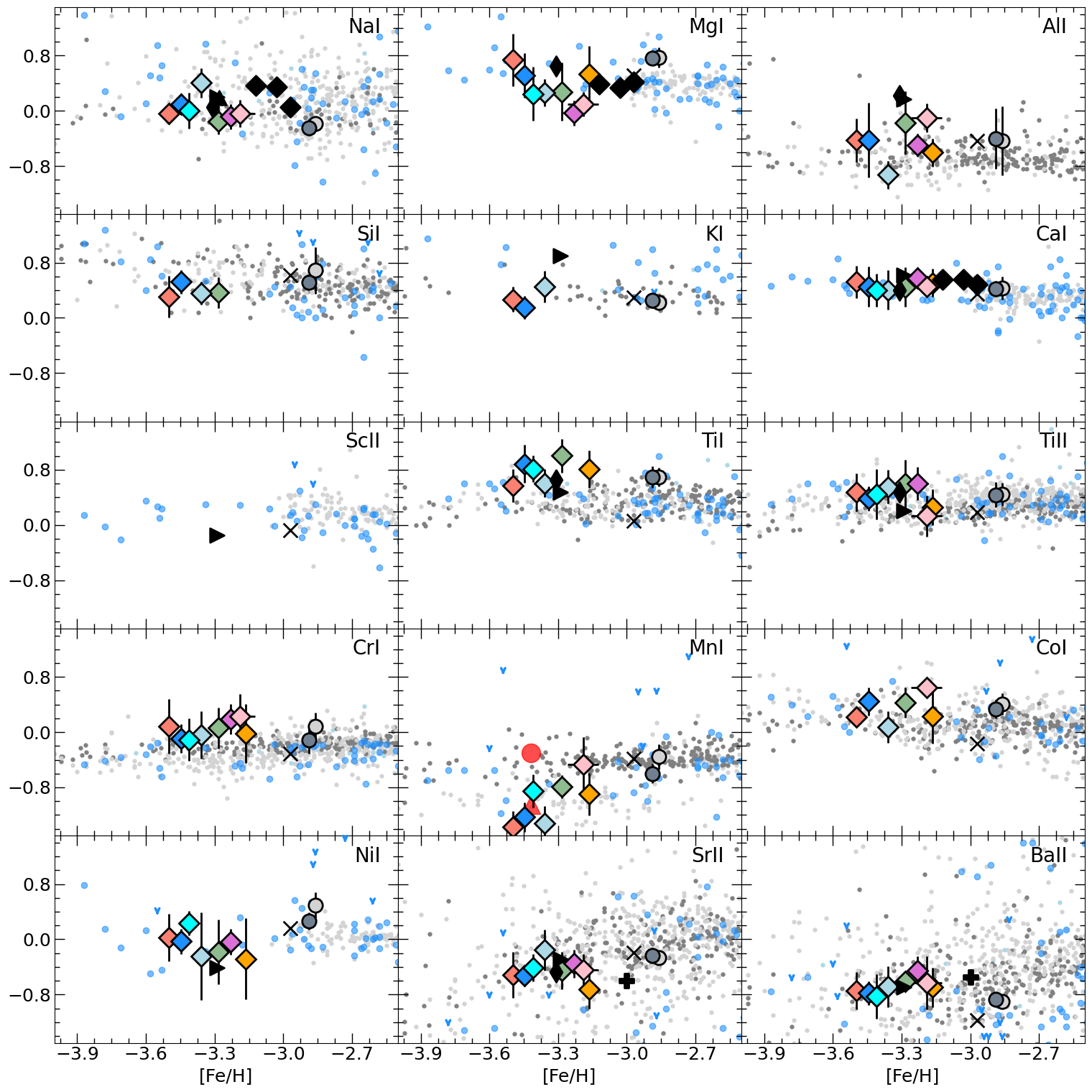}
    \\[0.15cm]
    \makebox[0pt][l]{\hspace{-6.5cm}\includegraphics[width=14cm]{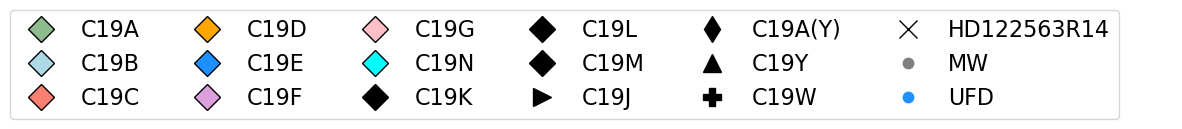}}
    \caption{1DNLTE [X/Fe] chemical abundances for 14 stars in the C-19 stream and HD122563, including the 8 stars from this paper and 7 stars in the literature (one star is analysed twice, C19A; see Legend).  
    These are compared with stars in the MW and UFDs (same symbols as in Fig.~\ref{fig:abus_LTE}). 
    Special analysis of \ion{Mn}{1} lines in BPS CS22925-0015 is shown in red, where the lower symbol is [Mn/Fe] from the 3 resonance lines used in this analysis and the upper symbol is [Mn/Fe] from 5 subordinate lines; this shows that the low [Mn/Fe] in the C-19 stars is unreliable (see Appendix~\ref{app:mn}).
    This view is zoomed in from Fig.~\ref{fig:abus_LTE}.
    }
    \label{fig:abus_NLTE}
\end{center}
\end{figure*}

Two C-19 members with detailed chemical abundance studies have been found serendipitously in the literature.  One was analysed from a Gemini/GRACES spectrum by \citet[][=C19J]{Jeong23} and the other from a Subaru/HDS spectrum by \citet[][=C19W]{Aoki13}. Their published chemical abundances are shown in Fig.~\ref{fig:abus_NLTE}.
These stars were not previously recognized as being member of C-19, and were observed in general follow-up studies of metal-poor stars.

This literature sample also includes our previous analysis of 3 stars in the core of C-19 from Gemini/GRACES spectra \citep[][=C19K, C19L, C19M]{MartinVenn2022}, and 2 outer stars analysed from VLT/UVES spectra \citep[][=C19J, C19Y]{Yuan2022}, shown in Fig.~\ref{fig:abus_NLTE}.  
The star C19J has been analysed in two studies, though only one included NLTE corrections -- as both are in good agreement in their stellar parameters and LTE abundances, then we adopt the NLTE results for C19J by \citet{Yuan2022}.

As seen in Fig.~\ref{fig:abus_NLTE},
the stellar abundances for the C-19 members in the literature are in good agreement with our new GHOST results, showing similar very low [Fe/H] values and element abundance ratios.
However, as the stars from the literature have not been analysed over the same wavelength regions, with the same stellar parameter determinations, nor with the same spectral analysis details as in this paper, then we may expect some differences.
Notably, the Fe results for \citet[C19J,][]{Jeong23} and \citet[C19W,][]{Aoki13} are slightly higher than ours (by $\sim$+0.2 only).
Also, higher [Al/Fe] and [Sr/Fe] abundances are found by \citet{Yuan2022} for C19J, and as both were calculated from strong lines in the blue, where their SNR was very low, we suggest those may actually be upper limits only, and/or have larger errors than reported\footnote{We note that \citet{Yuan2022} do not include uncertainties due to stellar parameters in their abundance analyses, and have therefore adopted $\sigma_{\rm par} = 0.2$ dex for all elements -- other than Al and Ni where we adopt $\sigma_{\rm par} = 0.3$ dex as those are typically more sensitive to the stellar parameters.}.  
The analyses of C19J between \citet{Jeong23} and \citet{Yuan2025} also show significant differences between their K, Ti, and Ba results; again, we adopt the NLTE results from \citet{Yuan2025} for C19J.

Similarly, our [Ni/Fe] for one  star (C19A) is somewhat lower than in \cite{Yuan2022}, and we suggest their result was likely an upper limit.
More positively, their NLTE abundances for [Ca/Fe] and [Ti/Fe] are in very good agreement with ours.

The published abundances for C19J-Y are included  in Tables~\ref{tab:abunds_stub} and \ref{tab:params_sums}.

\begin{table*}
%\centering
\scriptsize
\caption{Summary of the chemical abundances ([X/Fe]$_{\rm NLTE}$; CH is 1DLTE only) and total errors for each element in each member of C-19 and the standard stars.  
}
\label{tab:params_sums}
\setlength{\tabcolsep}{2.5pt}
%\resizebox{\textwidth}{!}{
\begin{tabular}{lccccccc}
\toprule
Target & [Fe/H]$_{\rm NLTE}^*$ & [CH/Fe] & [\ion{Na}{1}/Fe] & [\ion{Mg}{1}/Fe] &
[\ion{Al}{1}/Fe] & [\ion{Si}{1}/Fe] & [\ion{K}{1}/Fe] \\ [0.5ex]
% & {\hspace{1.5cm}} & NLTE & NLTE    &  NLTE   &  NLTE   &   NLTE  &  NLTE  & NLTE \\
\hline
HD122563 & $-2.88 \pm 0.09$ & $0.20 \pm 0.13$ & $-0.27 \pm 0.12$      & $0.73 \pm 0.06$ & $-0.43 \pm 0.26$ & $0.49 \pm 0.11$ & $0.23 \pm 0.11$ \\ 
%HD122563 & $-2.88 \pm 0.14$ & $-0.40 \pm 0.20$ & $-0.18 \pm 0.14$  & $0.81 \pm 0.16$ & $-0.42 \pm 0.46$ & $0.70 \pm 0.33$ & $0.24 \pm 0.15$ \\ # Full line list
$^\dagger$Offset ($-$R14) & $+0.07$ & $+0.13$ & ... &  $+0.22$ & $+0.01$ & $-0.13$ & $-0.07$   \\
HD222925 & $-1.44 \pm 0.11$ & $-0.22 \pm 0.15$ & ...               & $0.53 \pm 0.08$ & $-0.30 \pm 0.24$  
         & $0.08 \pm 0.22$ & $0.46 \pm 0.10$ \\ [2ex]
%HD222925 & $-1.41 \pm 0.08$ & $-0.10 \pm 0.20$ & ...               & $0.61 \pm 0.17$ & $-0.30 \pm 0.42$ & $0.38 \pm 0.19$ & $0.46 \pm 0.22$ \\ # Full line list
%
C19A & $-3.28 \pm 0.16$ & \ \ $0.29 \pm 0.27$ & $-0.16 \pm 0.17$  & $0.26 \pm 0.20$     & $-0.19 \pm 0.29$ & $0.36 \pm 0.17$ &     ...         \\
C19B & $-3.35 \pm 0.19$ & ...              & \ \ $0.40 \pm 0.21$  & $0.25 \pm 0.13$     & $-0.94 \pm 0.19$ & $0.35 \pm 0.23$ & $0.44 \pm 0.17$ \\
C19C & $-3.48 \pm 0.15$ & \ \ $0.19 \pm 0.18$ & $-0.06 \pm 0.14$     & $0.71 \pm 0.16$     & $-0.45 \pm 0.21$ & $0.28 \pm 0.19$ & $0.25 \pm 0.21$ \\
C19D & $-3.16 \pm 0.23$ & ...              & ...                  & $0.52 \pm 0.19$     & $-0.61 \pm 0.19$ &      ...        &     ...         \\
C19E & $-3.44 \pm 0.20$ & \ \ $0.09 \pm 0.32$ & \ \ $0.09 \pm 0.15$  & $0.51 \pm 0.15$     & $-0.43 \pm 0.32$ & $0.52 \pm 0.16$ & $0.14 \pm 0.17$ \\ 
C19F & $-3.23 \pm 0.19$ & ...              & $-0.23 \pm 0.16$   & $-0.03 \pm 0.18$ \ \  & $-0.16 \pm 0.17$ &      ...        &     ...         \\
C19G & $-3.18 \pm 0.22$ & ...              & $-0.05 \pm 0.20$     & $0.09 \pm 0.18$     & $-0.11 \pm 0.21$ &      ...        &     ...         \\
C19N & $-3.39 \pm 0.21$ & ...       & $-0.01 \pm 0.20$     & $0.23 \pm 0.20$     &     ...          &      ...        &     ...         \\
C19K & $-3.06 \pm 0.17$ & ...              & \ \ $0.33 \pm 0.22$  & $0.33 \pm 0.22$     &     ...          &      ...        &      ...        \\ 
C19L & $-3.14 \pm 0.15$ & ...              & \ \ $0.37 \pm 0.22$  & $0.39 \pm 0.20$     &     ...          &      ...        &      ...        \\ 
C19M & $-2.99 \pm 0.17$ & ...              & \ \ $0.05 \pm 0.17$  & $0.42 \pm 0.22$     &     ...          &      ...        &      ...        \\ 
C19J & $-3.31 \pm 0.21$ & \ \ $0.05 \pm0.36$   & \ \ $0.20 \pm 0.24$  &    ...          & \ \ $0.17 \pm 0.21$  &      ...        & $0.90 \pm 0.21$ \\ 
C19Y & $-3.28 \pm 0.20$ & ...   & \ \ $0.19 \pm 0.22$  &    ...              &      ...         &      ...        &      ...        \\ 
C19W & $-3.01 \pm 0.36$ & ...              & \ \ $0.50 \pm 0.31$  & $0.30 \pm 0.30$     &      ...         &       ...       &       ...       \\ [2ex]
$\bar{x} \pm \sigma$ & $-3.23 \pm 0.15$ & $0.16 \pm 0.09$ & $0.12 \pm 0.22$ & $0.33 \pm 0.19$ & $-0.34 \pm 0.32$ & $0.38 \pm 0.09$ & $0.43 \pm 0.29$ \\
$\bar{x}_w \pm \sigma_{\bar{x}_w}$ & $-3.25 \pm 0.05$ & $0.18 \pm 0.13$ & $0.06 \pm 0.05$ & $0.34 \pm 0.05$ & $-0.35 \pm 0.07$ & $0.39 \pm 0.09$ & $0.40 \pm 0.09$ \\
$\chi^2_\nu$        & 0.69             & 0.13             & 1.25            & 1.61            & 2.97             & 0.35            & 2.86        \\      
$\sigma_{\rm int}$  & 0.00             & 0.00             & 0.10            & 0.13            & 0.28             & 0.00            & 0.27        \\ 
\# targets        &  14              & 4                & 13              & 12              & 8                & 4               & 4             \\ [2ex]
\hline

\toprule
Target & [\ion{Ca}{1}/Fe] &  [\ion{Ti}{1}/Fe] & [\ion{Ti}{2}/Fe] & 
[\ion{Cr}{1}/Fe] & [\ion{Co}{1}/Fe] & [\ion{Sr}{2}/Fe] & [\ion{Ba}{2}/Fe] \\ [0.5ex]
\hline
HD122563 & $0.43 \pm 0.15$ & $0.67 \pm 0.16$ & $0.41 \pm 0.16$ & $-0.14 \pm 0.19$ & $0.30 \pm 0.18$  & $-0.26 \pm 0.17$ & $-0.91 \pm 0.15$ \\
$^\dagger$Offset ($-$R14) & $+0.08$ & $+0.61$ & $+0.22$ &  $+0.17$ & $+0.47$ & $-0.07$ & $+0.26$   \\
%HD122563 & $0.51 \pm 0.15$ & $0.71 \pm 0.16$ & $0.46 \pm 0.15$ & $0.10 \pm 0.17$  & $0.70 \pm 0.18$  & $-0.25 \pm 0.16$ & $-0.90 \pm 0.15$ \\  # Full line list
HD222925 & $0.30 \pm 0.16$ & $0.64 \pm 0.18$ & $0.38 \pm 0.17$ & $-0.17 \pm 0.17$ & $0.03 \pm 0.23$  &       ...        & $0.45 \pm 0.24$  \\ [2 ex]
%HD222925 & $0.24 \pm 0.15$ & $0.58 \pm 0.16$ & $0.32 \pm 0.16$ & $-0.04 \pm 0.15$ & $0.25 \pm 0.26$  &       ...        & $0.32 \pm 0.21$  \\  # Full line list
%
C19A & $0.48 \pm 0.20$ & $0.99 \pm 0.23$ & $0.58 \pm 0.21$ & $0.05 \pm 0.25$  & $0.42 \pm 0.22$  & $-0.47 \pm 0.30$ & $-0.63 \pm 0.20$ \\
C19B & $0.42 \pm 0.20$ & $0.60 \pm 0.21$ & $0.55 \pm 0.17$ & $-0.04 \pm 0.28$ & $0.07 \pm 0.22$  & $-0.17 \pm 0.33$ & $-0.70 \pm 0.26$ \\
C19C & $0.54 \pm 0.18$ & $0.55 \pm 0.20$ & $0.46 \pm 0.17$ & $0.07 \pm 0.26$  & $0.19 \pm 0.16$  & $-0.53 \pm 0.35$ & $-0.76 \pm 0.24$ \\
C19D & $0.53 \pm 0.19$ & $0.80 \pm 0.24$ & $0.26 \pm 0.19$ & $-0.04 \pm 0.31$ & $0.22 \pm 0.32$  & $-0.73 \pm 0.31$ & $-0.71 \pm 0.26$ \\
C19E & $0.50 \pm 0.24$ & $0.88 \pm 0.22$ & $0.38 \pm 0.17$ & $-0.10 \pm 0.21$ & $0.44 \pm 0.22$  & $-0.54 \pm 0.18$ & $-0.78 \pm 0.17$ \\
C19F & $0.58 \pm 0.18$ &     ...         & $0.60 \pm 0.20$ & $0.19 \pm 0.23$  &     ...          & $-0.40 \pm 0.23$ & $-0.30 \pm 0.19$ \\
C19G & $0.45 \pm 0.18$ &     ...         & $0.12 \pm 0.20$ & $0.22 \pm 0.26$  & $0.64 \pm 0.18$  & $-0.45 \pm 0.25$ & $-0.64 \pm 0.34$ \\
C19N & $0.39 \pm 0.22$ & $0.80 \pm 0.22$ & $0.44 \pm 0.22$ & $-0.13 \pm 0.27$ &     ...          & $-0.42 \pm 0.20$ & $-0.84 \pm 0.28$ \\
C19K & $0.56 \pm 0.24$ &    ...          &       ...       & $-0.16 \pm 0.21$ &     ...          &       ...        &    ...           \\
C19L & $0.56 \pm 0.16$ &     ...         &       ...       & $-0.17 \pm 0.21$ &     ...          &       ...        & $-0.22 \pm 0.22$ \\
C19M & $0.48 \pm 0.19$ &      ...        &       ...       & $-0.05 \pm 0.21$ &     ...          &       ...        &    ...           \\
C19J & $0.62 \pm 0.25$ & $0.48 \pm 0.26$ & $0.21 \pm 0.25$ &      ...         &     ...          & $-0.29 \pm 0.27$ & $-0.69 \pm 0.24$ \\
C19Y &      ...       &       ...        &       ...       &       ...        &     ...          &     ...          &      ...         \\
C19W &      ...       & $0.40 \pm 0.35$  &      ...        &       ...        &     ...          & $-0.60 \pm 0.42$ & $-0.55 \pm 0.44$ \\ [2ex]
$\bar{x} \pm \sigma$ & $0.51 \pm 0.07$ & $0.69 \pm 0.20$ & $0.40 \pm 0.16$ & $-0.01 \pm 0.13$ & $0.33 \pm 0.19$ & $-0.46 \pm 0.15$ & $-0.62 \pm 0.19$ \\
$\bar{x}_w \pm \sigma_{\bar{x}_w}$ & $0.51 \pm 0.06$ & $0.71 \pm 0.08$ & $0.41 \pm 0.06$ & $-0.03 \pm 0.07$ & $0.34 \pm 0.08$ & $-0.46 \pm 0.08$ & $-0.61 \pm 0.07$ \\
$\chi^2_\nu$        & 0.11             & 0.70             & 0.72            & 0.32            & 1.13             & 0.27            & 0.83   \\      
$\sigma_{\rm int}$  & 0.00             & 0.00             & 0.00            & 0.00            & 0.68             & 0.00            & 0.00   \\ 
\# targets          &  12              & 8                & 9               & 11              & 6                & 10              & 11     \\ [2ex]
\hline
% $<$C-19$>$ &  $0.52 \pm 0.06$ & $0.72 \pm 0.08$ & $0.42 \pm 0.06$ & $-0.02 \pm 0.07$ & $0.29 \pm 0.08$ & $-0.45 \pm 0.08$ & $-0.60 \pm 0.07$ \\
% Reduced $\chi^2$ & 0.14 & 0.62 & 0.79 & 0.35 & 2.25 & 0.26 & 0.80 \\
% \#  targets    &  12  & 8 & 9 & 11 & 6 &  10 & 11 \\ [2ex]
% \hline
\vspace{-2ex}
\end{tabular}
\\
$\bar{x} \pm \sigma$ = Arithmetic mean and standard deviation on the population.
\\
$\bar{x}_w \pm \sigma_{\bar{x}_w}$ = Weighted mean and error on the weighted mean.
\\
$\chi^2_\nu$ = Reduced $\chi^2$.  When $\chi^2_\nu < 1$, then the observed scatter is smaller than expected from the measurement uncertainties. 
% \[
% \chi^2_\nu =
% \frac{1}{N-1}
% \sum_{i=1}^{N}
% \frac{(x_i-\bar{x}_w)^2}{\sigma_i^2}
% \]
\\
$\sigma_{\rm int}$ =
Intrinsic abundance dispersion. When $\sigma_{\rm int}$ = 0.00, then no abundance variations remain after accounting for the measurement uncertainties.
% \[
% \sum_{i=1}^{N}
% \frac{(x_i-\bar{x}_w)^2}
% {\sigma_i^2+\sigma_{\rm int}^2}
% =N-1
% \]
\\
{$^*$} The mean 1DLTE iron abundances for these 14 members of C-19 is $\bar{x} \pm \sigma$([Fe/H]$_{\rm 1DLTE}$) $= -3.38 \pm 0.06$, with a similar reduced $\chi^2_\nu$ = 0.50.  
\\
{$^\dagger$} Offset ($-$R14) is the difference between our 1DNLTE abundances for HD122563 using the shorter C-19 line list compared to the abundances in \cite{Roederer14}. For more information, see Section~\ref{sect:diffabu}. 
%% C19J [Co/Fe] removed - clearly outlier in homogeneous analysis - was ($-0.41 \pm 0.25$).
\end{table*}

\subsection{Mean Abundances \& Dispersions in C-19}
\label{sect:meandisp}

The final NLTE abundances are presented in Table~\ref{tab:params_sums}.  Only CH is presented from a 1DLTE analysis.   Other elements are available in Table~\ref{tab:abunds_stub}, but not presented here due to missing detailed spectral line NLTE corrections (Sc, Ni) or other complications in the line formation (Mn, see Appendix~\ref{app:mn}).

Each abundance per star is a weighted mean, where the weighting has been by the number of spectral lines used (N).  [Fe/H] includes lines of \ion{Fe}{1} and \ion{Fe}{2}.
The weighted errors include the measurement errors (reduced by $\sqrt{N}$) and systematic errors (not reduced by N), combined in quadrature.

Treating C-19 as an ensemble of similar stars, we also have calculated several statistical properties for this sample; collected in Table~\ref{tab:params_sums}.
These include a simple arithmetic mean ($\bar{x}$) and standard deviation ($\sigma$), 
a weighted mean ($\bar{x}_w$) and error on the weighted mean ($\sigma_{\bar{x}_w}$). We also calculate 
a reduced $\chi^2_\nu$ to examine the scatter relative to the measurement uncertainties, and 
an intrinsic abundance dispersion $\sigma_{\rm int}$ to examine the star-to-star abundance variations after accounting for the measurement uncertainties.  
These latter two are calculated as:
\[
\chi^2_\nu =
\frac{1}{N-1}
\sum_{i=1}^{N}
\frac{(x_i-\bar{x}_w)^2}{\sigma_i^2}
\]
\[
\sum_{i=1}^{N}
\frac{(x_i-\bar{x}_w)^2}
{\sigma_i^2+\sigma_{\rm int}^2}
=N-1
\]
For any element, a reduced $\chi^2_\nu < 1$ indicates that the observed scatter is smaller than expected from the measurement uncertainties.
Similarly, when the intrinsic scatter $\sigma_{\rm int}$
is zero then there is no statistically detectable star-to-star abundance dispersion beyond the measurement errors.

\begin{figure}
    \setlength{\unitlength}{1cm}
    \includegraphics[width=8cm]{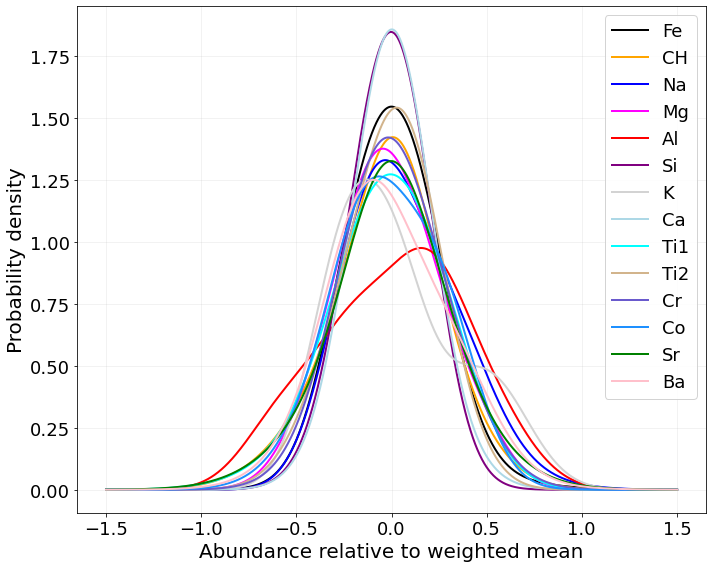}
    \caption{
    Probability density distribution of normalized abundances (relative to the weighted mean), estimated using an error convolved kernel density estimate (KDE). The convolution incorporates individual measurement uncertainties, in Table~\ref{tab:params_sums}, when characterizing the underlying distributions.  
    }
    \label{fig:KDE}
\end{figure}

An examination of the statistical values in Table~\ref{tab:params_sums} shows that most elements have a reduced $\chi^2_\nu < 1$, thus no intrinsic dispersion beyond the measurement errors.  Larger values are found for Al and K, although K is determined from  only 4  stars, making its elevated reduced $\chi^2_\nu$ value less robust.  Moderate values are found for Na, Mg, and Co.
Similarly, most elements have an intrinsic scatter $\sigma_{\rm int}$ equal or close to zero, other than Al and K, and to a lesser extent Na, Mg, and Co (the same elements above).  These results are again similar if only the 8 stars examined homogeneously with our GHOST spectra are considered (i.e., C19A-G from GHOST spectra, with or without C19N from HDS spectra);  the only exception is K where its $\chi^2_\nu < 1$ (0.8, but from only 3 stars).
We also tested the upper limits for CH in C19B and C19D, which are quite low (sub-solar); however, artificially treating those as data points still returns a low $\chi^2_\nu < 1$ (0.5) for CH.
A smooth representation of the distribution in stellar abundances incorporating the individual measurement errors is shown in Fig.~\ref{fig:KDE}.
The only element seen to vary significantly from the mean distribution is Al (in red). 

As a final test, we investigate whether the C-19 targets originate from a single chemical population by comparing a one- and two-component Gaussian mixture model for each element, incorporating individual measurement uncertainties through an error-convolved likelihood function. 
A Bayesian Information Criterion (BIC) was calculated 
for each element to evaluate the balance between goodness of fit and model complexity.
For all elements, the BIC comparison favoured the one-component model, including Al, indicating that the improvements in the two-component model were insufficient to justify its additional complexity. 
At the same time, the small sample sizes (subsets of the 14 stars and 14 elements) reduces the statistical power of these models to clearly distinguish between a single broad distribution and multiple underlying populations.

In summary, for this paper, we treat C-19 as a single stellar population, yet recognize that there appear to be significant aluminum variations.

\section{Discussion}
\label{sec:disc}

\subsection{MW comparisons}
\label{sec:litcomp}

The (1DLTE) chemical abundances for the C-19 stellar stream targets are shown in Fig.~\ref{fig:abus_LTE} and compared to other MW stars.
The MW halo stellar abundances are from \citet[][small dark grey]{Li2018, HLi2022} and selected data collected from the SAGA\footnote{SAGA database  available at: \url{http://sagadatabase.jp/} \\ \citep{Suda2008MW, Suda2017DG}.} database (i.e., \citealt{aoki2013hires, yong2013CEMP, Roederer14, Yong2021, Aoki2022, Sestito2023PIGSV}; all shown as small light grey points).
This plot of 1DLTE abundances is important as many of the MW comparison stars were analysed assuming no more than 1DLTE.  
This figure makes clear that the abundances estimated for C-19 stars are quite comparable to stars in the MW halo of similar metallicities.
The exceptions are \ion{K}{1}, \ion{Co}{1}, and \ion{Ni}{1}, which are all improved through NLTE corrections (next paragraph).  \ion{Mn}{1} in C-19 is also offset in LTE and NLTE, likely due to the break down of the line formation of the resonance lines in these EMP stars (see Appendix~\ref{app:mn}).

The 1DNLTE corrected abundances for our C-19 targets and standard stars are shown in Fig.~\ref{fig:abus_NLTE}, and compared to the 1DLTE MW star comparisons (i.e., small light grey dots are the same as in Fig.~\ref{fig:abus_LTE}).  
A few chemical abundance analyses were found in the literature for other stars that we now identify with C-19 (Section~\ref{sect:lit});  we cautiously add their published abundances to this figure (black symbols). It is important to note that the targets newly identified from the literature have not undergone the same homogeneous analysis that we have carried out here for our GHOST spectra.  For example, 
some of the C-19 targets taken from the literature do not have full wavelength coverage, e.g., four stars were analysed using Gemini-GRACES spectra, which only permits good spectral analyses above $\sim$480 nm in faint stars \citep[see][]{Kielty2021, Sestito2023PIGSV, Jeong23}. 

In general, we find good agreement between our chemical abundances from GHOST spectra of selected members of C-19, the published abundances of previously recognized and unrecognized members of C-19 in the literature, and the metal-poor MW halo stars. 
A more detailed comparison to stars in MW globular clusters will be presented in a forthcoming paper.

\subsection{UFD comparisons}
\label{sect:ufdcomp}

As the origin of C-19 is still uncertain,  we also compare the chemical abundances of its members to metal-poor stars in the ultra faint dwarf galaxies (UFDs), as shown in Fig.~\ref{fig:abus_LTE} (as blue dots).   
Abundance measurements (1DLTE) from the literature were initially selected from the SAGA database and updated with more recent results.  They include: Bootes I \citep{ Feltzing09,  Norris10, Gilmore13, Ishigaki14, Waller23}, 
Bootes II \citep{Ji16},
Carina II \citep{Ji20}, Carina III \citep{Ji20}, Cetus II \citep{Webber2023}, Coma Berenices \citep{Frebel10, Vargas13, Waller23}, Grus I \citep{Ji2019}, Grus II \citep{Hansen2020}, Hercules \citep{Koch08, Aden11, Vargas13,  Francois16}, Horologium I \citep{Nagasawa2018}, Leo IV \citep{Simon2010, Francois16, Vargas13}, Pisces II \citep{Spite2018Pis}, Reticulum II \citep{Ji16, Hayes23}, Segue 1 \citep{Norris10, Frebel14}, Segue 2 \citep{RoedererKirby2014}, Triangulum II \citep{Venn2017, Kirby2017, Ji2019}, Tucana II \citep{Ji16, Chiti18, Chiti2023}, Tucana III \citep{Hansen2017, Marshall2019}, Tucana V \citep{Hansen2024}, Ursa Major I \citep{Waller23}, and Ursa Major II \citep{Frebel10}.
The larger scatter in the UFD abundances are likely due to lower SNR, as UFD stars are fainter (farther) than similar objects in the MW.

While most chemical abundances of metal-poor stars in UFDs are similar to those of the stars in the MW, there are noticeable differences in the neutron-capture elements (Sr, Ba).  
In the MW halo, Sr and Ba in the EMP stars is expected to form primarily through the main-r, weak-r, and weak-s nucleosynthetic channels.
The main-r and weak-r processes build heavy elements through rapid neutron captures in high and moderate neutron flux environments.
The astrophysical sites of r-process nucleosynthesis remain debated, ranging from various types of core-collapse supernovae to neutron star mergers \cite[c.f.,][]{Nomoto13, Mashonkina2017,
Cowan21}. 
The main r-process produces the full range of r-process elements, including Ba and Eu, whereas weak r-process events primarily synthesize only the first r-process peak elements (like Sr).
In addition, a weak-s process occurs in metal-poor, fast rotating massive stars ($>15$M$_\odot$,  \citealt{Limongi2018, Prantzos2018, Sitnova2025}),
where rotation-induced mixing between the He-burning core and H-burning shell generates modest neutron densities that enable the synthesis of light neutron-capture elements (e.g., Sr).
Figs.~\ref{fig:abus_LTE} and \ref{fig:abus_NLTE} suggest the abundances of Sr and Ba in the C-19 stars are typical for their metallicity, and in better agreement with stars in the MW halo than those in the UFDs.  This is true for both the 1DLTE and 1DNLTE results, i.e., the NLTE corrections are small for this sample.

\cite{Mashonkina2017}
first noted that stars in UFDs typically have lower neutron-capture elements than similar metal-poor stars in the MW.  
This was thought to be due to 
stochastic and lower star formation efficiencies (e.g, \citealt{Tolstoy2009, Nomoto13, Revaz2018, Simon2019}), which leads to fewer events that can form the neutron-capture elements, thus lower Sr and Ba overall. 
The low total masses of the UFDs also make it difficult to retain Sr and Ba when they are both synthesized in and driven out by, e.g., core collapse supernovae.  
To emphasize this, both \cite{Mashonkina2017} and \cite{Ji2019} showed that the lower neutron-capture ratios are not typically seen in the metal-poor stars in classical dwarf galaxies, e.g., Sculptor \citep{Jablonka2015,
Skuladottir2015Scl, Hill19} or even Carina \citep{Venn12, Norris17}.

\begin{figure*}
\begin{center}
    \includegraphics[width=0.9\textwidth]{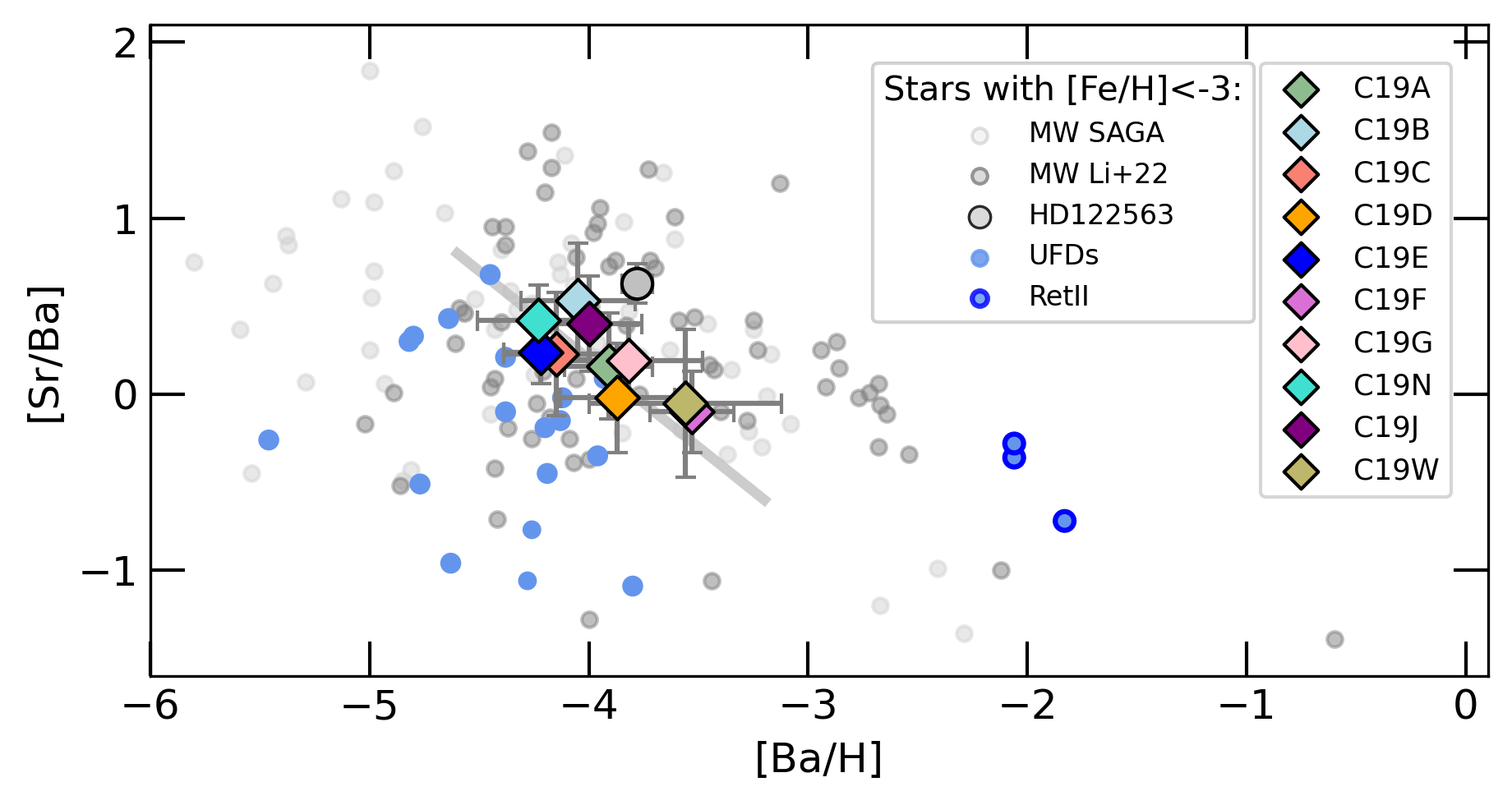}

\caption{
Sr and Ba NLTE abundances plotted as a function of [Ba/H], which includes a grey line to show a fixed [Sr/H] centered near the mean for C-19.
C-19 stellar abundances are compared to EMP stars in the Milky Way (SAGA by \citealt{Suda2008MW} in light grey; \citealt{HLi2022} in medium grey).
Abundances from individual stars in UFDs are also shown, collected from the SAGA database (\citealt{Suda2017DG}; see references in text) and the GHOULS survey (McConnachie \& Venn et al. 2026, Dovgal et al. 2026, medium blue). 
Only Ret II is offset due to a profound r-process enhancement event \citep{Ji2023Ret2}. 
We note that all C-19 members have [Sr/Ba]$>-0.4$,  the canonical r-process estimate from MW halo EMP stars \citep{Barklem05, Mashonkina2017}.
Our measurements for the standard star HD122563 are shown as a large filled grey circle with errorbar.
    } 
\label{fig:srbah}
\end{center}
\end{figure*}

In Fig.~\ref{fig:srbah}, we show that the ratio of [Sr/Ba] for EMP stars (recall, with [Fe/H]$<-3$) in UFDs tend towards the lower values also seen in MW halo stars.  Only Ret~II differs -- a unique UFD that has experienced a profound main r-process enrichment event \citep{Ji2016, Hayes2022, Ji2023Ret2} -- showing high barium ([Ba/H]$\sim-2$, and [Ba/Fe]$>+1$), and  [Sr/Ba]$<-0.4$, which is the canonical r-process estimate from metal-poor MW field stars (see \citealt{Barklem05, Mashonkina2017}.
That [Sr/Ba] can be higher in the UFDs than found in Ret~II itself, further emphasizes the need for 
distinct formation channels for these elements, and indicates that enrichment is not driven solely by the same r-process events.
Meanwhile, the stars in C-19 show higher [Sr/Ba] at a given [Ba/H] value than stars in the UFDs (see Fig.~\ref{fig:srbah}), though in good agreement with the abundance distribution of the MW halo EMP stars.
This chemical signature suggests that C-19 has more in common with the EMP stars in the MW halo than those in UFDs.  Of course, many EMP stars in the MW halo are likely to have formed in UFDs that were later accreted, thus this chemical signature and sample are not entirely independent.

Another interesting chemical signature distinguishing C-19 from EMP stars in UFDs is the lack of carbon-enhanced metal-poor (CEMP) stars.
While we know that our target selection was biased against finding the most strongly carbon-enhanced stars (see Section~\ref{sec:carb}), the low metallicity of C-19 means that stars with modest carbon enhancements might still have been selected. For example, \cite{Yuan2022, Yuan2025} excluded only stars with photometric metallicities above [Fe/H]$=-2.0$, leaving potential for extremely metal-poor stars with moderate carbon enhancements to remain in the sample.
Not finding any CEMP stars suggests that fraction may actually be quite low in C-19 (not just a selection bias). If so, this is very different from the fractions of CEMP stars in the MW halo and UFDs, where several studies (\citealt{yong2013CEMP, Placco14, Yoon16, Arentsen2022, Lucchesi2024}) have found 40 to 70\% of EMP stars with $-3.5 <$[Fe/H]$<-3.0$ in those systems are carbon-enhanced.
If the excess carbon is due to stochastic enrichment from faint Population III supernova at early times, as proposed in those studies, the lack of CEMP stars in C-19 supports that this was a well mixed system when its stars formed, regardless of its extremely low metallicity.

\section{Conclusions}

This paper demonstrates the remarkable performance of the new Gemini GHOST spectrograph, particularly when targeting faint (G $<17$) sources, such as the members of the C-19 stellar stream.
Both standard and high spectral resolution modes at the faint end were tested, yielding consistent results that are in excellent agreement with previous results in the literature.

The analysis of 7 newly identified C-19 members observed with GHOST, and 1 observed with Subaru/HDS, combined with published results for an additional 6 members, confirms the nearly negligible iron spread among these 14 stars, 
where [Fe/H]$_{\rm 1DNLTE} =-3.25 \pm 0.05$ or [Fe/H]$_{\rm 1DLTE} = -3.38\pm0.06$.
We also find that C-19 is highly homogeneous in its chemical abundances of other elements, consistent with a single stellar population.
The analysis of neutron-capture elements (Sr and Ba) suggests an enrichment history similar to the EMP stars in the MW halo, presumably forming in a system that could retain the heavy elements produced in massive stars and explosive events.  
However, the lack of CEMP stars in C-19 is unlike either the EMP stars in the MW halo or UFDs; possibly this is simply due to our selection biases, but it may also signal a well-mixed system when star formation occurred that is unusual at [Fe/H]$<-3$.
Thus, the chemistry of C-19 further underscores its uniqueness, offering new insights into the earliest stages of the formation of stars and galaxies. 

This paper demonstrates that precision spectroscopy with the GHOST spectrograph at Gemini-South offers a unique opportunity to explore the earliest stages of MW assembly, revealed through the chemical fingerprints of ancient stellar populations, such as the C-19 stellar stream, the most pristine stellar system found to date.

\section*{Acknowledgements}

We thank the anonymous reviewer for their careful reading of our manuscript and constructive comments which improved the quality of this paper.

KAV, DZ, and AD are grateful for funding through the National Science and Engineering Research Council Discovery Grant (RGPIN-2026-05879) and CREATE (2017-498006) programs, and the 
MITACS Globalink program. 
We also wish to acknowledge and respect the 
Lekwungen
%l\textschwa\textvbaraccent {k}$^{\rm w}$\textschwa\ng{}\textschwa n 
peoples on whose traditional territory the University of Victoria stands and the Songhees, Esquimalt and  WS\'ANE\'C peoples whose historical relationships with the land continue to this day.

ZY is supported by the National Key R\&D Program of China via  2024YFA1611601

NFM gratefully acknowledges support from the French National Research Agency (ANR) funded project ``Pristine" (ANR-18-CE31-0017) along with funding from CNRS/INSU through the Programme National Galaxies et Cosmologie and through the CNRS grant PICS07708. 

FG acknowledges support from the Research Fellow Gemini-ANID 2025 - 32RF250005 project.

ES acknowledges funding from the European Research Council (ERC, EARLYMW, Grant number 101170507).
Views and opinions expressed are those of the author(s) only and do not necessarily reflect those of the European Union or the European Research Council. Neither the European Union nor the granting authority can be held responsible for them.

GB acknowledges support under grant FOGALERA with reference PID2023-150319NB-
C21 funded by MICIU/AEI/10.13039/501100011033/FEDER, UE. This work is part of grant CEX2025-001609-S, awarded to the Instituto de Astrofísica de Canarias under the Severo Ochoa Centre of Excellence program and funded by MICIU/AEI/10.13039/501100011033.

CK acknowledges funding from the UK Science and Technology Facilities Council through grant ST/Y001443/1

This work is based on observations obtained with Gemini South/GHOST, during the commissioning runs in June and September 2022.
The international Gemini Observatory is a program of NSF’s NOIRLab, which is managed by the Association of Universities for Research in Astronomy (AURA) under a cooperative agreement with the NSF on behalf of the Gemini partnership: the National Science Foundation (United States), the National Research Council (Canada), CONICYT (Chile), Ministerio de Ciencia, Tecnolog\'{i}a e Innovaci\'{o}n Productiva (Argentina), Minist\'{e}rio da Ci\^{e}ncia, Tecnologia e Inova\c{c}\~{a}o (Brazil), and Korea Astronomy and Space Science Institute (Republic of Korea).

This work has made use of data from the European Space Agency mission Gaia \url{https://www.cosmos.esa.int/gaia}, processed by the Gaia Data Processing and Analysis Consortium (DPAC, \url{https://www.cosmos.esa.int/web/gaia/dpac/consortium}). Funding for the DPAC has been provided by national institutions, in particular the institutions participating in the Gaia Multilateral Agreement.

\section*{Data Availability}
  
GHOST spectra obtained via the SV, FT, and GO programs are available in Gemini Observatory Archive.
Only GHOST commissioning spectra are not available in the Gemini Observatory Archive.
Please contact us for access to the original raw data.  These files are listed in the Appendix.  All reduced data are incorporated into this paper.

\vspace{5mm}
\facilities{Gemini (GHOST), Gaia DR3}

\bibliography{ghost_planar,bibliography}{}
\bibliographystyle{aasjournal}

\appendix

\counterwithin{figure}{section}
\counterwithin{table}{section}

\section{\texttt{Galpy} Distance from Orbit Calculations}

\begin{figure}[h]
\begin{center}
    \setlength{\unitlength}{1cm}
    \includegraphics[width=16 cm]{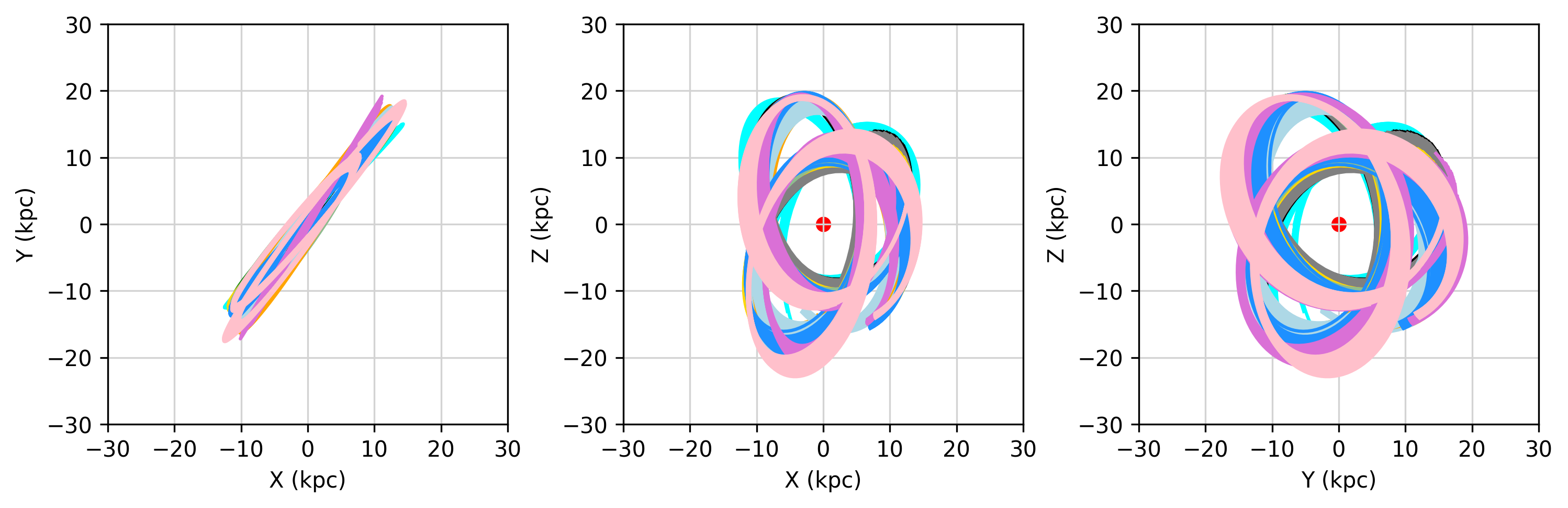}
    \caption{Orbits for all 14 C-19 members from \texttt{Galpy} (see text). Members in the main body of C-19 anchored at 18 kpc \citep{MartinVenn2022, Yuan2025}, and members in the C-19 extended stream have been adjusted slightly to match the main body.  New distances are listed as DIST(orbit) in Table~\ref{tab:targets}. The individual C-19 stream member orbits are identified using the same colours as in Fig.~\ref{fig:lb}.  A red dot marks the Galactic Centre.
    }
    \label{fig:orbits}
\end{center}
\end{figure}

Orbital parameters were determined using \texttt{galpy} \citep{Bovy15} and a Galactic halo stellar density distribution prior, i.e., \texttt{MWPotential2014} with an increased dark matter halo mass of $1.2 \times 10^{12}\rm M_{\odot}$  \citep{BlandHawthorn16}. 
Orbits were calculated using
Gaia DR3 proper motions \citep{GaiaDR3}, radial velocities from high resolution spectroscopy (see Table~\ref{tab:obs}), and a distance estimate.  Distance estimates for each C-19  stream member were initially set to the anchor distance for the main body of C-19 (18 kpc: \citealt{MartinVenn2022, Yuan2025}), and adjusted slightly until the orbit fits well with the stars in the main body (e.g., C19J -- C19M); see our values listed as DIST(orbit) in Table~\ref{tab:targets}.
Each star was backwards-integrated in the potential for 1 Gyr with 200 time-steps, producing the orbits shown in Fig.~\ref{fig:orbits} (using the same colours as in Fig.~\ref{fig:lb}).
Uncertainties on the orbital parameters were derived from a Monte Carlo simulation which draws 1000 times from a Gaussian distribution on uncertainties in the proper motions (estimated as $\Delta$PM $=0.01$ mas yr$^{-1}$), and the HRS radial velocities ($\Delta$RV $\leq$ 2 km\,s$^{-1}$).
The estimated uncertainties in our DIST(orbit) values are $\sim^{+1.0}_{-0.5}$ kpc, as noted in Table.~\ref{tab:targets}.

\section{DRAGONS Data Reduction Files per Target}

A full list of the GHOST data files used for the reduction of each target in C-19 is provided in Table~\ref{tab:exp}.
We include this Table here as some of our spectra are from commissioning spectra, which are not publicly available.

\begin{table*}
\caption{GHOST exposures for C-19 targets, including the calibration files used for the data reduction pipeline. These observations include both high resolution (HR) and standard resolution (SR) spectra observed in the single object mode. }
\label{tab:exp}
\centering
% \resizebox{0.47\textwidth}{!}
\hspace{-0.6cm}
\begin{tabular}{lllcc|llr}
\toprule
Target & Science  & Arm & t$_{\rm exp}$ & N$_{\rm exp}$ & 
Calibration Files & Type & t$_{\rm exp}$ \\
      & &     & (s) & & & & (s)\\
\hline
C19A & C19south\_hr\_1x4\_b3000\_ & Blue     &  3000 & x1   
& arc\_hr\_1x1\_thxe2\_20220628 & arc & 300 \\
   & \_r1000\_s80\_20220628 &  Red  & 1000 & x3 
& flats\_hr\_1x1\_20220628 & flat & 6 \\
   & (June 2022) &       &      &     
& bias\_1x4\_20220628 & 1x4 bias \\
  & &       &      &     
& bias\_1x1\_20220628  & 1x1 bias \\
  & &       &      &     
& HD122563\_hr\_1x1\_b60r10s1\_20220630 & slitview & 1 \\

\hline
C19B & C192544\_20220912\_ & Blue     &  600 & x3 
& arc\_SR\_1x1\_20220912 & arc & 300 \\
   & \_br600\_2x4\_sr &  Red  & 600 & x3 
   
& 20220914\_flat\_rb6s03\_1x1\_sr & flat & 6 \\
   & (Sept 2022) &       &      &     
& BIAS\_2x4\_20220915 & 2x4 bias \\
   &       &      &    & 
& BIAS\_1x1\_20220915  & 1x1 bias \\
   &       &      &    & 
& 20220914\_HR7596\_r1b2s01\_2x4\_sr & slitview & 0.1 \\

\hline
C19C & 20220913\_Gaia6944\_ & Blue     &  600 & x3  
& arc\_SR\_1x1\_20220912 & arc & 300 \\
   & \_br600\_2x4\_sr &  Red  & 600 & x3 
   
& 20220914\_flat\_rb6s03\_1x1\_sr.fits & flat & 6 \\ 
& (Sept 2022) &       &      &     
& BIAS\_2x4\_20220915 & 2x4 bias \\
  & &       &      &     
& BIAS\_1x1\_20220915  & 1x1 bias \\
  & &       &      &     
& 20220914\_HR7596\_r1b2s01\_2x4\_sr & slitview & 0.1 \\

\hline
C19D & 20220913\_Gaia7376\_ & Blue     &  900 & x3 
& arc\_SR\_1x1\_20220912 & arc & 300 \\
   & \_br900x3-s120\_2x4\_sr &  Red  & 900 & x3 
& 20220914\_flat\_rb6s03\_1x1\_sr & flat & 6 \\
   & (Sept 2022) &       &      &     
& BIAS\_2x4\_20220915 (blue) & 2x4 bias \\
     & &       &      &     
& BIAS\_2x8\_20220915 (red) & 2x8 bias \\
  & &       &      &     
& BIAS\_1x1\_20220915  & 1x1 bias \\
  & &       &      &    
& 20220914\_HR7596\_r1b2s01\_2x4\_sr & slitview & 0.1 \\
\hline
C19E & S20231212S0022 & Blue     &  900 & x3  
& S20231212S0027 & arc & 300 \\
   &  (2x4; sr; Dec 2023) &  Red  & 900 & x3 
& S20231210S0001 & flat & 6 \\
   &  &       &      &     
& S20231212S0009  & 2x4 bias \\
  & &       &      &     
& S20231212S0003  & 1x1 bias \\
  & &       &      &     
& S20231212S0022 & slitview & 70 \\
\hline

C19F & S20240907S0083 & Blue     &  1200 & x1 
& S20240906S0299 & arc &  \\
   & S20240907S0084  &  Red  & 1200 & x1 
&  S20240906S0300 & flat &  \\
   & S20240907S0085 &  (07 Sept 2024) & (sr) &  (2x8) 
& S20240906S0304  & 2x8 bias \\
  &  &       &      &     
& (science spectra) & slitview  & \\

\hline
C19G & S20240907S0085 & Blue     &  900 & x1  
& S20240906S0299 & arc &  \\
   & S20240907S0086  &  Red  & 900 & x1
&  S20240906S0300 & flat &  \\
   & S20240907S0098 &  (07 Sept 2024)  &  (sr)    &  (2x8)  
&  S20240906S0304 & 2x8 bias \\
  & S20240907S0099 &       &      &    & 
&  &    \\
 & S20240927S0031 & Blue     &  900 & x1 
& S20240927S0019 & arc &  \\
   & S20240927S0032  &  Red  & 900 & x1 
& S20240927S0020 & flat &  \\
   & S20240927S0033 & (27 Sept 2024)  &  (sr)    &  (2x8)  
& S20240927S0024  & 2x8 bias \\
  & S20240927S0034 &       &      &     
& (science spectra) & slitview  & \\
\hline

\end{tabular}
\end{table*}

\section{GHOST Spectra of C-19 Stars}

A sample of the GHOST spectra for the newly observed C-19 stars in this paper, and C19N from its Subaru/HDS spectrum.  All plots are Wavelength vs Normalized Flux.
Features shown include the spectral regions around the
CH 4300 feature, \ion{Sr}{2} 4215, \ion{Ba}{2} 4554, \ion{Na}{1}~D 5890, and \ion{Mg}{2}~b 5170 features; the latter includes \ion{Fe}{1} and Ti lines.

\begin{figure}[b]
\begin{center}
    \setlength{\unitlength}{1cm}
    \includegraphics[width=17cm]{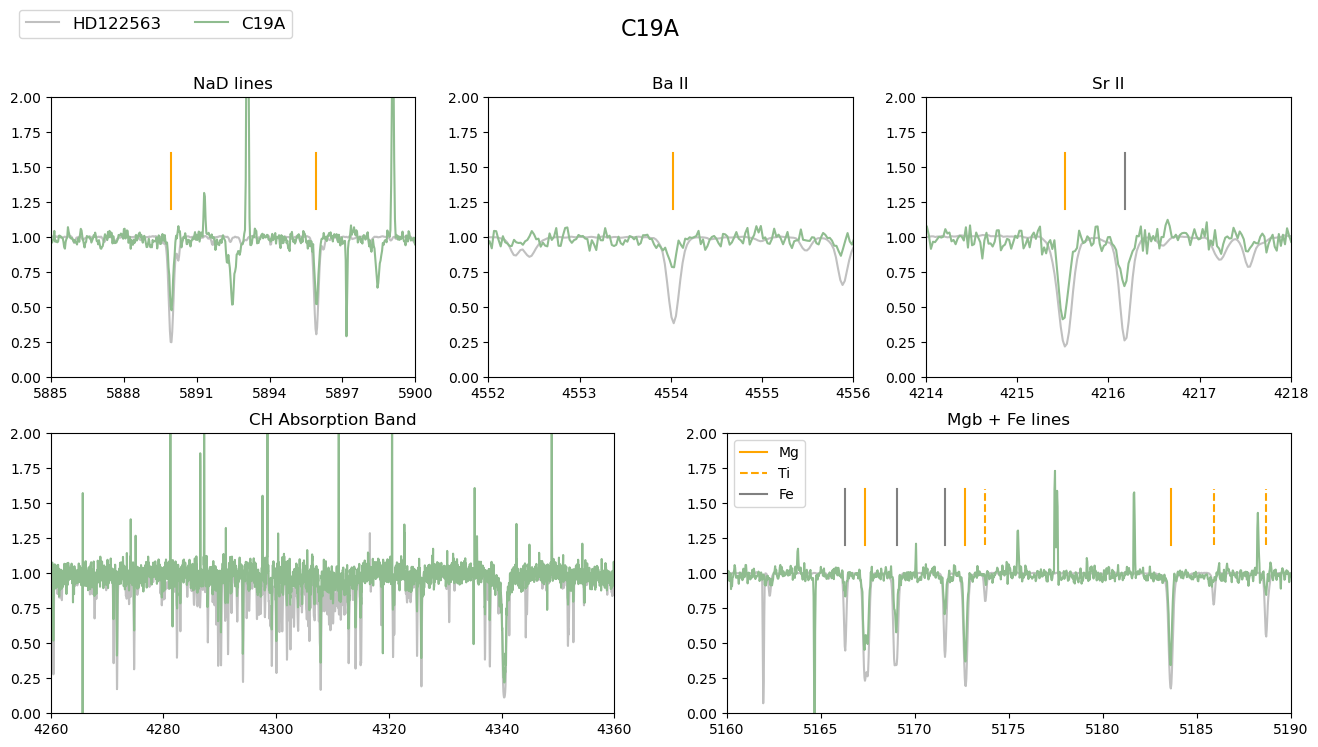}
    \caption{Sample of the GHOST spectra for C19A compared with two  EMP standard stars: HD122563 (light grey) and HE2340-6036 (dark grey, VLT/UVES). Key spectral features are identified (orange), as well as some Fe lines (grey). The NaD panel often includes interstellar absorption lines.
    All plots are Wavelength (\AA) vs Normalized Flux.}
    \label{fig:C19A}
\end{center}
\end{figure}

\begin{figure}
\begin{center}
    \setlength{\unitlength}{1cm}
    \includegraphics[width=17cm]{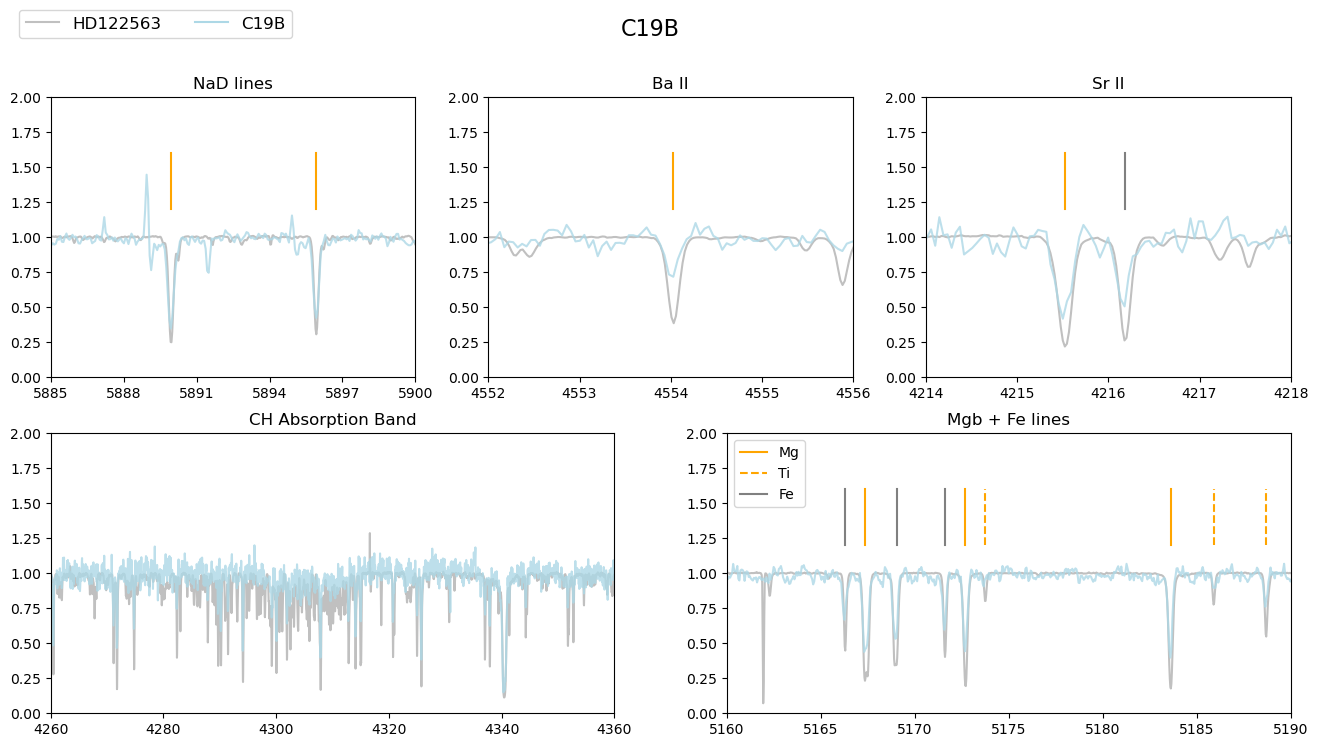}
    \caption{Same as Fig.~\ref{fig:C19A} for C19B. }
    \label{fig:C19B}
\end{center}
\end{figure}

\begin{figure}
\begin{center}
    \setlength{\unitlength}{1cm}
    \includegraphics[width=17cm]{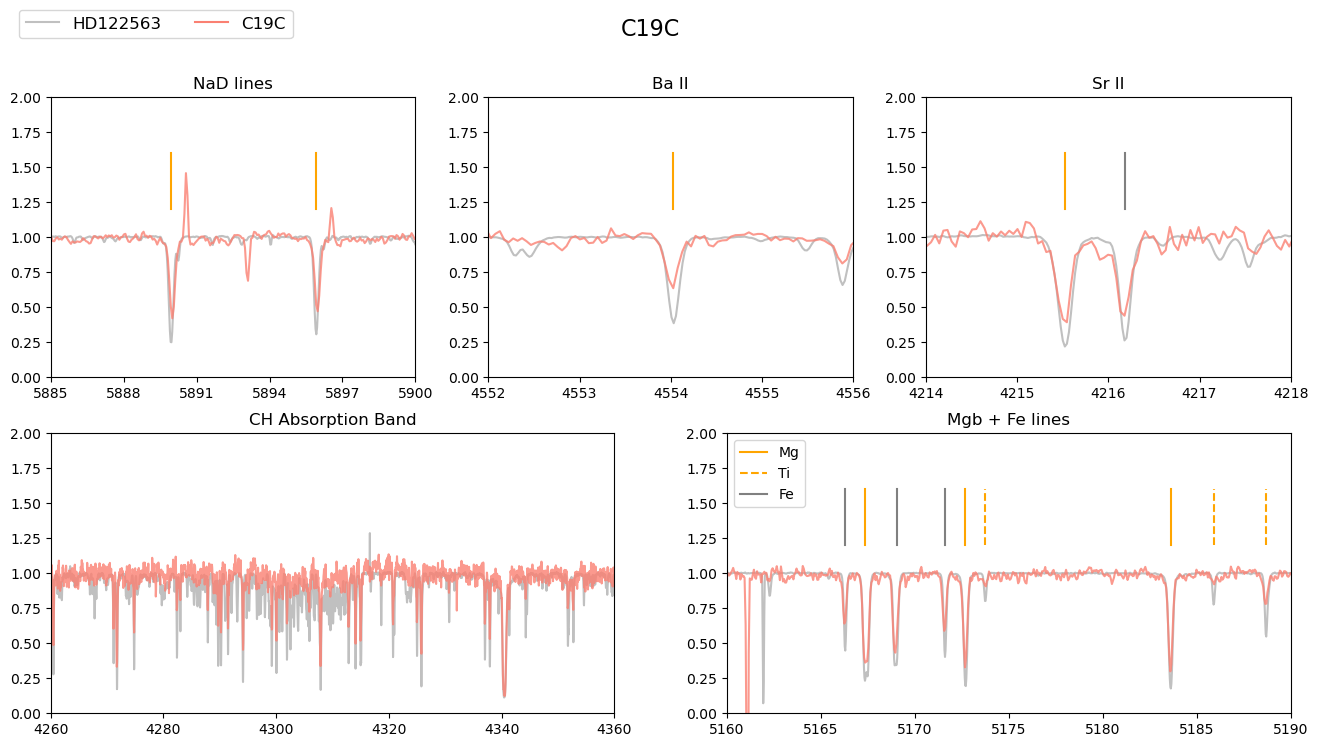}
    \caption{Same as Fig.~\ref{fig:C19A}  for C19C. }
    \label{fig:C19C}
\end{center}
\end{figure}

\begin{figure}
\begin{center}
    \setlength{\unitlength}{1cm}
    \includegraphics[width=17cm]{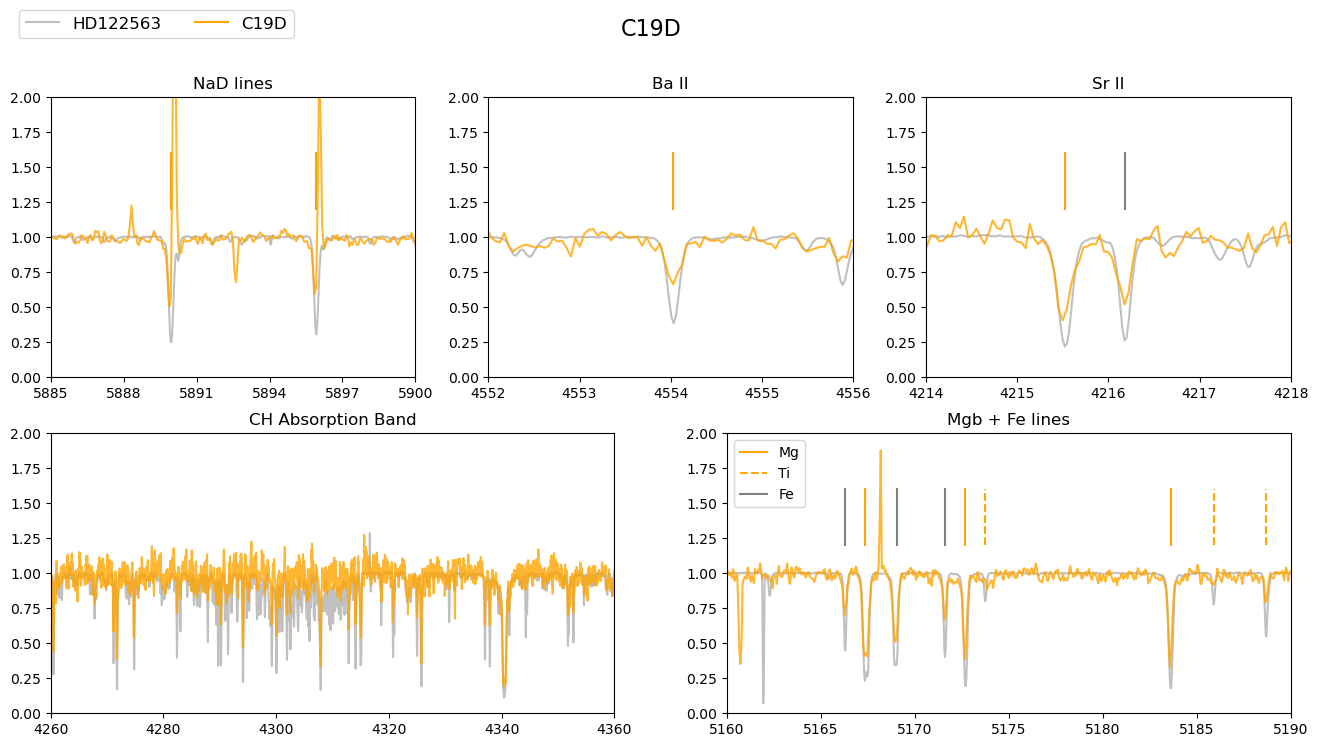}
    \caption{Same as above for C19D. 
    We note that this star is unusual as the \ion{Na}{1}~D lines have a stellar component in emission, possibly indicating a dust shell and an AGB star.  
    }
    \label{fig:C19D}
\end{center}
\end{figure}

\begin{figure}
\begin{center}
    \setlength{\unitlength}{1cm}
    \includegraphics[width=17cm]{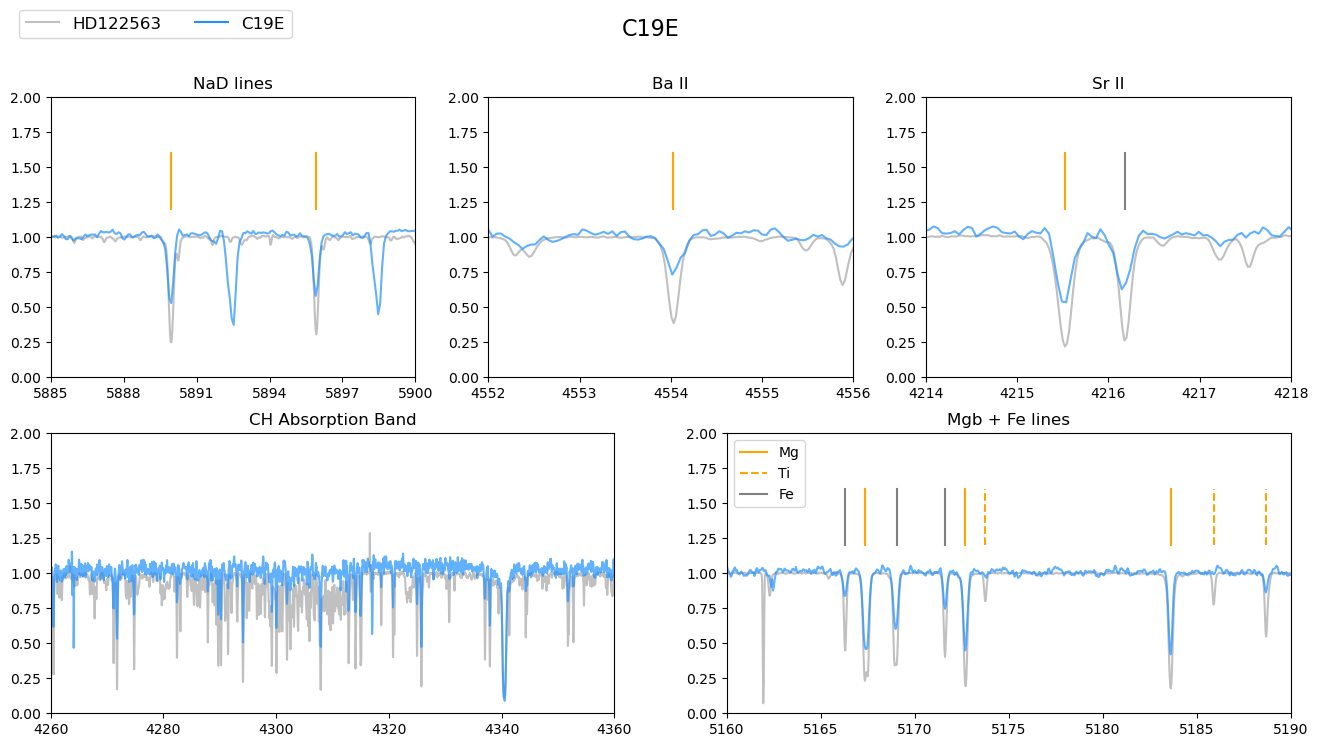}
    \caption{Same as Fig.~\ref{fig:C19A}  for C19E. }
    \label{fig:C19E}
\end{center}
\end{figure}

\begin{figure}
\begin{center}
    \setlength{\unitlength}{1cm}
    \includegraphics[width=17cm]{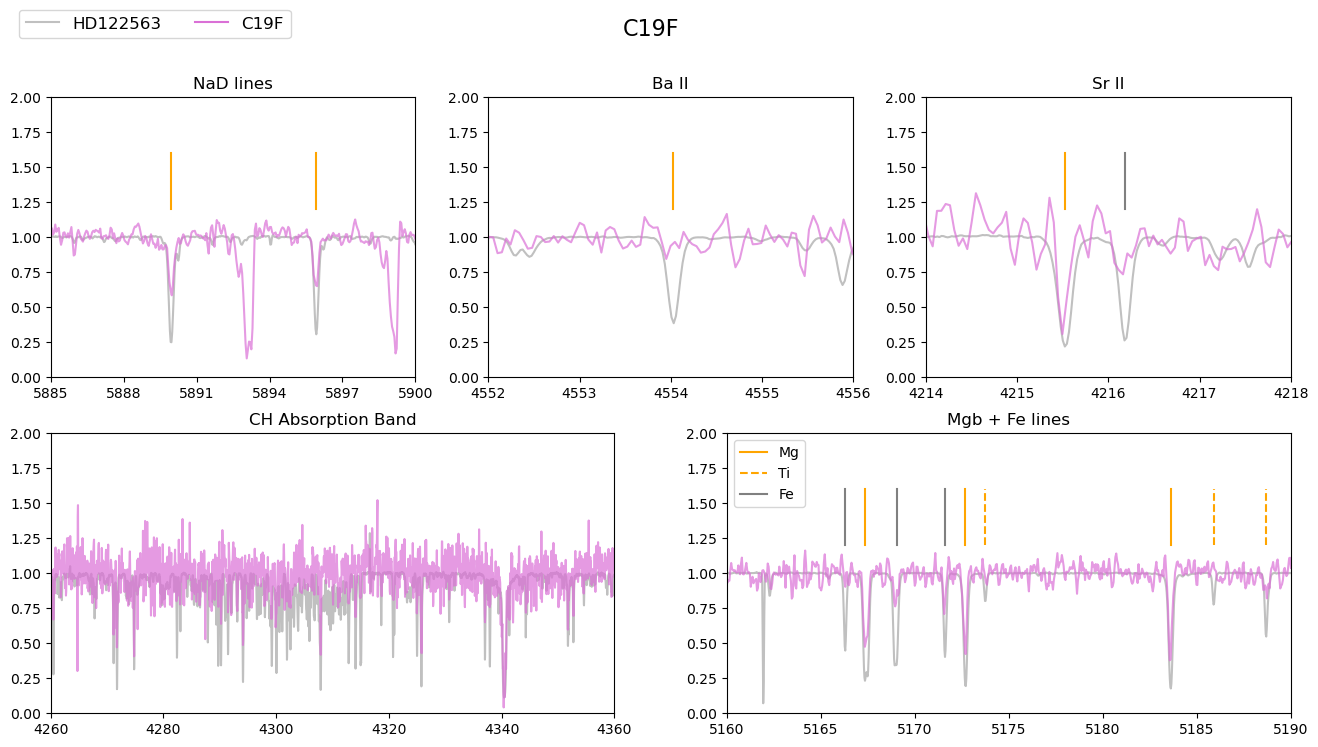}
    \caption{Same as Fig.~\ref{fig:C19A}  for C19F. }
    \label{fig:C19F}
\end{center}
\end{figure}

\begin{figure}
\begin{center}
    \setlength{\unitlength}{1cm}
    \includegraphics[width=17cm]{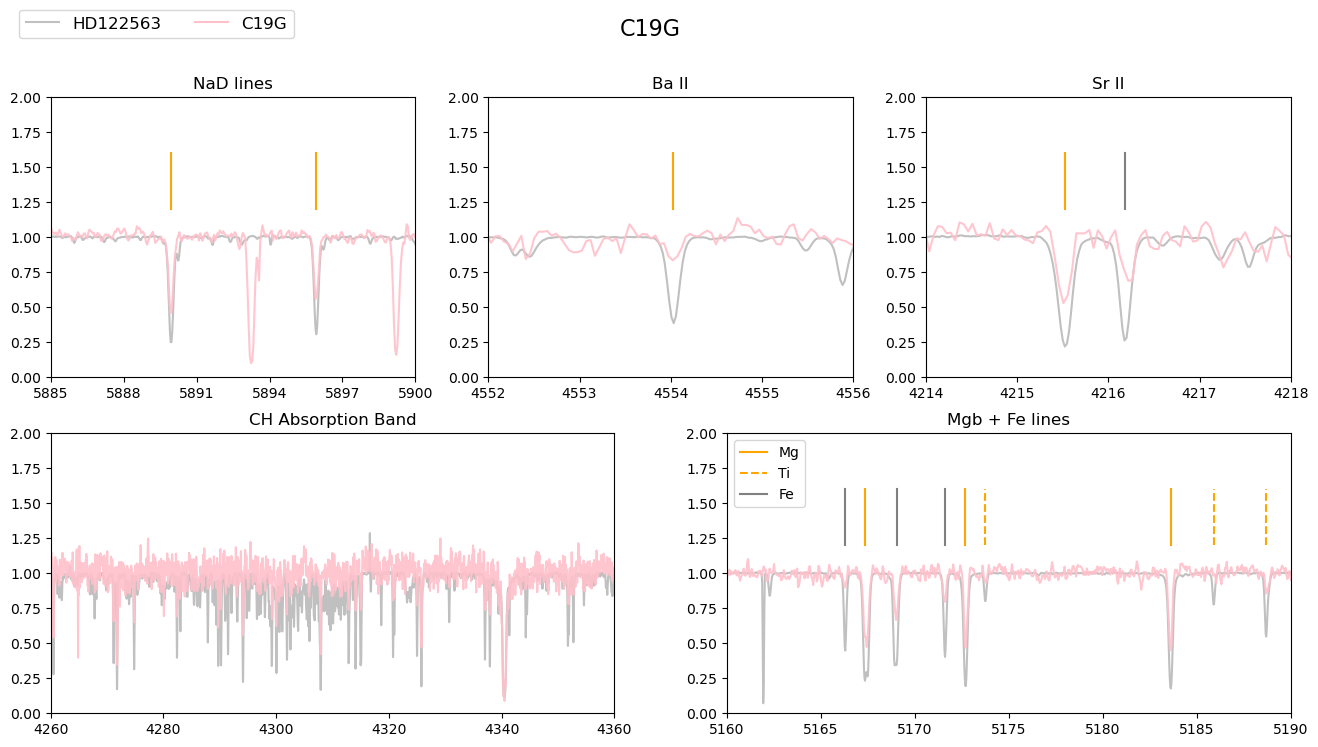}
    \caption{Same as Fig.~\ref{fig:C19A}  for C19G. }
    \label{fig:C19G}
\end{center}
\end{figure}

\begin{figure}
\begin{center}
    \setlength{\unitlength}{1cm}
    \includegraphics[width=17cm]{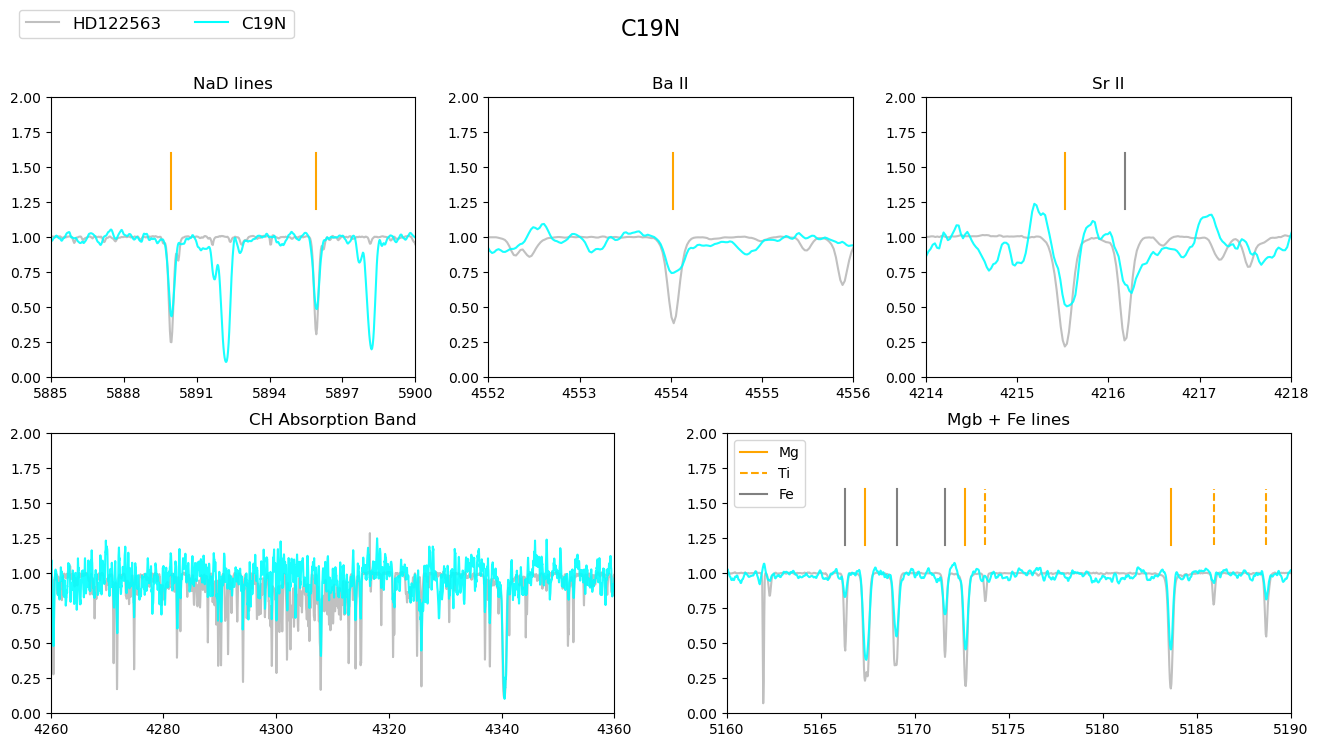}
    \caption{Same as Fig.~\ref{fig:C19A}  for C19N. }
    \label{fig:C19N}
\end{center}
\end{figure}

\clearpage

\section{Impact of stellar parameter errors on chemical abundances}
\label{app:errors}

In Table~\ref{tab:errsample2}, we provide a break down of the impact of the stellar parameter uncertainties on the chemical abundance errors for two representative stars: the standard star HD122563 and the target C19A.   
These two sample stars are selected as they bracket the range of our stellar parameters and the SNR of our GHOST spectra.
This Table is to show which stellar parameters dominate the abundance errors per species. We did not include an uncertainty due to the stellar metallicity as it was negligible in all cases when we used the error in the mean, i.e., $\sigma$(Fe)/$\sqrt{N}$.
These stellar parameter induced abundance errors were added in quadrature and listed as ``epar" in Table~\ref{tab:abunds_stub}.

\begin{table}
\caption{Sample abundance errors due to uncertainties in individual stellar parameters for the standard star HD122563 and the target star C19A. The uncertainties in effective temperature ($\Delta$T$_{\rm eff}$, in K), surface gravity ($\Delta$log$g$, in cgs units), and microturbulence ($\Delta\xi$, in \kms) are from Table~\ref{tab:params}. The effects on the mean abundances are shown per species, based on $N$ spectral lines (and the upper limit for Eu II, ``Eu\_UL").  HD12* is a label for the shorter line list analysis of HD122563.
}
\label{tab:errsample2}
\hspace{-1.5 cm}
\begin{tabular}{l|lcccl|lccc}
\hline
& \multicolumn{4}{c}{HD12*}   & & \multicolumn{4}{c}{C19A}     \\
 Elem &  N & $\Delta$T$_{\rm eff}$  & $\Delta$log$g$ & $\Delta\xi$  & &   N & $\Delta$T$_{\rm eff}$  & $\Delta$log$g$ & $\Delta\xi$  \\
&   & ($\pm$32)  &  ($\pm$0.13) & ($\pm$0.10)  & &  & ($\pm$111)  &  ($\pm$0.3) & ($\pm$0.1)   \\
\hline
CH          &  1 & 0.07  & 0.05  & 0.02  &  &  1 & 0.2   & 0.1   & 0.0   \\
\ion{Na}{1} &  1 & 0.053 & 0.013 & 0.048 &  &  2 & 0.120 & 0.029 & 0.036 \\
\ion{Mg}{1} &  4 & 0.029 & 0.019 & 0.015 &  &  5 & 0.095 & 0.050 & 0.023 \\
\ion{Al}{1} &  2 & 0.048 & 0.052 & 0.048 &  &  2 & 0.121 & 0.048 & 0.039 \\ 
\ion{Si}{1} &  1 & 0.033 & 0.021 & 0.027 &  &  2 & 0.121 & 0.037 & 0.023 \\
\ion{K}{1}  &  1 & 0.034 & 0.005 & 0.007 &  &    & ...   &  ...  &  ...  \\ 
\ion{Ca}{1} & 11 & 0.027 & 0.017 & 0.015 &  &  4 & 0.093 & 0.079 & 0.007 \\
\ion{Sc}{2} &  8 & 0.013 & 0.027 & 0.039 &  &  6 & 0.069 & 0.098 & 0.021 \\
\ion{Ti}{1} & 10 & 0.049 & 0.014 & 0.013 &  &  6 & 0.131 & 0.006 & 0.006 \\
\ion{Ti}{2} & 25 & 0.008 & 0.025 & 0.041 &  & 21 & 0.052 & 0.098 & 0.029 \\
\ion{Cr}{1} &  6 & 0.051 & 0.018 & 0.045 &  &  5 & 0.133 & 0.008 & 0.021 \\
\ion{Mn}{1} &  3 & 0.071 & 0.026 & 0.079 &  & 3 & 0.152 & 0.006 & 0.008 \\
\ion{Fe}{1} & 78 & 0.054 & 0.036 & 0.050 &  & 66 & 0.130 & 0.022 & 0.042 \\
\ion{Fe}{2} &  6 & 0.003 & 0.027 & 0.040 &  &  5 & 0.016 & 0.009 & 0.016 \\
\ion{Co}{1} &  3 & 0.049 & 0.030 & 0.050 &  &  4 & 0.137 & 0.003 & 0.011 \\
\ion{Ni}{1} &  3 & 0.057 & 0.023 & 0.061 &  &  3 & 0.134 & 0.001 & 0.013 \\
\ion{Sr}{2} &  2 & 0.025 & 0.013 & 0.070 &  &  2 & 0.086 & 0.078 & 0.088 \\
\ion{Ba}{2} &  4 & 0.022 & 0.039 & 0.029 &  &  2 & 0.090 & 0.101 & 0.011 \\
Eu\_UL &        3 & 0.018 & 0.035 & 0.002 &  &  1 & 0.083 & 0.092 & 0.020 \\
\hline
%\vspace{-2ex}
\end{tabular}
\end{table}

\clearpage

\section{Analysis of the Mn I lines}
\label{app:mn}

\begin{figure}[b]
\begin{center}
    \setlength{\unitlength}{1cm}
    \includegraphics[width=9cm]{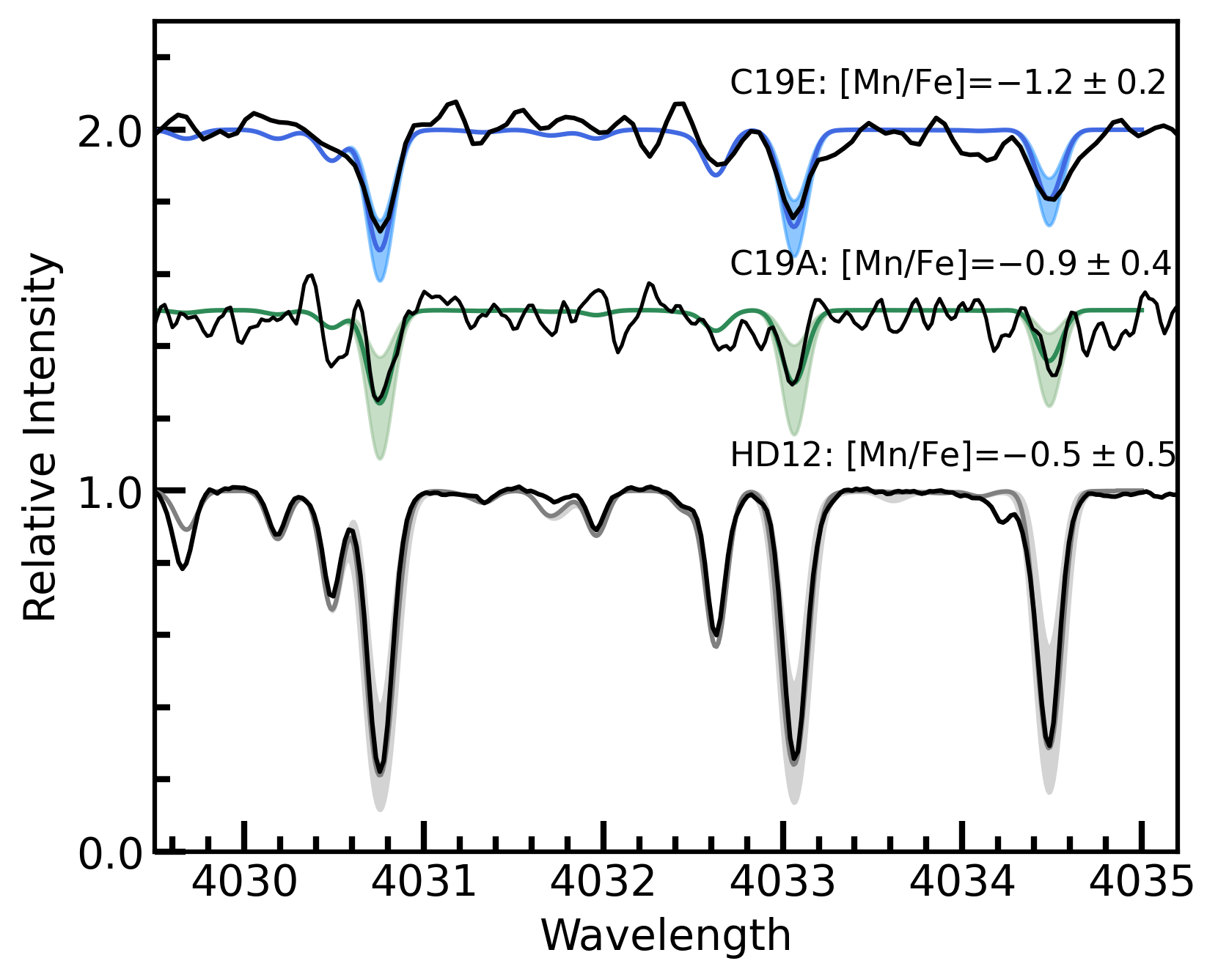}

    \caption{Syntheses of the \ion{Mn}{1} resonance lines in our GHOST spectra of HD122563 (=HD12 label), C19A (hotter) and C19E (cooler).  The normalized flux is shown, where spectra are offset for clarity.
    For each star, our best fit [Mn/Fe] are shown (coloured lines), and sample synthetic spectra offset by the uncertainties (grey lines). While our fits are excellent, the theoretical line formation in a 1DLTE analysis breaks down. See text and Table~\ref{tab:mn_cayrel}.}
    \label{fig:mnfe}
\end{center}
\end{figure}

The \ion{Mn}{1} abundances in the C-19 targets is much lower than seen in the metal-poor stars in the Milky Way.  We show the 1DLTE spectrum syntheses of the 
\ion{Mn}{1} resonance lines for three sample stars, HD122563, C19A, and C19E; see Fig.~\ref{fig:mnfe}.

To ascertain the importance of the apparent low abundance in [Mn/Fe] in the C-19 stars, we have also examined a VLT/UVES (R\,$\sim45$\,K) spectrum with extremely high SNR ($>500$) of the EMP standard star BPS~CS22952-0015 (Gaia~DR3 2440816964991039232), kindly provided by F.~Gran and V.~Hill (\textsl{priv.\,comms.}). 
This star has similar atmosphere parameters to our C-19 targets: \teff=4764\,K, \logg=1.18, [Fe/H]=$-3.39$ \citep{Soubiran2022}, or updated values based on Gaia DR3 data of \teff=4660\,K, \logg=1.77,  [Fe/H]=$-3.42$ \citep{Gran2026}.
Importantly this star has observations of both the strong \ion{Mn}{1} resonance lines and weak subordinate lines.
Analyses of both the resonance and subordinate lines shows an offset of $\sim$0.8 dex, see Table~\ref{tab:mn_cayrel},
which we do not see that in our analysis of the weak and strong lines in HD122563.

A similar result 
was seen for EMP stars in Sculptor by \citet{Skuladottir2024}. 
They found 3 Mn-poor stars in Sculptor, and suggested their [Mn/Fe] could be anti-correlated with temperature.  We do not find a relation with temperature here (i.e., our hotter C-19 stars have higher Mn, and our cooler stars have both high and low Mn).
More generally, we suggest that the line formation of the \ion{Mn}{1} resonance lines breaks down in the EMP stars, and requires a more detailed analysis, e.g., perhaps a 3DNLTE analyses such as the recent work on iron-group elements by \citet{Storm2025}.
%
%We do not include our [Mn/Fe] measurements throughout the rest of this analysis.
%
This analysis suggests that [Mn/Fe] in our C-19 stars is closer to solar, consistent with predicted yields from CCSN simulations, and in agreement with other EMP stars in both the MW and UFD galaxies.

\begin{table*}
\caption{Study of \ion{Mn}{1} lines in BPS CS 22925-0015 (adopting \teff=4660 K, \logg=1.2, [Fe/H]=-3.42) and HD122563.
}
\label{tab:mn_cayrel}
\hspace{-1.5 cm}
\begin{tabular}{lrr|rrrr|rrrr}
\hline
 &    &  . &   \multicolumn{4}{c}{BPS CS 22925-0015} & \multicolumn{4}{|c}{HD122563}    \\
Wavel & $\chi$ & loggf & EW  & A(\ion{Mn}{1}) &  [Mn/Fe] & [Mn/Fe] &  EW & A(\ion{Mn}{1}) & [Mn/Fe] & [Mn/Fe] \\
 (\AA) & (eV)   & . & m\AA &  & LTE & NLTE  & m\AA &  &  LTE & NLTE \\
\hline
4030.753  &  0.0  & $-0.47$ & 73.1 & 0.99 & $-1.02$ & -0.79 & 134.4 & 1.87 & $-0.66$ & $-0.53$ \\
4033.062  &  0.0  & $-0.62$ & 61.9 & 0.91 & $-1.10$ & -0.83 & 120.8 & 2.12 & $-0.41$ & $-0.29$ \\
4034.483  &  0.0  & $-0.81$ & 51.1 & 0.89 & $-1.12$ & -0.81 & 112.7 & 2.23 &  $-0.30$ & $-0.19$ \\
4041.355  &  2.11 &  $0.28$ & 10.2 & 1.35 & $-0.66$ & -0.20 &  38.3 & 2.21 & $-0.32$ & $0.02$ \\
4055.544  &  2.14 & $-0.07$ & 4.5  & 1.58 & $-0.43$ &  0.02 &  17.1 & 2.12 & $-0.41$ & $-0.10$ \\
4079.234  &  2.14 & $-0.42$ & ...  & ...  & ...     & ...   &  11.7 & 2.28 & $-0.25$ & $0.06$ \\
4082.939  &  2.18 & $-0.35$ & ...  & ...  & ...     & ...   &  10.3 & 2.19 & $-0.34$ & $-0.03$ \\
4754.042  &  2.28 & $-0.09$ & 5.0  & 1.46 & $-0.55$ & -0.10 &  ...  & ...  &     ... \\
4783.427  &  2.30 &  $0.04$ & 5.8  & 1.66 & $-0.35$ &  ...  &  23.0 & 1.97 & $-0.56$ & $-0.22$ \\
4823.524  &  2.32 &  $0.14$ & 6.4  & 1.62 & $-0.39$ &  0.08 &  25.0 & 2.21 &  $-0.32$ & $0.02$ \\
\hline
\vspace{-2ex}
\end{tabular}
\end{table*}

% \begin{table}
% \caption{Study of \ion{Mn}{1} lines in BPS CS 22925-0015 (adopting \teff=4660, \logg=1.2, [Fe/H]=-3.42).
% }
% \label{tab:mn_cayrel}
% \hspace{-1.5 cm}
% \begin{tabular}{llrrrrr}
% \hline
% Wavel & $\chi$ & loggf & EW  & A(MnI) &  [Mn/Fe] & [Mn/Fe]  \\
%       & (eV)   &       &     &        & (LTE)    & (NLTE) \\
% \hline
% 4030.753  &  0.0  & -0.47 &  73.1  &  0.99  &  -1.02 & -0.79 \\
% 4033.062  &  0.0  & -0.62 &  61.9  &  0.91  &  -1.10 & -0.83 \\
% 4034.483  &  0.0  & -0.81 &  51.1  &  0.89  &  -1.12 & -0.81 \\
% 4041.355  &  2.11 &  0.28 &  10.2  &  1.35  &  -0.66 & -0.20 \\
% 4055.544  &  2.14 & -0.07 &  4.5   &  1.58  &  -0.43 &  0.02 \\
% 4754.042  &  2.28 & -0.09 &  5.0   &  1.46  &  -0.55 & -0.10 \\
% 4783.427  &  2.30 &  0.04 &  5.8   &  1.66  &  -0.35 &  ... \\
% 4823.524  &  2.32 &  0.14 &  6.4   &  1.62  &  -0.39 &  0.08 \\
% \hline
% \vspace{-2ex}
% \end{tabular}
% \end{table}

\clearpage

\section{Spectral Linelists}

The spectral line list was carefully selected to maximize the number of lines observed in the C-19 targets, while also providing a homogeneous data set.
In Tables~\ref{tab:felines} and \ref{tab:otherlines}, the elements are identified as Elem using the code where the atomic number is followed by the excitation state after the decimal (e.g., \ion{Fe}{1} is 26.0).  Each spectral line is then identified by its wavelength (\AA), excitation potential ($\chi$ in eV), and oscillator strength (log\,gf), each as listed in \texttt{linemake} (see Main text), and equivalent width measurement (in m\AA).  
To squeeze the tables, HD22 = HD222925 and HD12=HD122563).
Lines that are used for the 
C-19 stellar abundances but are too strong for reliable analyses in the standard stars HD122563 and HD222925 are noted in italics.
These lines are kept in our analysis, but used with caution, mainly as comparisons with other weaker lines.
The very strong Ca II triplet lines are identified at end of these Tables, including their wavelengths and the stellar equivalent widths, each shown in italics.
Missing data, extremely strong or weak lines, or very noisy data are not measured and noted as ellipses.

\begin{table*}
\caption{Equivalent widths of iron lines in C-19 members}
\label{tab:felines} 
\hspace{-2.5 cm}
\begin{tabular}{lllrrrrrrrrrrr} 
\toprule 
Wave & Elem & \, $\chi$ & loggf &  HD22 & HD12 & C19A &  C19B &  C19C & C19D & C19E  & C19F & C19G & C19N \\
\AA  &   & eV   &            & m\AA  & m\AA  & m\AA &  m\AA &  m\AA & m\AA & m\AA  & m\AA & m\AA & m\AA \\ 
\hline 
3878.018 & 26.0 & 0.96 & -0.91 &    ...  & 148.9 &  ...  & 132.0 & 115.0 &  ...  & 120.9 & ... & ... & ...\\
3902.945 & 26.0 & 1.56 & -0.47 &    ...  & 127.0 & 80.0 & 102.0 & 110.0 & 110.0 & 92.8 & ... & ... & ... \\
3920.258 & 26.0 & 0.12 & -1.75 & ...  & {\it 160.0} &  ...  &  ...  & 150.0 & 130.0 & 121.6 & ... & ... & ... \\
3922.912 & 26.0 & 0.05 & -1.65 &     ...  & {\it 175.0} & 100.0 & 120.0 & 138.0 & 145.0 &  ...  & ... & ... & ... \\
4005.242 & 26.0 & 1.56 & -0.61 &  ...  & 127.9 & 67.0 & 78.0 & 140.0 & 113.3 & 100.2 & 98.0 & 78.2 & ... \\
4063.594 & 26.0 & 1.56 & 0.06 &    ...  &  ...  & 90.0 & 107.6 & 144.0 & 122.7 & 133.0 & 160.0 & ... & ... \\
4071.738 & 26.0 & 1.61  & -0.02 &   ...  &  ...  & 85.0 & 104.9 & 144.0 & 120.0 & 134.0 & 145.0 & ... & 140.0 \\
4132.058 & 26.0 & 1.61 & -0.68 & 
...  & 131.8 & 60.0 & 110.0 & 114.0 & 135.0 & 106.0 & 100.0 & ... & 148.1 \\
4143.868 & 26.0 & 1.56 & -0.51 & 
...  & 130.5 & 71.6 & 105.0 & 130.0 & 113.3 & 99.0 & 82.0 & 76.6 & 134.1 \\
4181.754 & 26.0 & 2.83 & -0.37 & 
100.5 & 68.8 & 24.0 &  ...  & 33.6 &  ...  & 38.7 & ... & ... & ... \\
4187.039 & 26.0 & 2.41 & -0.55 & 
110.5 & 85.2 & 35.0 & 55.0 & 59.5 &  87.0 & 38.1 & 60.0 & 60.0 & ... \\
4187.795 & 26.0 & 2.43 & -0.55 & 
120.8 & 90.6 & 24.0 & 44.0 & 76.2 & 84.1 & 50.1 & 70.0 & ... & ... \\
4191.430 & 26.0 & 2.47 & -0.67 & 
103.7 & 76.6 &  ...  & 45.0 &  48.0 &  ...  & 28.9 & 60.0 & ... & ... \\
4199.095 & 26.0 & 3.05 & 0.16 & 
111.9 & 78.0 & 34.0 & 43.0 & 66.3 &  55.0 & 26.7 & 60.0 & 48.7 & ... \\
4202.029 & 26.0 & 1.49 & -0.71 & 
...  & 128.2 &  ...  & 109.0 & 103.0 & 115.1 & 89.6 & 110.0 & 73.1 & ... \\
4216.184 & 26.0 & 0.0 & -3.36 & 96.9 & 111.5 & 39.0 & 83.0 & 104.0 & 90.0 & 69.8 & ... & ... & 94.6 \\
4222.213 & 26.0 & 2.45 & -0.97 & 89.2 & 68.9 &  19.0 &  30.0 & 41.3 & 40.0 & 36.2 & ... & ... & ... \\
4227.426 & 26.0 & 3.33 & 0.27 &   ...  & 88.0 & 30.0 & 30.0 & 65.0 & 35.0 & 26.3 & ... & ... & ... \\
4233.603 & 26.0 & 2.48 & -0.60 &   102.3 & 83.0 & 36.0 & 43.0 & 76.3 & 70.0 & 37.6 & ... & ... & ... \\
4250.119 & 26.0 & 2.47 & -0.41  & 116.2 & 91.2 & 29.0 & 90.0 & 65.0 & 58.7 & 31.9 & 40.0 & ... & ... \\
4250.787 & 26.0 & 1.56 & -0.71 &    ...  & 123.2 & 72.5 & 90.0 & 108.2 & 124.1 & 94.0 & 60.0 & 71.9 & 132.0 \\
4260.474 & 26.0 & 2.40 & 0.11 &    ...  & 112.4 & 65.0 & 70.0 & 92.1 & 105.0 & 96.1 & 60.0 & 76.7 & 151.0 \\
4271.153 & 26.0 & 2.45 & -0.35 &   116.2 & 99.3 & 38.6 & 90.0 & 75.6 & 76.7 & 74.8 & 70.0 & 72.4 & ...  \\
4271.760 & 26.0 & 1.49 & -0.16 &   ...  &  ...  & 85.0 & 135.0 & 137.5 & 136.7 & 107.2 & 135.7 & ... & 150.0 \\
4282.403 & 26.0 & 2.18 & -0.78 &   116.4 & 87.4 & 49.0 &  69.0 & 50.0 & 60.0 & 39.9 & ... & 45.0 & ... \\
4325.762 & 26.0 & 1.61 & 0.01 &    ...  & 90.0 & 116.0 & 138.6 & 140.0 & 113.9 & ... & ... & ... \\
4337.045 & 26.0 & 1.56 & -1.70 &   91.0 & 40.0 &  ...  & 60.2 & 78.7 & 49.8 & 60.0 & ... & 86.0 \\
4352.735 & 26.0 & 2.22 & -1.28 &   87.8 & 70.1 &  20.0 & 45.0 & 41.6 & 37.8 &  28.0 & ... & ... & 63.0 \\
4375.930 & 26.0 & 0.0 & -3.03 &   114.5 & 124.4 & 64.0 & 111.0 & 118.5 & 110.0 & 87.9 & 70.0 & ... & 110.0 \\
4404.750 & 26.0 & 1.56 & -0.14 &    ...  & {\it 155.0} &  ...  & 115.0 & 141.5 & 130.0 & 110.5 & 120.0 & ... & 147.8 \\
4415.122 & 26.0 & 1.61 & -0.62 &    ...  & 131.7 & 69.8 &  88.0 & 114.5 & 132.4 & 88.8 & 91.0 & 96.0 & ... \\
4427.310 & 26.0 & 0.05 & -2.92 &    119.8 & 125.8 & 62.5 & 98.0 & 130.0 & 120.0 & 106.6 & 50.0 & ... & 115.0 \\
4430.614 & 26.0 & 2.22 & -1.66  & 68.9 & 51.4 &  ...  &  ...  & 37.0 & 38.0 &  ...  & ... & ... & ... \\
4442.339 & 26.0 & 2.20 & -1.26   & 95.3 & 76.7 &  16.0 & 55.0 &  33.0 &  ...  & 31.3 & ... & ... & 39.0 \\
4447.717 & 26.0 & 2.22 & -1.34   & 87.8 & 69.4 &  16.0 & 34.0  & 40.0 & 37.0 & 36.4 & ... & 37.0 & ... \\
4459.117 & 26.0 & 2.18 & -1.28   &  ...  & 90.1 &  25.0 & 40.0 & 50.0 & 78.3 & 32.3 & ... & ... & ... \\
4461.653 & 26.0 & 0.09 & -3.21   & 105.5 & 116.5 & 60.0 &  ...  & 120.0 & 110.0 & 84.6 & 75.0 & ...& 102.4  \\
4466.551 & 26.0 & 2.83 & -0.60   & 89.4 & 82.9 & 23.0 & 49.0 & 58.0 & 60.0 &  36.0 & ... & ... & ... \\
4489.739 & 26.0 & 0.12 & -3.97  & 62.4 & 82.4 &  20.0 & 40.0 & 64.4 & 48.0 & 34.0 & ... & ... & ... \\
4494.563 & 26.0 & 2.20 & -1.14   & 102.9 & 80.6 & 28.0 & 51.0 &  ...  &  34.0 & 32.5 & 38.5 & ... & ... \\
4528.614 & 26.0 & 2.18 & -0.82   &  ...  & 99.1 & 44.0 &  62.0 & 80.0 &  62.0 & 56.2 & ... & 45.0 & 54.5 \\
4531.148 & 26.0 & 1.49 & -2.16   & 89.5 & 81.2 & 28.9 &  40.0 & 71.7 & 50.0 & 42.9 & ... & ... & ... \\
4602.941 & 26.0 & 1.49 & -2.21   & 81.9 & 79.1 &  15.0 & 55.0 & 51.6 &  27.0 & 30.0 & ... & ... & 59.1 \\
4871.318 & 26.0 & 2.87 & -0.34   & 104.1 & 76.6 &  ...  & 40.0 & 43.3 & 59.2 &  32.0 & ... & 30.0 & 36.9 \\
4872.138 & 26.0 & 2.88 & -0.60  & 92.8 & 64.9 &  ...  & 44.0 & 26.0 &  21.0 &  20.0 & ... & ... & ... \\

\hline 
\end{tabular} 
\end{table*} 
\begin{table*}
\hspace{-2.5 cm}
\begin{tabular}{lllrrrrrrrrrrr} 
\toprule 
Wave & Elem & \, $\chi$ & loggf &   HD22 & HD12 & C19A &  C19B &  C19C & C19D & C19E  & C19F  & C19G & C19N \\
\hline 
4890.755 & 26.0 & 2.88 & -0.38 &   110.2 & 74.6 &  28.0 & 68.0 & 33.0 & 44.0 & 30.9 & 48.0 & ... & 48.0 \\
4891.492 & 26.0 & 2.85 & -0.14 &    116.0 & 87.4 & 43.3 & 53.0 & 50.9 & 52.0 & 42.2 & 50.0 & 30.0 & 62.6 \\
4918.994 & 26.0 & 2.86 & -0.37 &    105.0 & 77.7 & 31.5 & 58.0 & 36.0 & 45.6 & 25.0  & 32.0 & 30.0 & 40.0 \\
4920.502 & 26.0 & 2.83 & 0.06 &   139.0 & 97.9 & 44.0 & 55.0 & 70.0 & 82.8 & 55.0 & 30.0 & 48.0 & 60.0 \\
4939.687 & 26.0 & 0.86 & -3.25   & 56.5 & 68.0 &  15.0 & 60.0 & 45.0 & 28.0 &  ...  & ... & ... & 48.1 \\
4994.130 & 26.0 & 0.91 & -2.97   & 72.7 & 78.8 &  ...  & 50.0 & 60.2 & 60.0 & 30.0 & ... & ... & ... \\
5006.119 & 26.0 & 2.83 & -0.61   & 92.2 & 67.6 &  20.0 & 50.0 & 37.0 &  26.0 &  22.0 & ... & ... & ... \\
5012.068 & 26.0 & 0.86 & -2.60   & 101.4 & 105.7 & 44.0 & 94.0 & 80.0 & 89.4 & 56.7 & 43.0 & ... & 81.4 \\
5041.756 & 26.0 & 1.48 & -2.20   &  ...  & 86.7 &  30.0 & 45.0 & 53.3 & 84.0 & 33.5 & 33.0 & ... & ... \\
5049.820 & 26.0 & 2.28 & -1.35   & 83.3 & 68.9 &  ...  &  30.0 & 53.0 & 32.9 &  20.0 & ... & ... & ... \\
5051.635 & 26.0 & 0.91 & -2.76   & 88.1 & 94.7 &  34.0 & 60.0 & 82.2 & 92.6 & 39.0 & ... & ... & 75.1 \\
5083.339 & 26.0 & 0.96 & -2.84   & 73.2 & 82.7 &  20.0 & 53.0 & 72.0 &  40.0 & 32.9 & ... & ... & 33.8 \\
5110.413 & 26.0 & 0.0 & -3.76   & 93.4 & 111.4 & 42.1 &  92.0 & 101.3 & 96.0 & 60.4 & 42.0 & ... & 105.0 \\
5123.720 & 26.0 & 1.01 & -3.06   & 76.0 & 72.0 &  ...  & 37.0 & 52.2 & 48.0 &  22.0 & ... & ... & 37.0 \\
5127.360 & 26.0 & 0.91 & -3.25   & 58.6 & 66.6 & 20.0  &  ...  & 55.0 & 54.8 &  27.0 & ... & ... & 30.2 \\
5150.840 & 26.0 & 0.99 & -3.04   & 62.1 & 69.5 &  ...  & 42.0 & 46.7 &  17.0 & 44.0 & ... & ... & 30.0 \\
5166.282 & 26.0 & 0.0 & -4.12   & 60.5 & 89.8 & 20.5 & 77.0 & 77.1 & 65.0 & 37.5 & ... & ... & 36.9 \\
5171.596 & 26.0 & 1.48 & -1.72   & 108.2 & 104.8 & 42.0 & 93.1 & 95.0 & 58.0 & 56.5 & 54.8 & 48.0 & 78.7 \\
5192.344 & 26.0 & 2.99 & -0.42   & 93.7 & 65.5 &  16.0 & 53.2 & 27.9 &  30.0 & 35.8 & ... & ... & \\
5194.940 & 26.0 & 1.56 & -2.09   & 88.4 & 85.0 &  34.0 & 60.2 & 55.3 & 48.0 & 40.8 & ... & 30.0 & 54.1 \\
5216.274 & 26.0 & 1.61 & -2.08   & 79.6 & 78.7 & 19.0  & 52.1 & 55.0 & 52.0 &  27.0 & ... & ... & 32.0 \\
5232.940 & 26.0 & 2.94 & -0.19   & 113.4 & 88.0 & 27.0 & 70.0 & 49.6 & 45.0 & 46.3 & 35.0 & 35.0 & ... \\
5266.555 & 26.0 & 2.99 & -0.49   & 94.7 & 68.7 &  ...  &  ...  & 40.0 & 44.5 & 26.3 & ... & ... & ... \\
5269.537 & 26.0 & 0.86 & -1.33   &  ...  & 170.0 & 96.0 & 147.9 & 170.0 & 130.0 & 138.6 & 97.0 & ... & 149.6 \\
5324.180 & 26.0 & 3.21 & -0.11  & 95.8 & 68.0 &  ...  & 25.0 & 55.7 & 31.5 &  31.0 & ... & 30.0 & 28.6 \\
5328.039 & 26.0 & 0.91 & -1.47   & ...  & 155.0 & 86.0 & 148.2 & 145.0 & 125.0 & 120.6 & 102.0 & ... & 149.4 \\
5341.020 & 26.0 & 1.61 & -1.95   & 95.4 & 91.6 &  24.0 & 65.0 &  ...  & 57.0 & 37.2 & ... & 30.0 & 47.8 \\
5371.489 & 26.0 & 0.96 & -1.64   & ...  & 144.9 & 75.0 & 140.0 & 145.9 & 118.0 & 119.1 & 85.0 & ... & ...\\
5397.128 & 26.0 & 0.91 & -1.98   & 130.2 & 132.2 & 70.0 & 125.0 & 145.0 & 105.6 & 116.2 & 75.0 & ... & ... \\
5405.775 & 26.0 & 0.99 & -1.85   & 132.6 & 133.3 & 78.0 & 104.0 & 120.0 & 100.0 & 102.5 & 63.0 & ... & ... \\
5429.696 & 26.0 & 0.96 & -1.88   & 138.8 & 137.0 & 74.0 & 113.1 & 134.6 & 110.0 & 99.3 & 65.0 & ... & ... \\
5434.524 & 26.0 & 1.01 & -2.13   & 116.4 & 121.6 & 50.0 & 104.0 & 113.5 & 95.0 & 87.4 & 60.0 & ... & ... \\
5497.516 & 26.0 & 1.01 & -2.83   & 89.9 & 88.6 &  20.0 & 60.0 & 78.3 & 51.5 & 40.3 & ... & ... & ... \\
5501.465 & 26.0 & 0.96 & -3.05   & 70.2 & 82.3 & 20.0 & 49.0 & 54.4 & 42.5 &  35.0 & ... & ... & 45.0 \\
5506.779 & 26.0 & 0.99 & -2.79   & 82.8 & 91.8 &  21.0 & 71.0 & 65.9 & 57.0 & 32.4 & ... & ... & 45.0 \\
6494.980 & 26.0 & 2.40 & -1.27   & 83.2 & 75.7 &  21.0 & 42.0 & 39.0 &  ...  & 35.0 & ... & ... & 33.3 \\
8688.624 & 26.0 & 2.17 & -1.20   & 125.0 & 117.7 & 39.0 &  85.0 & 110.0 & 85.8 & 100.0 & ... & 40.0 & ... \\
4178.854 & 26.1 & 2.58 & -2.44   & 99.1 & 55.0 &  ...  &  ...  & 25.0 & 25.0 &  ...  & ... & ... & ... \\
4233.163 & 26.1 & 2.58 & -1.81   &  133.7 & 83.5 & 38.0 & 43.0 & 59.4 & 70.0 & 73.5 & ... & 30.0 & ... \\
4515.334 & 26.1 & 2.84 & -2.36   & 89.1 & 39.7 &  ...  &  ...  &  ...  &  ...  &  ...  & ... & ... & ... \\
4522.628 & 26.1 & 2.84 & -1.99   &  ...  & 60.4 &  25.0 &  30.0 & 50.0 & 30.0 & 25.9 & ... & ... & ... \\
4555.888 & 26.1 & 2.83 & -2.25   & 98.2 & 47.9 &  ...  &  ...  &  25.0 &  35.0  &  ...  & ... & ... & ... \\
4583.829 & 26.1 & 2.81 & -1.74   & 129.5 & 77.4 & 23.0 & 50.0 & 63.0 & 68.0 &  32.0 & 35.0 & 40.0 & 55.0 \\
4923.922 & 26.1 & 2.89 & -1.21   & 170.0 & 105.0 & 43.0 & 65.0 & 66.0 & 70.0 & 58.9 & 60.0 & ... & 80.0 \\
5018.435 & 26.1 & 2.89 & -1.35   &  ...  & 110.2 & 50.0 & 80.0 & 84.6 & 90.0 & 67.7   & 38.0 & 70.0 & 91.0 \\
5234.624 & 26.1 & 3.22 & -2.21   & 87.5 & 39.4 &  ...  &  ...  &  ...  &  ...  &  ...  & ... & ... & ... \\
5275.997 & 26.1 & 3.20 & -1.90   & 101.0 & 45.2 &  ...  &  22.0  &  ...  &  ...  &  ...  & ... & ... & 15.0 \\
%
% 4629.332 & 26.1 & 2.807 & -2.26 & ... & 94.8 & 50.0 &  ...  &  ...  &  ...  &  ...  &  ... & ...  \\
% 5197.568 & 26.1 & 3.23 & -2.05 & ... & 86.2 & 35.9 &  ...  &  ...  &  ...  &  ...  &  ...  & ... \\
\hline 
\end{tabular} 
\end{table*}

\begin{table*}
\caption{Equivalent widths of other (non-iron) lines in C-19 members}
\label{tab:otherlines}
\hspace{-2.5 cm}
\begin{tabular}{lllrrrrrrrrrrr}
\toprule
Wave & Elem & $\chi$ & loggf & HD22 & HD12 & C19A  & C19B & C19C & C19D & C19E & C19F & C19G & C19N \\
\AA  &      & eV     &   &   m\AA  & m\AA  &  m\AA & m\AA & m\AA & m\AA & m\AA & m\AA & m\AA & m\AA \\ 
\hline
5889.951 & 11.0 & 0.00 &  0.12  &  ... & ... & 98.6 & {\it 197.8} & {\it 156.9} & ... & 142.9 & 110.0 & 125.6 & {\it 155.3} \\
5895.924 & 11.0 & 0.00 & -0.18 &  ... & {\it 153.3} & 90.5 & {\it 168.2} & 143.0 & ... & 122.7 & 99.0 & 105.7 & {\it 158.5} \\
3829.355 & 12.0 & 2.71 & -0.23 &  ... & ... & 84.3 & 121.9 & {\it 156.8} & ... & ... & ... & ... & ... \\
4167.271 & 12.0 & 4.34 & -1.00 &  97.2 & 46.5 & ... & ... & ... & 21.0 & ... & ... & ... & ... \\
4571.096 & 12.0 & 0.00 & -5.69 & 62.2 & 82.2 & ... & 21.2 & 55.0 & 52.0 & 14.7 & ... & ... & ... \\
4702.991 & 12.0 & 4.34 & -0.67 &  118.6 & 71.3 & 23.0 & 23.0 & 60.0 & 55.0 & 35.0 & ... & ... & 26.5 \\
5172.684 & 12.0 & 2.71 & -0.40 &  ... & ... & 122.5 & 142.4 & 162.7 & 140.7 & 143.2 & 140.0 & 120.0 & 150.0 \\
5183.604 & 12.0 & 2.71 & -0.18 &  ... & ... & 124.6 & {\it 163.2} & {\it 188.8} & {\it 167.6} & 149.3 & {\it 155.0} & 135.0 & {\it 160.0} \\
5528.405 & 12.0 & 4.34 & -0.62 & 121.7 & 72.4 & 29.9 & 24.0 & 50.0 & 42.0 & 32.0  & ... & ... & 30.6 \\
3944.006 & 13.0 & 0.00 & -0.62 &   {\it 160.6} & {\it 155.0} & 93.0 & 82.1 & 118.1 & 93.4 & 100.0 & 100.0 & ... & ... \\
3961.520 & 13.0 & 0.01 & -0.32 &  139.7 & 133.4 & 73.3 & 89.2 & 108.0 & 101.4 & 72.7 & 120.0 & 100.0 & ... \\
3905.523 & 14.0 & 1.91 & -1.09 &  ... & ... & 118.2 & {\it 152.5} & 128.0 & ... & {\it 153.1} & ... & ... & ... \\
4102.936 & 14.0 & 1.91 & -3.14 &  76.9 & 79.8 & 29.5 & ... & 63.8 & ... & 37.9 & ... & ... & ... \\
7664.911 & 19.0 & 0.00 &  0.13 &  114.5 & ... & ... & ... & 33.7 & ... & 27.7 & ... & ... & ... \\
7698.974 & 19.0 & 0.00 & -0.17 &  80.8 & 43.9 & ... & 33.0 & 30.9 & ... & ... & ... & ... & ... \\
%4226.728 & 20.0 & 0.00 &  0.24 &   ... & ... & ... & 156.2 & 183.1 & 134.8 & 152.9 & ... & ... & 145.3 \\
4283.011 & 20.0 & 1.88 & -0.22 &   92.3 & 61.0 & 45.0 & ... & 35.0 & 42.0 & ... & ... & ... & ... \\
4318.652 & 20.0 & 1.90 & -0.21 &  94.3 & 52.7 & 22.0 & 25.0 & 30.0 & ... & 19.0 & ... & ...& ...  \\
4425.437 & 20.0 & 1.88 & -0.39 & 85.0 & 45.9 & 20.0 & ... & 24.4 & 30.3 & ... & ... & ... & ... \\
4434.957 & 20.0 & 1.88 & -0.03 &   119.1 & 66.4 & 30.4 & ... & 30.0 & ... & ... & 40.0 & ... & ... \\
4435.679 & 20.0 & 1.88 & -0.50 &  ... & 40.2 & ... & ... & ... & 27.0 & 31.2 & ... & ... & ... \\
4454.779 & 20.0 & 1.90 &  0.25 &  146.3 & 76.5 & 30.0 & 45.0 & 35.4 & 65.8 & 28.8 & 50.0 & 50.0 & 35.7 \\
5588.749 & 20.0 & 2.52 &  0.21 &  91.8 & 46.3 & ... & 28.0 & 30.0 & 25.5 & ... & ... & ... & 22.4 \\
6102.723 & 20.0 & 1.88 & -0.89 &  67.4 & 36.9 & ... & 26.2 & ... & ... & ... & ... & ... & ... \\
6122.217 & 20.0 & 1.88 & -0.41 &  104.5 & 63.8 & 15.0 & 25.3 & 40.0 & 51.5 & 24.0 & ... & ... & 31.2 \\
6162.173 & 20.0 & 1.90 &  0.10 &   115.7 & 75.9 & 23.3 & 55.5 & 36.8 & 54.1 & 34.0 & 40.0 & 33.0 & 44.1 \\
6439.075 & 20.0 & 2.52 &  0.47 &  102.9 & 57.3 & 20.0 & 25.0 & 35.0 & 25.1 & ... & 40.0 & 24.0 & 16.0 \\
4246.822 & 21.1 & 0.32 &  0.32 &  {\it 177.7} & 129.5 & 84.9 & 108.7 & 105.9 & ... & 106.5 & 93.0 & 75.8 & 120.6 \\
4314.083 & 21.1 & 0.62 & -0.10 &  143.4 & 114.1 & 43.9 & 88.2 & 81.9 & 86.8 & 59.4 & 49.1 & 75.0 & 90.0 \\
4320.732 & 21.1 & 0.60 & -0.26 &  134.1 & 91.4 & 31.0 & 53.7 & 71.1 & 51.5 & ... & 75.0 & 35.0 & 52.0 \\
4324.996 & 21.1 & 0.59 & -0.44 &  113.1 & 84.9 & ... & 47.6 & 47.0 & 47.2 & 51.8 & 61.2 & 34.0 & ... \\
4374.457 & 21.1 & 0.62 & -0.44 &  109.4 & 80.1 & 28.9 & 52.2 & 48.7 & 43.1 & 45.9 & 33.0  & 30.0 & ... \\
4400.389 & 21.1 & 0.60 & -0.51 &  107.3 & 77.5 & 23.8 & 51.7 & 31.8 & 24.9 & 37.0  & ... & ... & ... \\
4415.557 & 21.1 & 0.59 & -0.64 &  98.8 & 74.6 & 26.4 & 29.2 & 60.0 & 47.2 & 40.1  & ... & ... & ... \\
5031.021 & 21.1 & 1.36 & -0.26 &  64.4 & 40.1 & ... & 26.0 & 35.0 & 25.0 & ...  & ... & ... & ... \\
%5526.790 & 21.1 & 1.77 &  0.13 &  68.8 & 38.9 & ... & ... & ... & 10.7 & 10.0  & ... & ... \\
3989.758 & 22.0 & 0.02 & -0.13 &  108.7 & 64.8 & 22.0 & 22.0 & 22.0 & 71.5 & 22.9  & ... & ... & ... \\
3998.636 & 22.0 & 0.05 &  0.02 &   87.5 & 68.9 & 33.0 & 50.0 & ... & 42.7 & 31.0  & ... & ... & ... \\
4533.239 & 22.0 & 0.85 &  0.54 &  75.5 & 51.5 & 19.0 & 26.6 & 21.5 & ... & 20.8  & ... & ... & 42.0 \\
4681.909 & 22.0 & 0.05 & -1.07 &  32.4 & 27.8 & ... & ... & ... & ... & 16.8  & ... & ... & ... \\
4981.731 & 22.0 & 0.85 &  0.57 &  78.8 & 59.6 & 20.6 & 25.0 & 28.9 & 30.0 & 26.6  & ... & ... & 34.0 \\
4991.066 & 22.0 & 0.83 &  0.45 &  77.1 & 54.2 & 18.0 & 35.0 & 37.0 & 32.0 & 21.0  & ... & ... & 43.0 \\
4999.503 & 22.0 & 0.82 &  0.32 & 76.4 & 47.2 & ... & 24.6 & 24.0 & 29.0 & ...  & ... & ... & 28.6 \\
5014.276 & 22.0 & 0.81 &  0.04 &  72.3 & 58.7 & ... & ... & 25.0 & 24.5 & ...  & ... & ... & 20.0 \\
5192.969 & 22.0 & 0.02 & -0.95 &  61.2 & 35.9 & ... & ... & ... & ... & 20.0  & ... & ... & ... \\
5210.384 & 22.0 & 0.05 & -0.82 &  44.6 & 41.6 & 21.5 & 22.0 & ... & ... & ...  & ... & ... & ... \\
\hline
\end{tabular}
\end{table*}
\begin{table*}
\hspace{-2.5 cm}
\begin{tabular}{lllrrrrrrrrrrr}
\toprule
Wave & Elem & $\chi$ & loggf &  HD22 & HD12 & C19A  & C19B & C19C & C19D & C19E & C19F & C19G & C19N \\
\hline
3913.461 & 22.1 & 1.11 & -0.36 &  ... & 130.6 & 62.5 & 100.0 & 86.2 & 112.7 & 102.4 & ... & ... & ... \\
4012.384 & 22.1 & 0.57 & -1.78 &  ... & 99.6 & 70.7 & 70.0 & ... & ... & 73.3 & 60.0 & 50.8 & ... \\
4028.338 & 22.1 & 1.89 & -0.92 &  107.5 & 51.6 & 50.0 & 29.0 & 29.7 & 30.0 & 22.0 & ... & ... & ... \\
4290.215 & 22.1 & 1.16 & -0.87 &  {\it 166.7} & 96.0 & 45.0 & 88.0 & 83.0 & 60.0 & 65.0 & 61.5 & 34.1 & ... \\
4300.042 & 22.1 & 1.18 & -0.46 &  {\it 151.6} & 105.1 & 54.8 & 80.0 & 73.8 & 86.0 & ... & 99.0 & 55.4 & 90.0 \\
4337.915 & 22.1 & 1.08 & -1.13 &  143.7 & 100.5 & 47.6 & 61.7 & 75.3 & 67.5 & 74.2 & 70.8 & 44.7 & 117.1 \\
4394.059 & 22.1 & 1.22 & -1.77 &  90.4 & 62.5 & 25.0 & 19.0 & 44.3 & 45.0 & 20.0 & ... & ... & ... \\
4395.031 & 22.1 & 1.08 & -0.54 &  ... & 124.5 & 90.0 & 105.0 & 95.3 & 101.0 & 92.8 & 78.5 & 61.8 & ... \\
4395.839 & 22.1 & 1.24 & -1.93 &  82.3 & 49.4 & ... & ... & 18.0 & 16.1 & 15.0 & ... & ... & ... \\
4399.765 & 22.1 & 1.24 & -1.20 &  123.6 & 87.5 & 40.0 & 68.8 & 57.2 & 55.1 & 50.0 & 40.0 & 57.0 & 80.0 \\
4417.719 & 22.1 & 1.16 & -1.43 &  133.5 & 93.7 & 52.0 & 70.0 & 75.0 & 43.8 & 49.4 & ... & 31.5 & 50.2 \\
4418.331 & 22.1 & 1.24 & -1.99 &  76.8 & 49.2 & ... & 30.0 & 30.0 & 26.0 & 20.0 & ... & ... & ... \\
4443.801 & 22.1 & 1.08 & -0.71 &  {\it 169.2} & 117.4 & 62.9 & 107.1 & 76.1 & 80.0 & 78.5 & 115.0 & 66.1 & 92.1 \\
4450.482 & 22.1 & 1.08 & -1.52 &  119.9 & 84.1 & 30.2 & 54.6 & 55.3 & 47.5 & 43.3 & 35.0 & ... & 65.0 \\
4464.450 & 22.1 & 1.16 & -2.08 &  97.9 & 65.2 & 25.0 & 32.6 & 42.2 & 17.3 & 25.0 & ... & ... & ... \\
4468.493 & 22.1 & 1.13 & -0.63 &  {\it 174.6} & 119.0 & 61.7 & 94.5 & 90.5 & 106.1 & 82.6 & 75.0 & 75.3 & 137.6 \\
4470.857 & 22.1 & 1.16 & -2.28 &  75.8 & 46.8 & ... & 24.8 & 26.2 & 25.0 & ... & ... & ... & 36.0 \\
4501.270 & 22.1 & 1.11 & -0.77 &  {\it 162.8} & 115.7 & 62.6 & 91.3 & 77.6 & 81.3 & 75.8 & 110.0 & 89.0 & 80.0 \\
4533.969 & 22.1 & 1.24 & -0.77 &  ... & 119.8 & 62.4 & 88.7 & 79.2 & 65.5 & 77.8 & 72.0 & 59.8 & 88.0 \\
4563.761 & 22.1 & 1.22 & -0.96 &  {\it 154.6} & 108.4 & 56.2 & 76.1 & 85.3 & 70.3 & 70.6 & 75.0 & 42.8 & 57.3 \\
4571.971 & 22.1 & 1.57 & -0.31 &  {\it 168.4} & 108.2 & 55.7 & 55.6 & 83.0 & 72.7 & 74.3 & 85.0 & 49.4 & 85.0 \\
5129.156 & 22.1 & 1.89 & -1.34 &  77.3 & 39.6 & 17.0 & 30.0 & 15.0 & 12.0 & ... & ... & ... & ... \\
5154.070 & 22.1 & 1.56 & -1.92 &  71.8 & 42.5 & ... & 25.0 & 27.0 & 21.6 & ... & ... & ... & 20.0 \\
5188.680 & 22.1 & 1.58 & -1.21 &  135.2 & 78.2 & 25.0 & 52.5 & 66.4 & 40.6 & 26.7 & ... & ... & 40.0 \\
5226.543 & 22.1 & 1.56 & -1.30 &  107.3 & 70.0 & 35.0 & 44.2 & 36.4 & 35.0 & ... & ... & ... & 22.2 \\
4254.352 & 24.0 & 0.00 & -0.09 &  135.9 & 108.7 & 58.0 & 97.2 & 76.9 & 73.0 & 66.5 & 65.0 & 55.5 & 75.0 \\
4274.812 & 24.0 & 0.00 & -0.22 &  133.3 & 107.9 & 41.5 & 70.0 & 69.5 & 76.9 & 68.0 & 65.0 & 63.0 & 75.0 \\
4289.731 & 24.0 & 0.00 & -0.37 &  131.5 & 97.9 & 41.6 & 47.1 & 60.6 & ... & 47.6 & ... & 56.9 & 80.0 \\
5206.023 & 24.0 & 0.94 &  0.02 &  112.4 & 82.6 & 24.8 & 40.7 & 69.9 & 48.4 & 41.0 & 40.0 & 26.0 & 43.1 \\
5208.409 & 24.0 & 0.94 &  0.17 &  ... & 95.0 & 47.3 & 61.5 & 76.3 & 77.4 & 39.0 & 30.0 & 63.0 & 70.0 \\
%5296.691 & 24.0 & 0.98 & -1.36 &  26.0 & 15.5 & ... & ... & ... & 11.7 & ... & ... & ... & ... \\
%5298.272 & 24.0 & 0.98 & -1.14 &  39.8 & 22.8 & ... & ... & ... & 34.8 & ... & ... & ... & ... \\
5409.783 & 24.0 & 1.03 & -0.67 &  59.9 & 41.3 & ... & ... & 22.0 & 30.4 & ... & ... & ... & ... \\
4030.753 & 25.0 & 0.00 & -0.50 &  {\it 170.3} & 134.4 & 30.0 & 60.2 & 87.2 & 105.7 & 62.5 & ... & 93.0 & ... \\
4033.062 & 25.0 & 0.00 & -0.65 &  139.0 & 120.8 & 23.0 & 35.0 & 51.5 & 71.9 & 40.8 & ... & 53.3 & 100.0 \\
4034.483 & 25.0 & 0.00 & -0.84 &  115.0 & 112.7 & 20.0 & ... & 51.6 & 68.8 & 46.4 & ... & 36.1 & 70.0 \\
%4783.42 & 25.0 & 2.30 & 0.04 &  23.0 & ... & ... & ... & ... & ... & ... & ... & ... \\
%4823.52 & 25.0 & 2.32 & 0.14 &  25.0 & ... & ... & ... & ... & ... & ... & ... & ... \\ 
3845.461 & 27.0 & 0.92 &  0.01 &  79.5 & 87.6 & 27.0 & 37.0 & 45.8 & 37.0 & 65.0 & ... & ... & ... \\
3894.073 & 27.0 & 1.05 &  0.10 &  ... & ... & 29.9 & 45.0 & 46.3 & 41.6 & 38.2 & ... & ... & ... \\
3995.302 & 27.0 & 0.92 & -0.22 &  88.9 & 80.8 & 31.9 & 50.0 & 43.5 & 58.5 & 50.0 & ... & 40.0 & ... \\
4121.311 & 27.0 & 0.92 & -0.32 &  84.9 & 89.7 & 28.6 & 28.4 & 41.8 & 68.8 & 39.9 & ... & ... & ... \\
3807.140 & 28.0 & 0.42 & -1.23 &  123.2 & 111.8 & 30.0 & ... & 94.0 & 60.0 & ... & ... & ... & ... \\
3858.297 & 28.0 & 0.42 & -0.96 &  145.4 & 121.2 & 38.0 & 54.9 & 76.5 & 53.9 & ... & ... & ... & ... \\
%4980.17  & 28.0 & 3.60 & 0.07 &  32.9 & 15.0 & ... & ... & ... & ... & ... & ... &  ... \\
5476.904 & 28.0 & 1.82 & -0.78 &  93.5 & 78.8 & 21.6 & 45.0 & 39.6 & 49.2 & 17.0 & $<$25 & $<$25 & 35.0\\
4077.714 & 38.1 & 0.00 &  0.15 &  ... & 158.5 & 73.6 & 116.2 & 109.6 & 103.3 & 112.0 & 120.0 & 104.8 & 142.9 \\
4215.524 & 38.1 & 0.00 & -0.17 &  ... & 149.4 & 76.8 & 127.6 & 125.1 & 110.4 & 95.9 & 95.0 & 89.1 & 130.0 \\
%3982.600 & 39.1 & 0.13 & -0.49 &  57.2 & ... & ... & ... & 14.1 & ... & ... & ... & ... \\
%4496.962 & 40.1 & 0.71 & -0.89 &  50.2 & ... & 25.0 & ... & 24.1 & ... & ... & ... & ... \\
4554.029 & 56.1 & 0.00 &  0.17 &  ... & 96.1 & 30.2 & 50.0 & 58.1 & 51.2 & 52.4 & ... & 35.0 & 52.7 \\
4934.076 & 56.1 & 0.00 & -0.15 &  ... & 89.9 & 19.0 & 54.4 & 53.1 & 71.3 & ... & $<$35 & 18.0 & 58.3 \\
6141.713 & 56.1 & 0.70 & -0.08 &  {\it 160.8} & 39.0 & ... & 33.0 & 40.0 & 25.9 & ... & ... & ... & 35.0 \\
6496.897 & 56.1 & 0.60 & -0.38 &  ... & 33.3 & ... & 25.0 & 16.0 & 12.6 & ... & ... & 15.0 & 12.0 \\
%4433.890 & 62.1 & 0.43 & -0.19 &  ... & ... & ... & 10.0 & 12.5 & ... & ... & ... & ... \\
4129.725 &  63.1 &  0.00 &  0.20 &  {\it 180.0} & $<$10 & $<$30 & $<$40 & $<$40 & $<$60 & $<$40 & $<$60 & $<$60 & $<$30 \\
\textit{8498.023} &  20.1 &  1.69 & -1.31 & ... & ... & \textit{181.0} & \textit{409.0} & \textit{462.0} & \textit{407.0} & \textit{338.0} & \textit{160.0} & \textit{290.0} & ... \\
\textit{8542.091} &  20.1 &  1.70 & -0.36 & ... & ... & \textit{493.0} & \textit{823.0} & \textit{827.0} & \textit{733.0} & \textit{563.0} & \textit{470.0} & \textit{507.0} & ... \\
\textit{8662.141} &  20.1 &  1.69 & -0.62 & ... & ... & \textit{422.0} & \textit{666.0} & \textit{681.0} & \textit{645.0} & \textit{504.0} & \textit{396.0} & \textit{419.0} & ...\\

\hline
\vspace{-2ex}
\end{tabular}
% TODAY: 16 Feb 2025 : added C19F
\end{table*}

\end{document}